\documentclass[%
 reprint,
 nofootinbib,
 amsmath,amssymb,
 aps,
]{revtex4-2}

\usepackage{graphicx}
\usepackage{dcolumn}
\usepackage{bm}
\usepackage[hidelinks, colorlinks=true]{hyperref}

\newcommand{\sout}{\bgroup \color{red} \ULdepth=-.5ex \ULset}

\renewcommand{\Re}{\operatorname{Re}}
\renewcommand{\Im}{\operatorname{Im}}

\usepackage{siunitx}
\usepackage{times}

\allowdisplaybreaks[1]

\begin{document}


\title{Model analysis on the effectiveness of the HAL QCD method for
  hadron-hadron interactions}

\author{Takayasu Sekihara} 
\email{sekihara@kpu.ac.jp}
\affiliation{Graduate School of Life and Environmental Sciences,
  Kyoto Prefectural University, Sakyo-ku, Kyoto 606-8522, Japan}

\author{Kei Fujiwara} 
\affiliation{Graduate School of Life and Environmental Sciences,
  Kyoto Prefectural University, Sakyo-ku, Kyoto 606-8522, Japan}

\date{\today}

\begin{abstract}

  The HAL QCD method has been one of the powerful tools to extract
  hadron-hadron interactions directly from lattice QCD simulation data.
  In this paper, we aim at examining the effectiveness of the HAL QCD
  method by deriving a formula to calculate quantities in the HAL QCD
  method, such as the so-called R-correlators and HAL QCD local
  potentials, from the hadron-hadron scattering amplitudes within
  effective models.  In this framework, we can judge whether the HAL
  QCD local potential, evaluated in the present formula, reproduces
  the properties of the original hadron-hadron interaction or not via
  the scattering amplitude, which is a solution of the
  Lippmann--Schwinger equation with the original hadron-hadron
  interaction as an input.
  In an analysis within a simple model of elastic scattering, we show
  that, when the original interaction is predominantly local, the HAL
  QCD local potentials quantitatively reproduce phase shifts of the
  hadron-hadron scatterings and correctly indicate the
  existence/absence of the bound state with its binding energy $\sim
  \si{MeV}$.  Higher-order corrections in the derivative expansion of
  the HAL QCD nonlocal potentials can provide a better description of
  the binding energies of the bound states.  The lattice
  discretization and finite-volume effects of spacetime modify the
  results only slightly, as long as their scale is not comparable to
  the interaction range.
  Furthermore, we consider the $\bar{K} N$ potential in a bare to
  $\bar{K} N$ transition amplitude, which shows singular behavior
  around the origin in the recent HAL QCD results of the lattice QCD
  simulation data, and discuss the cause of such singular behavior in
  the HAL QCD method through our model analysis of the $\bar{K} N$
  scattering.
  
\end{abstract}

\maketitle


\section{Introduction}

Hadron-hadron interaction has been one of the most important topics in
hadron physics, because it contains information on the quark dynamics
inside hadrons, which can serve as a hint to understanding the
fundamental theory of strong interaction, quantum chromodynamics
(QCD).  Besides recent model analyses, e.g.,~\cite{Sekihara:2023ihc},
the hadron-hadron interaction has been studied in lattice QCD
simulations.  In this line, the HAL QCD method~\cite{Ishii:2006ec,
  Aoki:2009ji, Ishii:2012ssm, Aoki:2012tk} and the Hamiltonian
effective field theory~\cite{Hall:2013qba, Hall:2014uca, Liu:2015ktc,
  Liu:2016wxq, Liu:2016uzk, Wu:2017qve, Abell:2023nex, Liu:2023xvy,
  Hockley:2024ipz, Han:2025gkp} have been powerful tools to
investigate hadron-hadron interactions.  In particular, the HAL QCD
method enables us to directly extract hadron-hadron potentials via
hadron-hadron correlations: for example, $N
\Omega$~\cite{HALQCD:2018qyu}, $\Lambda \Lambda$-$N
\Xi$~\cite{HALQCD:2019wsz}, $N \phi$~\cite{Lyu:2022imf}, $\Omega
_{ccc} \Omega _{ccc}$~\cite{Lyu:2021qsh}, $D^{\ast} D$ for
$T_{cc}^{+}$~\cite{Lyu:2023xro},
nucleon-charmonium~\cite{Lyu:2024ttm}, $K N$~\cite{Murakami:2025owk},
and $\bar{D} N$~\cite{Yamada:2026bfl}.  Furthermore, improvements in
experiments provide opportunities for the study of various
hadron-hadron interactions.  Traditionally, scattering experiments
have been one of the most important ways, and in this line the
hadron-hadron interaction was experimentally investigated in, e.g.,
Ref.~\cite{J-PARCE40:2021qxa}.  In addition, experimentalists recently
use the so-called femtoscopy~\cite{STAR:2014dcy, STAR:2018uho,
  ALICE:2020mfd, ALICE:2019gcn, ALICE:2021cpv, ALICE:2023wjz,
  ALICE:2022yyh, ALICE:2025kma} to extract information on the
hadron-hadron interaction via the correlation of the outgoing hadrons
from relativistic ion collisions (see also theoretical
studies~\cite{Lednicky:1981su, Morita:2014kza, Morita:2016auo,
  Morita:2019rph, Kamiya:2019uiw}).

In the present study, we focus on the HAL QCD method.  The HAL QCD
method provides various hadron-hadron potentials that can be accessed
in experiments as well.  For example, both the $p \Xi
^{-}$~\cite{HALQCD:2019wsz} and $p \Omega ^{-}$~\cite{HALQCD:2018qyu}
potentials in the HAL QCD analyses were used to compare with the
experimental data of relativistic ion collisions via
femtoscopy~\cite{ALICE:2020mfd}.  The consistency with the
$T_{cc}^{+}$ data from LHCb~\cite{LHCb:2021auc} was discussed through
the $D^{\ast} D$ potential in the HAL QCD method~\cite{Lyu:2023xro}.
The resulting consistencies strongly indicate that the HAL QCD method
is indeed reliable enough to evaluate hadron-hadron potentials.
However, in some cases, the HAL QCD method shows discrepancies with
experimental data.  The HAL QCD $K N$
potential~\cite{Murakami:2025owk} produces a repulsive interaction
that is almost quantitatively consistent with experiments, but the
values of the $K^{+} p$ phase shifts above $p_{\rm lab} =
\SI{300}{MeV}/c$ obtained from the HAL QCD local potential deviate
from the experimental data.  Furthermore, the HAL QCD $\bar{K} N$
potential in Refs.~\cite{Murakami:2023phq, Murakami:2025oig}, which
was evaluated from the three-point correlation function of the
negative-parity $\Lambda$ baryon source operator and meson-baryon sink
operators, shows singular behavior around the origin.  Because the HAL
QCD method works fairly well in several cases, it is necessary to
develop a way to judge whether the HAL QCD method is applicable or not
to a certain hadron-hadron scattering system.

Our motivation in the present study originates precisely from this
point.  In the following, we derive a formula to calculate quantities
in the HAL QCD method, such as the so-called R-correlators and HAL QCD
local potentials, from hadron-hadron scattering amplitudes.  With this
formula, one can calculate the R-correlators and HAL QCD local
potentials with their own scattering amplitude, which will serve as
a good basis for how the properties of the hadron-hadron interaction
in their model are reflected in the HAL QCD local potential.  We note
that analyses on the effectiveness of the HAL QCD method have been
done in, e.g., Refs.~\cite{Gongyo:2018gou, HALQCD:2018gyl,
  Aoki:2019gqt, Aoki:2021ahj}.  In Ref.~\cite{Gongyo:2018gou}, they
discussed the asymptotic behavior of the so-called
Nambu--Bethe--Salpeter wave function, especially focusing on the
theoretical treatment of bound states.  In Refs.~\cite{HALQCD:2018gyl,
  Aoki:2019gqt, Aoki:2021ahj}, they mainly discussed the effectiveness
of the derivative expansion in the HAL QCD method.  Compared to these
preceding studies, our study provides a formulation from a more global
viewpoint, which enables anyone to calculate HAL QCD quantities in all
ranges by using their own interaction and scattering amplitude in a
model.

The paper is organized as follows.  In Sec.~\ref{sec:HAL}, we describe
the details of the quantities in the HAL QCD method: the correlation
functions, so-called R-correlators, derivative expansion of the HAL
QCD nonlocal potentials, and HAL QCD local potentials.  In this
section, we also derive the formula to calculate the quantities in the
HAL QCD method from hadron-hadron scattering amplitudes.  Next, in
Sec.~\ref{sec:amp}, we formulate our effective model to calculate the
hadron-hadron scattering amplitudes.  In this model, in addition to
meson-baryon elastic scattering, we include the transition from a bare
excited baryon state to the meson-baryon two-body state.  Through this
mixing of the bare and meson-baryon states, we can examine whether we
can evaluate the meson-baryon potential in the HAL QCD method or not,
even when the bound state is a superposition of the bare and
meson-baryon molecular states.  Then, in Sec.~\ref{sec:modelA}, we
focus on meson-baryon elastic scattering and numerically examine
whether or not the HAL QCD method can reproduce the original
meson-baryon interactions by using the formula derived in
Sec.~\ref{sec:HAL}.  In Sec.~\ref{sec:modelB}, keeping the recent HAL
QCD results on the $\bar{K} N$ potential~\cite{Murakami:2023phq,
  Murakami:2025oig} in mind, we extend our discussion to the case of
the transition from a bare state to the meson-baryon two-body state
and try to extract the local meson-baryon potentials in this case.
Section~\ref{sec:sum} is devoted to the summary and outlook of the
present study.

\section{Correlation functions and HAL QCD method}
\label{sec:HAL}

\begin{figure}[b]
  \centering
  \includegraphics[scale=0.95]{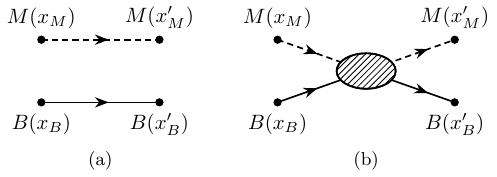}
  \caption{Diagrams for the correlation function of the meson ($M$)
    -baryon ($B$) elastic scattering: (a) disconnected and (b)
    connected.}
  \label{fig:Corr_MB}
\end{figure}
\begin{figure}[b]
  \centering
  \includegraphics[scale=0.95]{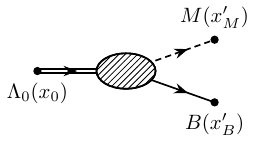}
  \caption{Diagram for the correlation function of the bare state
    ($\Lambda _{0}$) to the meson ($M$) -baryon ($B$) scattering
    state.}
  \label{fig:Corr_bare}
\end{figure}

First, we explain what quantities are evaluated in the HAL QCD method.
For simplicity, throughout this study we focus on $s$-wave meson ($M$)
-baryon ($B$) scattering in the center-of-mass frame and introduce the
interaction between them, while we neglect couplings to other
meson-baryon channels.  In this case, the HAL QCD method measures the
correlation in Fig.~\ref{fig:Corr_MB}.  Furthermore, we allow the
$s$-wave meson-baryon system to couple to one bare excited baryon
state $\Lambda _{0}$, which enables us to consider the transition from
a bare state $\Lambda _{0}$ to the meson-baryon scattering state.
This bare state $\Lambda _{0}$ corresponds to the negative-parity
$\Lambda$ baryon source operator in Refs.~\cite{Murakami:2023phq,
  Murakami:2025oig}.  In the present framework, the physical
negative-parity $\Lambda$ is a superposition of the bare $\Lambda
_{0}$ and meson-baryon molecular states.  In this $\Lambda _{0}$-$M B$
transition case, the HAL QCD method measures the three-point
correlation in Fig.~\ref{fig:Corr_bare}.  The extension to general
hadron-hadron scatterings is straightforward.

In this paper, we mainly employ the semirelativistic formulation in
which, for example, the meson propagator of mass $m_{M}$ has the
following form:
\begin{equation}
  \langle 0 | \mathrm{T} \phi _{M} ( x^{\prime} )
  \phi _{M}^{\dagger} ( x ) | 0 \rangle
  = \int \frac{d^{4} p}{( 2 \pi )^{4}} 
  \frac{m_{M}}{\mathcal{E}_{M} ( p ) }
  \frac{i e^{- i p \cdot ( x^{\prime} - x )}}{p^{0}
    - \mathcal{E}_{M} ( p ) + i 0} ,
  \label{eq:prop}
\end{equation}
where $\phi _{M} ( x )$ is the meson field, T means to take the time-ordered
product, and $\mathcal{E}_{M} ( p ) \equiv \sqrt{m_{M}^{2} +
  \bm{p}^{2}}$.  We use the same form of propagator for the baryon,
because the baryon spin is irrelevant in $s$-wave scatterings.
Furthermore, for simplicity, we neglect the internal structure of each
hadron in the present study, in contrast to lattice QCD
simulations in which hadrons are composed of products of quark
operators.

Here we emphasize that in our framework, besides the bare excited
baryon state $\Lambda _{0}$, we take into account the elastic
meson-baryon scattering with all possible values of relative momentum.
In this sense, our framework is compatible with the so-called
time-dependent HAL QCD method developed in Ref.~\cite{Ishii:2012ssm}.
On the other hand, we do not consider inelastic contributions such as
the two-meson--baryon system $M M B$ or systems with excited hadrons
($M^{\ast} B$, $M B^{\ast}$, and $M^{\ast} B^{\ast}$), which are
automatically included in actual lattice QCD simulations.  In
principle, one could incorporate these contributions by extending the
model space in the present framework.

Below, we first derive the formula in continuum spacetime, and
then extend it to lattice spacetime.  Throughout this study,
continuum spacetime has infinite volume, while lattice spacetime
has finite volume with periodic boundary conditions, as done in
lattice simulations.

\subsection{Meson-baryon elastic scattering}

The key quantity of the HAL QCD method is the correlation function of
meson-baryon scattering, as depicted in Fig.~\ref{fig:Corr_MB}:
\begin{align}
  C ( t , r ) \equiv & \int d^{3} x_{M} \int d^{3} x_{B}
  \langle 0 | \phi _{B} ( x_{B}^{\prime} ) \phi _{M} ( x_{M}^{\prime} )
  \notag \\
  & \times
  \phi _{M}^{\dagger} ( x_{M} ) \phi _{B}^{\dagger} ( x_{B} ) | 0 \rangle .
  \label{eq:Corr_MB}
\end{align}
This means that the meson and baryon are created at $x_{M}^{\mu} = (
t_{0} , \bm{x}_{M} )$ and $x_{B}^{\mu} = ( t_{0} , \bm{x}_{B} )$,
respectively, and are annihilated at $x_{M}^{\prime \mu} = ( t_{0} + t
, \bm{x}_{M}^{\prime} )$ and $x_{B}^{\prime \mu} = ( t_{0} + t ,
\bm{x}_{B}^{\prime} )$, respectively.  We assume that the creations
(and similarly the annihilations) take place at the same time, the
annihilations take place after the creation, $t > 0$, and the spatial
distance of the two annihilation points, $r \equiv | \bm{r} |$ with
$\bm{r} \equiv \bm{x}_{M}^{\prime} - \bm{x}_{B}^{\prime}$, is fixed.
In addition, we perform the $\bm{x}_{M}$ and $\bm{x}_{B}$ integrals.
This is motivated by the so-called wall source for quarks used in
lattice QCD simulations, and we extend this idea to hadrons.

The correlation function~\eqref{eq:Corr_MB} can be evaluated in the
standard way of quantum field theory.  In particular, elastic
scattering brings disconnected [Fig.~\ref{fig:Corr_MB}(a)] and
connected [Fig.~\ref{fig:Corr_MB}(b)] contributions to the correlation
function:
\begin{equation}
  C ( t ,r )
  = C ( t , r )_{\rm disc} + C ( t , r )_{\rm conn} .
\end{equation}

Because the disconnected contribution becomes the product of the meson and
baryon propagators, we can easily evaluate it as
\begin{align}
  & C ( t , r )_{\rm disc}
  \notag \\
  & = \int d^{3} x_{M} \int \frac{d^{4} p_{M}}{( 2 \pi )^{4}} 
  \frac{m_{M}}{\mathcal{E}_{M} ( p_{M} ) }
  \frac{i e^{- i p_{M} \cdot ( x_{M}^{\prime} - x_{M} )}}{p_{M}^{0}
    - \mathcal{E}_{M} ( p_{M} ) + i 0}
  \notag \\
  & \phantom{=} \times \int d^{3} x_{B} \int \frac{d^{4} p_{B}}{( 2 \pi )^{4}} 
  \frac{m_{B}}{\mathcal{E}_{B} ( p_{B} ) }
  \frac{i e^{- i p_{B} \cdot ( x_{B}^{\prime} - x_{B} )}}{p_{B}^{0}
    - \mathcal{E}_{B} ( p_{B} ) + i 0} 
  \notag \\
  & = \int \frac{d p_{M}^{0}}{2 \pi}
  \frac{i e^{- i p_{M}^{0} t}}{p_{M}^{0} - m_{M} + i 0}
  \int \frac{d p_{B}^{0}}{2 \pi}
  \frac{i e^{- i p_{B}^{0} t}}{p_{B}^{0} - m_{B} + i 0}
  \notag \\
  & = e^{- i M_{\rm th} t} ,
\end{align}
where $m_{B}$ is the baryon mass, $M_{\rm th} \equiv m_{M} + m_{B}$ is
the threshold energy of the system, and we performed the coordinate integral
\begin{equation}
  \int d^{3} x_{M} e^{- i \bm{p}_{M} \cdot \bm{x}_{M}}
  = ( 2 \pi )^{3} \delta ^{3} ( \bm{p}_{M} )
\end{equation}
for the meson, and similarly for the baryon.

As for the connected diagram, on the other hand, we have the following
expression according to the standard technique of quantum field
theory:
\begin{widetext}
\begin{align}
  C ( t , r )_{\rm conn}
  = & \int d^{3} x_{M} \int d^{3} x_{B} \int d^{4} x
  \int \frac{d^{4} p_{M}}{( 2 \pi )^{4}}
  \frac{m_{M}}{\mathcal{E}_{M} ( p_{M} ) }
  \frac{i e^{- i p_{M} \cdot ( x - x_{M} )}}{p_{M}^{0}
    - \mathcal{E}_{M} ( p_{M} ) + i 0}
  \int \frac{d^{4} p_{B}}{( 2 \pi )^{4}} 
  \frac{m_{B}}{\mathcal{E}_{B} ( p_{B} ) }
  \frac{i e^{- i p_{B} \cdot ( x - x_{B} )}}{p_{B}^{0}
    - \mathcal{E}_{B} ( p_{B} ) + i 0}
  \notag \\
  & \times
  \int \frac{d^{4} p_{M}^{\prime}}{( 2 \pi )^{4}}
  \frac{m_{M}}{\mathcal{E}_{M} ( p_{M}^{\prime} ) }
  \frac{i e^{- i p_{M}^{\prime} \cdot ( x_{M}^{\prime} - x )}}{p_{M}^{\prime 0}
    - \mathcal{E}_{M} ( p_{M}^{\prime} ) + i 0}
  \int \frac{d^{4} p_{B}^{\prime}}{( 2 \pi )^{4}}
  \frac{m_{B}}{\mathcal{E}_{B} ( p_{B}^{\prime} ) }
  \frac{i e^{- i p_{B}^{\prime} \cdot ( x_{B}^{\prime} - x )}}{p_{B}^{\prime 0}
    - \mathcal{E}_{B} ( p_{B}^{\prime} ) + i 0}
  [ - i T ( p_{M}^{0} + p_{B}^{0} , | \bm{p}_{M} | , | \bm{p}_{M}^{\prime} | ) ] .
\end{align}
Here, $x^{\mu}$ is the coordinate where the scattering takes place, and
$T$ is the meson-baryon scattering amplitude whose variables are the
center-of-mass energy $p_{M}^{0} + p_{B}^{0}$ and the absolute values of
the relative momenta in the initial state $| \bm{p}_{M} |$ and in the
final state $| \bm{p}_{M}^{\prime} |$.  Due to the $s$-wave nature,
the scattering amplitude does not depend on the angles of the momenta.
In Sec.~\ref{sec:amp}, we formulate the scattering amplitude in our
model.  Then, as in the disconnected case, we perform the integrals
with respect to the coordinates and simplify the expression as
\begin{align}
  C ( t , r )_{\rm conn}
  = & \int \frac{d p_{M}^{0}}{2 \pi}
  \frac{i e^{i p_{M}^{0} t_{0}}}{p_{M}^{0} - m_{M} + i 0}
  \int \frac{d p_{B}^{0}}{2 \pi}
  \frac{i e^{i p_{B}^{0} t_{0}}}{p_{B}^{0} - m_{B} + i 0}
  \int \frac{d^{4} p_{M}^{\prime}}{( 2 \pi )^{4}}
  \frac{m_{M}}{\mathcal{E}_{M} ( p_{M}^{\prime} ) }
  \frac{i e^{- i p_{M}^{\prime} \cdot x_{M}^{\prime}}}{p_{M}^{\prime 0}
    - \mathcal{E}_{M} ( p_{M}^{\prime} ) + i 0}
  \notag \\
  & \times
  \int \frac{d^{4} p_{B}^{\prime}}{( 2 \pi )^{4}}
  \frac{m_{B}}{\mathcal{E}_{B} ( p_{B}^{\prime} ) }
  \frac{i e^{- i p_{B}^{\prime} \cdot x_{B}^{\prime}}}{p_{B}^{\prime 0}
    - \mathcal{E}_{B} ( p_{B}^{\prime} ) + i 0}
  [ - i T ( p_{M}^{0} + p_{B}^{0} , 0 , | \bm{p}_{M}^{\prime} | ) ]
  ( 2 \pi )^{4} \delta ^{4} ( p_{M}^{\prime} + p_{B}^{\prime} - p_{M} - p_{B} )
  \notag \\
  = & \int \frac{d p_{M}^{0}}{2 \pi}
  \frac{i e^{i p_{M}^{0} t_{0}}}{p_{M}^{0} - m_{M} + i 0}
  \int \frac{d p_{B}^{0}}{2 \pi}
  \frac{i e^{i p_{B}^{0} t_{0}}}{p_{B}^{0} - m_{B} + i 0}
  \notag \\
  & \times \int \frac{d^{4} p^{\prime}}{( 2 \pi )^{4}}
  \frac{m_{M} m_{B}}{\mathcal{E}_{M} ( p^{\prime} )
    \mathcal{E}_{B} ( p^{\prime} ) }
  \frac{i e^{- i ( p_{M}^{0} + p_{B}^{0} ) ( t + t_{0} )}
    e^{i \bm{p}^{\prime} \cdot \bm{r}} T ( p_{M}^{0} + p_{B}^{0} , 0 , | \bm{p}^{\prime} | )}{[ p^{\prime 0}
      - \mathcal{E}_{M} ( p^{\prime} ) + i 0 ]
  [ p_{M}^{0} + p_{B}^{0} - p^{\prime 0}
    - \mathcal{E}_{B} ( p^{\prime} ) + i 0 ]} ,
\end{align}
\end{widetext}
where the integral variable $p_{M}^{\prime \mu}$ was renamed as
$p^{\prime \mu}$.  Note that, owing to the wall source for hadrons,
both the initial meson and baryon have zero momentum in the present
formulation.  To further perform the integrals, we introduce the
center-of-mass energy $W \equiv p_{M}^{0} + p_{B}^{0}$ and integrate
by substitution:
\begin{align}
  & \int \frac{d p_{M}^{0}}{2 \pi}
  \frac{i e^{- i p_{M}^{0} t}}{p_{M}^{0} - m_{M} + i 0}
  \int \frac{d p_{B}^{0}}{2 \pi}
  \frac{i e^{- i p_{B}^{0} t}}{p_{B}^{0} - m_{B} + i 0}
  \notag \\
  & = \int \frac{d W}{2 \pi}
  \int \frac{d p_{B}^{0}}{2 \pi}
  \frac{i^{2} e^{- i W t}}{[ W - p_{B}^{0} - m_{M}
      + i 0 ][ p_{B}^{0} - m_{B} + i 0 ]}
  \notag \\
  & = \int \frac{d W}{2 \pi}
  \frac{i e^{- i W t}}{W - M_{\rm th} + i 0} ,
\end{align}
and hence, together with performing the $p^{\prime 0}$ integral, we
have
\begin{align}
  & C ( t , r )_{\rm conn}
  \notag \\
  & = \int _{- \infty}^{\infty} \frac{d W}{2 \pi}
  \frac{i e^{- i W t}}{W - M_{\rm th} + i 0}
  \int \frac{d^{3} p^{\prime}}{( 2 \pi )^{3}}
  \frac{m_{M} m_{B}}{\mathcal{E}_{M} ( p^{\prime} )
    \mathcal{E}_{B} ( p^{\prime} ) }
  \notag \\
  & \phantom{=} \times 
  \frac{e^{i \bm{p}^{\prime} \cdot \bm{r}} T ( W , 0 , p^{\prime} )}{W
    - W_{M B} ( p^{\prime} ) + i 0} ,
\end{align}
where $W_{M B} ( p ) \equiv \mathcal{E}_{M} ( p ) + \mathcal{E}_{B} (
p )$.  For later convenience, we introduce a function
$\Phi ( W , r )$:
\begin{align}
  & \Phi ( W , r )
  \notag \\
  & \equiv \int \frac{d^{3} p^{\prime}}{( 2 \pi )^{3}}
  \frac{m_{M} m_{B}}{\mathcal{E}_{M} ( p^{\prime} )
    \mathcal{E}_{B} ( p^{\prime} ) }
  \frac{e^{i \bm{p}^{\prime} \cdot \bm{r}} T ( W , 0 , p^{\prime} )}{W
    - W_{M B} ( p^{\prime} )
    + i 0}
  \notag \\
  & =
  \int _{0}^{\infty} d p^{\prime} \frac{p^{\prime 2}}{2 \pi ^{2}}
  \frac{m_{M} m_{B}}{\mathcal{E}_{M} ( p^{\prime} )
    \mathcal{E}_{B} ( p^{\prime} ) }
  \frac{j_{0} ( p^{\prime} r ) T ( W , 0 , p^{\prime} )}{W
    - W_{M B} ( p^{\prime} ) + i 0} ,
  \label{eq:Phi}
\end{align}
where $j_{0} ( x ) \equiv \sin ( x ) / x$ is the spherical Bessel
function arising from the $s$-wave projection of $e^{i \bm{p}^{\prime}
  \cdot \bm{r}}$:
\begin{equation}
  \int \frac{d \Omega _{\bm{p}^{\prime}}}{4 \pi} e^{i \bm{p}^{\prime} \cdot \bm{r}}
  = \frac{1}{2} \int _{-1}^{1} d C e^{i p^{\prime} r C}
  = j_{0} ( p^{\prime} r ) .
\end{equation}
With the function $\Phi ( W , r )$, we can simplify the expression of the
connected contribution as
\begin{equation}
  C ( t , r )_{\rm conn}
  = \int _{- \infty}^{\infty} \frac{d W}{2 \pi}
  \frac{i e^{- i W t}}{W - M_{\rm th} + i 0}
  \Phi ( W , r ) .
  \label{eq:C_conn}
\end{equation}

\begin{figure}[t]
  \centering
  \includegraphics[scale=0.95]{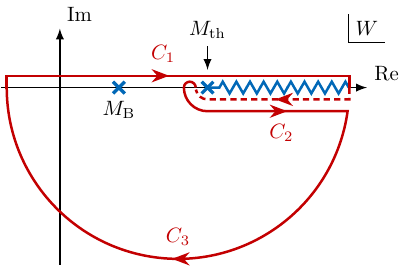} 
  \caption{The contour to calculate the integral in
    Eq.~\eqref{eq:C_conn}.  The cross symbols indicate the poles
    of the integrand, which are located at $W = M_{\rm B}$ and $M_{\rm
      th}$, while the zigzag line indicates the branch cut.  The
    dashed line indicates that the line is in the unphysical Riemann
    sheet.}
  \label{fig:contour}
\end{figure}

To perform the $W$ integral in Eq.~\eqref{eq:C_conn}, we extend the
contour of the integral as shown in Fig.~\ref{fig:contour}.  Here, we
divide the contour into three parts: $C_{1}$, $C_{2}$, and $C_{3}$.
The first part $C_{1}$ is the straight line infinitesimally above the
real $W$ axis in the range $( - \infty, \infty )$, i.e., the $W$
integral in Eq.~\eqref{eq:C_conn}.  The second part $C_{2}$ is the
back-and-forth line on the real $W$ axis at $M_{\rm th} \leq W < \infty$.
Finally, the third part $C_{3}$ is the lower-half circle in the
physical Riemann sheet of the complex $W$ plane.  When we close the
contour with these three parts, the contour surrounds the poles of the
integrand $\Phi ( W , r ) / ( W - M_{\rm th} + i 0 )$: the pole at $W
= M_{\rm th}$ and a bound-state pole at $W = M_{\rm B} < M_{\rm th}$
if it exists in the scattering amplitude\footnote{We assume that there
does not exist any resonance pole below the real $W$ axis in the
physical Riemann sheet.  Indeed, we do not have such a resonance pole
in our model of Secs.~\ref{sec:modelA} and \ref{sec:modelB}.}:
\begin{equation}
  T ( W , 0 , p^{\prime} )
  \simeq \frac{\gamma ( 0 ) \gamma ( p^{\prime} )}{W - M_{\rm B} + i 0} ,
  \label{eq:T_pole}
\end{equation}
where the meaning of the residue $\gamma ( 0 ) \gamma ( p^{\prime} )$
will be discussed in Sec.~\ref{sec:composite}.  As a result, according
to the residue theorem, we have
\begin{align}
  & \oint _{C_{1} + C_{2} + C_{3}}
  \frac{d W}{2 \pi} \frac{i e^{- i W t}}{W - M_{\rm th} + i 0}
  \Phi ( W , r )
  \notag \\
  & = e^{- i M_{\rm B} t} \Psi _{\rm B} ( r )
  + e^{- i M_{\rm th} t} \Phi ( M_{\rm th} , r ) ,
  \label{eq:C1C2C3}
\end{align}
where $\Psi _{\rm B} ( r )$ is defined as
\begin{align}
  \Psi _{\rm B} ( r )
  \equiv & \frac{\gamma ( 0 )}{M_{\rm B} - M_{\rm th}}
  \int _{0}^{\infty} d p^{\prime} \frac{p^{\prime 2}}{2 \pi ^{2}}
  \frac{m_{M} m_{B}}{\mathcal{E}_{M} ( p^{\prime} )
    \mathcal{E}_{B} ( p^{\prime} ) }
  \notag \\ & \times
  \frac{j_{0} ( p^{\prime} r ) \gamma ( p^{\prime} )}{M_{\rm B}
    - W_{M B} ( p^{\prime} )} .
  \label{eq:Psi_B}
\end{align}
We will see in Sec.~\ref{sec:composite} that this $\Psi _{\rm B} ( r
)$ is, except for a constant factor, the bound-state wave function in
coordinate space.  If the bound state does not exist, we simply
take $\gamma ( p^{\prime} ) = 0$ and $\Psi _{\rm B} ( r ) = 0$.  On
the left-hand side of Eq.~\eqref{eq:C1C2C3}, on the other hand, the
$C_{1}$ integral coincides with $C ( t , r )_{\rm
  conn}$~\eqref{eq:C_conn}, and the $C_{3}$ integral vanishes when $|
W | \to \infty$ thanks to the factor $e^{- i W t}$.  In addition, the
$C_{2}$ integral takes the branch cut contribution\footnote{We note
that the dashed backward line in Fig.~\ref{fig:contour} is in the
unphysical Riemann sheet infinitesimally below the real $W$ axis.
This line exactly coincides with the backward line in the physical
Riemann sheet infinitesimally above the real $W$ axis.}:
\begin{align}
  & \int _{C_{2}} \frac{d W}{2 \pi} \frac{i e^{- i W t}}{W - M_{\rm th}}
  \Phi ( W , r )
  \notag \\
  & = \int _{M_{\rm th}}^{\infty}
  \frac{d W}{2 \pi} \frac{i e^{- i W t}}{W - M_{\rm th}}
  \left [ - \Phi ( W + i 0 , r ) + \Phi ( W - i 0 , r ) \right ]
  \notag \\
  & = \frac{1}{\pi} \int _{M_{\rm th}}^{\infty} d W
  \frac{e^{- i W t} \Im \Phi ( W , r )}{W - M_{\rm th}} ,
\end{align}
where we used
\begin{equation}
  \Phi ( W - i 0 , r ) = \Phi ( W + i 0 , r )^{\ast} ,
\end{equation}
and hence
\begin{equation}
  \Phi ( W - i 0 , r ) - \Phi ( W + i 0 , r )
  = - 2 i \Im \Phi ( W , r ).
\end{equation}
Based on the above results, we obtain the formula for the connected
contribution:
\begin{align}
  C ( t , r )_{\rm conn}
  = & e^{- i M_{\rm B} t} \Psi _{\rm B} ( r )
  + e^{- i M_{\rm th} t} \Phi ( M_{\rm th} , r )
  \notag \\
  & - \frac{1}{\pi} \int _{M_{\rm th}}^{\infty} d W
  \frac{e^{- i W t} \Im \Phi ( W , r )}{W - M_{\rm th}} .
\end{align}

We finally obtain the formula for the correlation function $C ( t , r
)$~\eqref{eq:Corr_MB}:
\begin{align}
  C ( t , r )
  = & C ( t , r )_{\rm disc} + C ( t , r )_{\rm conn}
  \notag \\
  = & e^{- i M_{\rm B} t} \Psi _{\rm B} ( r )
  + e^{- i M_{\rm th} t} \left [ 1 + \Phi ( M_{\rm th} , r ) \right ]
  \notag \\
  & - \frac{1}{\pi} \int _{M_{\rm th}}^{\infty} d W
  \frac{e^{- i W t} \Im \Phi ( W , r )}{W - M_{\rm th}} .
  \label{eq:Corr_MB_final}
\end{align}

When the HAL QCD method is adopted to evaluate quantities from lattice
QCD simulations, Euclidean time is used.  Therefore, to prepare the
same conditions, we perform the analytic continuation $i t \to t$.
Then, dividing by the irrelevant factor $e^{- M_{\rm th} t}$, we
derive the so-called R-correlator $R ( t , r )$:
\begin{align}
  R ( t , r )
  \equiv & \frac{C ( - i t , r )}{e^{- M_{\rm th} t}}
  \notag \\
  = & e^{- ( M_{\rm B} - M_{\rm th} ) t} \Psi _{\rm B} ( r )
  + 1 + \Phi ( M_{\rm th} , r ) 
  \notag \\
  & - \frac{1}{\pi} \int _{M_{\rm th}}^{\infty} d W
  \frac{e^{- ( W - M_{\rm th} ) t}
    \Im \Phi ( W , r )}{W - M_{\rm th}} .
  \label{eq:Rcorr_MB}
\end{align}
With this formula, we can calculate the R-correlator once we obtain
the meson-baryon scattering amplitude $T ( W , p , p^{\prime} )$ in a
given model.  The R-correlator satisfies the Scr\"{o}dinger-like
equation with nonlocal potential $U ( \bm{r} , \bm{r}^{\prime}
)$~\cite{Ishii:2012ssm}
\begin{align}
  & - \frac{1}{2 \mu} \bm{\nabla}^{2} R ( t , r )
  + \int d^{3} r^{\prime} U ( \bm{r} , \bm{r}^{\prime} ) R ( t , r^{\prime} )
  \notag \\ &
  = \left ( 
  - \frac{\partial}{\partial t}
  + \frac{1 + 3 \delta ^{2}}{8 \mu}
  \frac{\partial ^{2}}{\partial t^{2}} \right ) R ( t , r ) .
\end{align}
In the leading-order of the derivative expansion (or velocity
expansion), nonlocal potential becomes
\begin{equation}
  U ( \bm{r} , \bm{r}^{\prime} )
  = V_{\rm HAL} ( r ) \delta ^{3} ( \bm{r} - \bm{r}^{\prime} ) ,
  \label{eq:U_LO}
\end{equation}
then the R-correlator provides us with the local potential according
to the formula
\begin{align}
  & V_{\rm HAL} ( r )
  \notag \\ &
  \equiv \frac{1}{R ( t , r )} \left ( \frac{1}{2 \mu r}
  \frac{\partial ^{2}}{\partial r^{2}} r
  - \frac{\partial}{\partial t}
  + \frac{1 + 3 \delta ^{2}}{8 \mu}
  \frac{\partial ^{2}}{\partial t^{2}} \right ) R ( t , r ) ,
  \label{eq:V_HAL_cont}
\end{align}
where $\mu \equiv m_{M} m_{B} / ( m_{M} + m_{B} )$ is the reduced mass
of the meson-baryon system and $\delta \equiv ( m_{M} - m_{B} ) / (
m_{M} + m_{B} )$.  This is the so-called time-dependent HAL QCD method
developed in Ref.~\cite{Ishii:2012ssm}.  Because we can perform the
exact $s$-wave projection in the continuum space case, we can simply
differentiate the R-correlator with respect to $r$.  In the present
study, we refer to the local potential $V_{\rm HAL} ( r )$ obtained
from the formula~\eqref{eq:V_HAL_cont} as the HAL QCD potential.

Furthermore, we may consider higher-order terms of the derivative
expansion for $U ( \bm{r} , \bm{r}^{\prime} )$.  In particular,
for $s$-wave meson-baryon scattering, the next-to-leading order
term $\mathcal{O} ( \bm{\nabla} )$ is absent, leaving the
next-to-next-to-leading order ($\mathrm{N}^{2}\mathrm{LO}$)
term:
\begin{equation}
  U ( \bm{r} , \bm{r}^{\prime} )
  = \left [ V_{0}^{\mathrm{N}{}^{2}\mathrm{LO}} ( r )
    + V_{2}^{\mathrm{N}{}^{2}\mathrm{LO}} ( r ) \bm{\nabla}^{2} \right ]
  \delta ^{3} ( \bm{r} - \bm{r}^{\prime} ) .
  \label{eq:U_N2LO}
\end{equation}
These $V_{0, 2}^{\mathrm{N}{}^{2}\mathrm{LO}} ( r )$ terms, referred
to as the HAL QCD N${}^{2}$LO potentials, are related to $V_{\rm HAL} (
r )$ as follows:
\begin{equation}
  V_{0}^{\mathrm{N}{}^{2}\mathrm{LO}} ( r )
  + V_{2}^{\mathrm{N}{}^{2}\mathrm{LO}} ( r )
  \frac{\bm{\nabla}^{2} R ( t , r )}{R ( t , r )}
  = V_{\rm HAL} ( r ) .
  \label{eq:LO_N2LO}
\end{equation}
In the present study, we mainly consider the HAL QCD local potential
$V_{\rm HAL} ( r )$, while we examine the HAL QCD
$\mathrm{N}{}^{2}\mathrm{LO}$ potentials $V_{0,
  2}^{\mathrm{N}{}^{2}\mathrm{LO}} ( r )$ in Sec.~\ref{sec:N2LO}.

The HAL QCD potential $V_{\rm HAL} ( r )$ is expected to contain
information on the meson-baryon interaction, because the
R-correlator~\eqref{eq:Rcorr_MB} can be decomposed into physical
eigenstates $| \Psi _{W_{n}} \rangle$ of eigenenergies $W = W_{n}$ in
the following manner (see, e.g., Ref.~\cite{Iritani:2018vfn}):
\begin{align}
  R ( t , r )
  = & \sum _{n} e^{- (W_{n} - M_{\rm th}) t } \langle r | \Psi _{W_{n}} \rangle
  \langle \Psi _{W_{n}} | M B ( p = 0 ) \rangle
  \notag \\
  = & e^{- (M_{\rm B} - M_{\rm th}) t } \langle r | \Psi _{\rm B} \rangle
  \langle \Psi _{\rm B} | M B ( p = 0 ) \rangle
  \notag \\
  & + \int _{0}^{\infty} d p \frac{p^{2}}{2 \pi ^{2}}
  \frac{m_{M} m_{B}}{\mathcal{E}_{M} ( p ) \mathcal{E}_{B} ( p )}
  e^{- [ W_{M B} ( p ) - M_{\rm th} ] t}
  \notag \\
  & \phantom{+}
  \times 
  \langle r | p_{\rm phys} \rangle
  \langle p_{\rm phys} | M B ( p = 0 ) \rangle ,
  \label{eq:Rcorr_MB_phys}
\end{align}
where $| M B ( p = 0 ) \rangle$ is the zero-momentum noninteracting
meson-baryon state and $| r \rangle$ is the noninteracting
meson-baryon state with spatial distance $r$.  Furthermore, $|
\Psi _{\rm B} \rangle$ is the physical bound state and $| p_{\rm phys}
\rangle$ is a physical scattering state of relative momentum $p$.
Both $| \Psi _{\rm B} \rangle$ and $| p_{\rm phys} \rangle$ are
eigenstates of the full Hamiltonian $\hat{H}$ and correspond to the
physical eigenstates $| \Psi _{W_{n}} \rangle$.  We prove that this
equation is truly correct in Appendix~\ref{app:phys}.  Because the
factor $\langle r | \Psi _{W_{n}} \rangle$, which corresponds to the
Nambu--Bethe--Salpeter wave function, is indeed a wave function as a
solution of a wave equation, the HAL QCD
potential~\eqref{eq:V_HAL_cont} is expected to reproduce the phase
shift of meson-baryon scattering at large $r$.

\subsection{Transition from bare state to meson-baryon scattering state}

Next, we develop the formula for the R-correlator in the case of the
transition from a bare excited baryon state, which we denote as
$\Lambda _{0}$, to the meson-baryon scattering state.  According to
the diagram shown in Fig.~\ref{fig:Corr_bare}, the correlation
function we calculate is
\begin{equation}
  C_{0} ( t , r )
  \equiv \int d^{3} x_{0}
  \langle 0 | \phi _{B} ( x_{B}^{\prime} ) \phi _{M} ( x_{M}^{\prime} )
  \phi _{\Lambda _{0}}^{\dagger} ( x_{0} ) | 0 \rangle .
  \label{eq:Corr_bare}
\end{equation}
This means that the bare state is created at $x_{0}^{\mu} = ( t_{0} ,
\bm{x}_{0} )$, the transition $\Lambda _{0} \to M B$ takes place, and
the meson and baryon are annihilated at $x_{M}^{\mu \prime} = ( t_{0}
+ t , \bm{x}_{M}^{\prime} )$ and $x_{B}^{\mu \prime} = ( t_{0} + t ,
\bm{x}_{B}^{\prime} )$, respectively.  As in the elastic scattering
case, we assume $t > 0$ and define $r \equiv | \bm{r} |$, $\bm{r}
\equiv \bm{x}_{M}^{\prime} - \bm{x}_{B}^{\prime}$.

Again, the correlation function can be evaluated in the standard way of
quantum field theory.  Because a disconnected diagram is absent in the
$\Lambda _{0} \to M B$ transition, we have
\begin{align}
  & C_{0} ( t , r )
  \notag \\
  & = \int d^{3} x_{0} \int d^{4} x
  \int \frac{d^{4} P}{( 2 \pi )^{4}}
  \frac{m_{\Lambda _{0}}}{\mathcal{E}_{\Lambda _{0}} ( P ) }
  \frac{i e^{- i P \cdot ( x - x_{0} )}}{P^{0}
    - \mathcal{E}_{\Lambda _{0}} ( P ) + i 0}
  \notag \\
  & \phantom{=} \times
  \int \frac{d^{4} p_{M}^{\prime}}{( 2 \pi )^{4}}
  \frac{m_{M}}{\mathcal{E}_{M} ( p_{M}^{\prime} ) }
  \frac{i e^{- i p_{M}^{\prime} \cdot ( x_{M}^{\prime} - x )}}{p_{M}^{\prime 0}
    - \mathcal{E}_{M} ( p_{M}^{\prime} ) + i 0}
  \notag \\
  & \phantom{=} \times
  \int \frac{d^{4} p_{B}^{\prime}}{( 2 \pi )^{4}}
  \frac{m_{B}}{\mathcal{E}_{B} ( p_{B}^{\prime} ) }
  \frac{i e^{- i p_{B}^{\prime} \cdot ( x_{B}^{\prime} - x )}}{p_{B}^{\prime 0}
    - \mathcal{E}_{B} ( p_{B}^{\prime} ) + i 0}
  [ - i \Gamma ( P^{0} , | \bm{p}_{M}^{\prime} | ) ] ,
\end{align}
where the bare state mass and momentum are $M_{0}$ and $P^{\mu}$,
respectively, $\mathcal{E}_{\Lambda _{0}} ( P ) \equiv \sqrt{M_{0}^{2}
  + \bm{P}^{2}}$, $x^{\mu}$ is the coordinate where the transition
takes place, and $\Gamma$ is the $\Lambda _{0} \to M B$ transition
amplitude depending on the center-of-mass energy $P^{0}$ and the
absolute value of the relative momentum $| \bm{p}_{M}^{\prime} |$ in
the final state.  Due to the $s$-wave nature, the transition amplitude
does not depend on the angle of the momentum (see Sec.~\ref{sec:amp}).
Evaluation of $C_{0} ( t , r )$ can be performed in a similar manner
to the elastic scattering case.  Performing the integrals with respect
to the coordinates, we have
\begin{align}
  C_{0} ( t , r ) 
  = & \int _{- \infty}^{\infty} \frac{d W}{2 \pi}
  \frac{i e^{- i W t}}{W - M_{0}}
  \int \frac{d^{3} p^{\prime}}{( 2 \pi )^{3}} 
  \frac{m_{M} m_{B}}{\mathcal{E}_{M} ( p^{\prime} )
    \mathcal{E}_{B} ( p^{\prime} )}
  \notag \\
  & \times \frac{e^{i \bm{p}^{\prime} \cdot \bm{r}}
    \Gamma ( W , p^{\prime} )}{W - W_{M B} ( p^{\prime} ) + i 0} ,
\end{align}
where we introduced $W \equiv P^{0}$ and the integral variable
$p_{M}^{\prime \mu}$ was renamed as $p^{\prime \mu}$.  As in the
elastic scattering case, we introduce
\begin{align}
  & \Phi _{0} ( W , r )
  \notag \\
  & \equiv
  \int \frac{d^{3} p^{\prime}}{( 2 \pi )^{3}} 
  \frac{m_{M} m_{B}}{\mathcal{E}_{M} ( p^{\prime} )
    \mathcal{E}_{B} ( p^{\prime} )}
  \frac{e^{i \bm{p}^{\prime} \cdot \bm{r}} \Gamma ( W , p^{\prime} )}{W
    - W_{M B} ( p^{\prime} ) + i 0}
  \notag \\
  & = \int _{0}^{\infty} d p^{\prime} \frac{p^{\prime 2}}{2 \pi ^{2}} 
  \frac{m_{M} m_{B}}{\mathcal{E}_{M} ( p^{\prime} ) \mathcal{E}_{B} ( p^{\prime} )}
  \frac{j_{0} ( p^{\prime} r ) \Gamma ( W , p^{\prime} )}{W
    - W_{M B} ( p^{\prime} ) + i 0} ,
  \label{eq:def_Phi}
\end{align}
then the correlation function becomes
\begin{equation}
  C_{0} ( t , r ) 
  = \int _{- \infty}^{\infty} \frac{d W}{2 \pi}
  \frac{i e^{- i W t}}{W - M_{0}} \Phi _{0} ( W , r ) .
  \label{eq:Wint}
\end{equation}

To perform the $W$ integral in Eq.~\eqref{eq:Wint}, we consider the
contour of the integral in Fig.~\ref{fig:contour}, as in the previous
case.  When a bound state exists, we have
\begin{equation}
  \Gamma ( W , p^{\prime} )
  \simeq \frac{\gamma ( p^{\prime} ) \gamma _{0}}{W - M_{\rm B} + i 0}
\end{equation}
near the bound-state energy\footnote{The factor $\gamma ( p^{\prime}
)$ of the residue has essentially the same meaning as that of the
residue in the meson-baryon scattering amplitude~\eqref{eq:T_pole}.
See Sec.~\ref{sec:composite}.}, and hence it contributes to the
integral via the residue theorem.  On the other hand, because the
transition amplitude satisfies $\Gamma ( M_{0} , p^{\prime} ) = 0$,
the point $W = M_{0}$ of the function $\Phi _{0} ( W , r ) / ( W -
M_{0} )$ is a removable singularity\footnote{The bare state $\Lambda
_{0}$ is not a physical state, i.e., not an eigenstate of the full
Hamiltonian $\hat{H}$, and hence the pole at $W = M_{0}$ does not
emerge in the correlation function.}.  As a result, we have
\begin{align}
  C_{0} ( t , r ) = e^{- i M_{\rm B} t} \Psi _{\rm B0} ( r )
  - \frac{1}{\pi} \int _{M_{\rm th}}^{\infty} d W
  \frac{e^{- i W t} \Im \Phi _{0} ( W , r )}{W - M_{0}} ,
  \label{eq:Corr_bare_final}
\end{align}
where
\begin{align}
  \Psi _{\rm B0} ( r )
  \equiv & \frac{\gamma _{0}}{M_{\rm B} - M_{0}}
  \int _{0}^{\infty} d p^{\prime} \frac{p^{\prime 2}}{2 \pi ^{2}} 
  \frac{m_{M} m_{B}}{\mathcal{E}_{M} ( p^{\prime} ) \mathcal{E}_{B} ( p^{\prime} )}
  \notag \\ & \times
  \frac{j_{0} ( p^{\prime} r ) \gamma ( p^{\prime} )}{M_{\rm B}
    - W_{M B} ( p^{\prime} )} .
  \label{eq:Psi_Bzero}
\end{align}
This $\Psi _{\rm B0} ( r )$ is again, except for a constant factor,
the bound-state wave function in coordinate space, as seen in
Sec.~\ref{sec:composite}.

From the correlation function $C_{0} ( t , r )$, we can calculate
the R-correlator in the HAL QCD method: 
\begin{align}
  R_{0} ( t , r )
  \equiv & \frac{C_{0} ( - i t , r )}{e^{- M_{\rm th} t}}
  \notag \\
  = & e^{- ( M_{\rm B} - M_{\rm th} ) t} \Psi _{\rm B0} ( r )
  \notag \\
  & - \frac{1}{\pi} \int _{M_{\rm th}}^{\infty} d W
  \frac{e^{- ( W - M_{\rm th} ) t}
    \Im \Phi _{0} ( W , r )}{W - M_{0}} .
  \label{eq:Rcorr_bare}
\end{align}
Then, with the same formula~\eqref{eq:V_HAL_cont}, we can calculate
the HAL QCD potential $V_{\rm HAL} ( r )$, once we obtain the
transition amplitude $\Gamma ( W , p )$ in a given model and the
R-correlator $R_{0} ( t , r )$ from $\Gamma ( W , p )$.  We also note
that, as in the elastic scattering case, the R-correlator $R_{0} ( t ,
r )$ satisfies the following equation:
\begin{align}
  R_{0} ( t , r )
  = & \sum _{n} e^{- (W_{n} - M_{\rm th}) t } \langle r | \Psi _{W_{n}} \rangle
  \langle \Psi _{W_{n}} | \Lambda _{0} \rangle
  \notag \\
  = & e^{- (M_{\rm B} - M_{\rm th}) t } \langle r | \Psi _{\rm B} \rangle
  \langle \Psi _{\rm B} | \Lambda _{0} \rangle
  \notag \\
  & + \int _{0}^{\infty} d p \frac{p^{2}}{2 \pi ^{2}}
  \frac{m_{M} m_{B}}{\mathcal{E}_{M} ( p ) \mathcal{E}_{B} ( p )}
  e^{- [ W_{M B} ( p ) - M_{\rm th} ] t}
  \notag \\
  & \phantom{+}
  \times 
  \langle r | p_{\rm phys} \rangle
  \langle p_{\rm phys} | \Lambda _{0} \rangle ,
  \label{eq:Rcorr_bare_phys}
\end{align}
where $| \Lambda _{0} \rangle$ indicates the bare state at rest.  We
prove this in Appendix~\ref{app:phys}.

\subsection{Lattice discretization and finite-volume effects of spacetime}
\label{sec:lattice}

Up to here, we considered continuum spacetime in infinite volume.
Next, we turn to the case of lattice spacetime, in which the lattice
spacing $a$, lattice size $N^{4}$, and length of a side of the box $L
\equiv N a$ are key quantities.  Throughout this study, we employ $a =
\SI{0.121}{fm}$ and $N = 32$ both for spatial and temporal
configurations according to Ref.~\cite{Murakami:2023phq} and take
periodic boundary conditions.  As we will see, spatially, lattice
discretization and finite volume affect both differentials and
integrals.  In particular, the scattering amplitude, which contains
momentum integrals, must be modified.  Temporally, on the other hand,
only the time derivative is relevant in the present study.  The
time-opposite propagation of particles due to periodic boundary
conditions would be strongly suppressed by the factor $e^{- L W_{\rm
    typ}} \sim 10^{-9}$ with a typical energy $W_{\rm typ} \sim
\si{GeV}$.

On the lattice, the spatial coordinate of a particle is discretized as
\begin{equation}
  \bm{r} = a \bm{n}_{r} ,
  \label{eq:r_an}
\end{equation}
where each component of the vector $\bm{n}_{r} \equiv ( n_{x} , n_{y}
, n_{z} )^{\rm T}$ takes
\begin{equation}
  n_{i} = 1 , 2 , \ldots , N
  \quad
  ( i = x , y , z ) .
  \label{eq:nrx}
\end{equation}
In a similar manner, the relative momentum of two particles
becomes
\begin{equation}
  \bm{p} = \frac{2 \pi}{L} \bm{n}_{p} ,
  \label{eq:p_2Ln}
\end{equation}
with $\bm{n}_{p} \equiv ( n_{p x} , n_{p y} , n_{p z} )^{\rm T}$ and
\begin{equation}
  n_{p i} = - \frac{N}{2} + 1 , - \frac{N}{2} + 2 , \ldots ,
  \frac{N}{2}
  \quad ( i = x , y , z ) .
  \label{eq:npx}
\end{equation}
As can be seen from Eq.~\eqref{eq:p_2Ln}, the finite-volume effect
with periodic boundary conditions restricts the minimum momentum step
via $L$, whereas the lattice spacing $a$ limits the maximum momentum.
Due to discretization of the momentum, the momentum integral becomes a
summation:
\begin{equation}
  \int \frac{d^{3} p}{( 2 \pi )^{3}}
  \to 
  \frac{1}{L^{3}}
  \sum _{\bm{n}_{p}} .
  \label{eq:corresp_mom}
\end{equation}

\begin{table}[!t]
  \caption{Examples of the weights $v_{\alpha}$ and representatives
    $\bm{n}_{\alpha} $ in the $A_{1}^{+}$ projection.}
  \label{tab:A1plus}
  \centering
  \begin{ruledtabular}
    \begin{tabular}{rrl}
      \multicolumn{1}{l}{$\alpha$} &
      \multicolumn{1}{l}{$v_{\alpha}$} &
      $\bm{n}_{\alpha}$
      \\
      \hline
      $1$ & $1$ & $(0, 0, 0)^{\rm T}$
      \\
      $2$ & $6$ & $(1, 0, 0)^{\rm T}$
      \\
      $3$ & $12$ & $(1, 1, 0)^{\rm T}$
      \\
      $4$ & $8$ & $(1, 1, 1)^{\rm T}$
      \\
      $\vdots$ & $\vdots$ & $\vdots$
      \\
      $1528$ & $24$ & $(15, 16, 16)^{\rm T}$
      \\
      $1529$ & $8$ & $(16, 16, 16)^{\rm T}$
    \end{tabular}
  \end{ruledtabular}
\end{table}

Then, we consider the partial wave projection in lattice
discretization.  While the usual $s$-wave projection is valid in
continuum space, it becomes the $A_{1}^{+}$ projection in lattice
space.  In the $A_{1}^{+}$ projection, a solid angle integral of a
function $f ( \bm{p} )$ becomes
\begin{equation}
  \int \frac{d \Omega _{\bm{p}}}{4 \pi} f ( \bm{p} )
  \to
  \frac{1}{24}
  \sum _{g \in G}
  f_{g \bm{n}_{p}} ,
  \quad
  f_{\bm{n}_{p}} \equiv f ( 2 \pi \bm{n}_{p} / L ) ,
\end{equation}
where $g \in G$ represents the 24 elements of the cubic transformation
group, whose explicit form is given in Appendix~\ref{app:projection}.
The $A_{1}^{+}$ projection introduces contamination by higher orbital
waves $l \ge 4$ with a negligible amount at low
energy~\cite{Ishii:2006ec}.  Furthermore, all the momenta $\bm{n}_{p}$
can be classified into sets labeled by $\alpha$ whose elements can be
transformed into each other by $g \in G$.  As a consequence, we have
\begin{equation}
  \sum_{\bm{n}_{p}} f_{\bm{n}_{p}}
  =
  \sum _{\alpha = 1}^{A} \frac{v_{\alpha}}{24}
  \sum _{g \in G}
  f_{g \bm{n}_{\alpha}} ,
  \label{eq:formula_1}
\end{equation}
with the weights $v_{\alpha}$, i.e., the number of elements in set
$\alpha$, and representatives $\bm{n}_{\alpha}$, examples of which are
listed in Table~\ref{tab:A1plus}.  Because $g \in G$ represents the 24
elements of the cubic transformation group, we have the relation $| g
\bm{n}_{\alpha} | = | \bm{n}_{\alpha} |$.  In particular, the absolute
value of the momentum can be uniquely specified by $\alpha$:
$p_{\alpha} \equiv 2 \pi | \bm{n}_{\alpha} | / L$.  The total
number of sets $A$ depends on $N$, and in the case of $N = 32$, we have
$A = 1529$.\footnote{As one can see from Table~\ref{tab:A1plus}, we
take the weight $v_{\alpha}$ of the last set of the representative
$(16, 16, 16)^{\rm T}$ to be $8$.  However, strictly speaking, this is
not correct, because we do not have $n_{p i} = - 16$ ($i = x , y , z$)
in the present formulation, as in Eq.~\eqref{eq:npx}, and hence $(16,
-16, 16)^{\rm T}$, for example, does not appear in the summation on
the left-hand side of Eq.~\eqref{eq:formula_1}.  The same overcounting
happens when $n_{p i} = - 16$ ($i = x , y , z$) is concerned.
Nevertheless, such high-momentum contributions are negligible in the
following calculations, so we leave them as they are.}

As a consequence of lattice discretization, instead of the
spherical Bessel function $j_{0} ( x )$, we have
\begin{equation}
  \tilde{j}_{0} ( p_{\alpha} , \bm{r} )
  \equiv \frac{1}{24}
  \sum _{g \in G}
  e^{i ( g \bm{p}_{\alpha} ) \cdot \bm{r}} .
\end{equation}
Here and below, we put a tilde on quantities evaluated in lattice
spacetime.  We note that $\tilde{j}_{0}$, and hence the R-correlators
and HAL QCD potentials, is strictly speaking not a function of $r
\equiv | \bm{r} |$ but depends on the angle of $\bm{r}$.  However, as
we will see later, this angular dependence is not significant in the
present study.

Similarly, we replace the scattering amplitude $T ( W , 0 , p^{\prime}
)$ with its lattice counterpart $\tilde{T} ( W , p_{\alpha = 1},
p_{\alpha ^{\prime}} )$, whose formulation will be given in
Sec.~\ref{eq:amp_lat}. From Eq.~\eqref{eq:Phi} in continuum space,
we can simply extend $\Phi ( W , r )$ to its lattice counterpart
$\tilde{\Phi} ( W , \bm{r} )$ by replacing the momentum integral with a
summation~\eqref{eq:corresp_mom} as
\begin{equation}
  \tilde{\Phi} ( W , \bm{r} )
  = \frac{1}{L^{3}} \sum _{\alpha = 1}^{A}
  \frac{v_{\alpha} m_{M} m_{B}}{\mathcal{E}_{M} ( p_{\alpha} )
    \mathcal{E}_{B} ( p_{\alpha} )}
  \frac{\tilde{j}_{0} ( p_{\alpha} , \bm{r} )
    \tilde{T} ( W , p_{1} , p_{\alpha} )}{W - W_{M B} ( p_{\alpha} )} .
\end{equation}
This $\tilde{\Phi} ( W , \bm{r} )$ function contributes to the
connected correlation function as in Eq.~\eqref{eq:C_conn}.  In
lattice spacetime, because a branch cut does not exist, the integral
in Eq.~\eqref{eq:C_conn} simply picks up the poles of $\tilde{\Phi} (
W , \bm{r} ) / ( W - M_{\rm th} )$.  As is well known, the scattering
amplitude in lattice space has poles at $W = \tilde{W}_{n}$ ($n =
1$, $2$, $\ldots$) on the real $W$ axis instead of an explicit
imaginary part above the threshold: in the infinite-volume limit,
states at $\tilde{W}_{n}$ become the bound state, if it exists, as
well as the scattering states.  Therefore, we express the scattering
amplitude as
\begin{equation} 
  \tilde{T} ( W , p_{1} , p_{\alpha} )
  = \sum _{n} 
  \frac{\tilde{\gamma}_{n} ( p_{1} ) \tilde{\gamma}_{n} ( p_{\alpha} )}{W
    - \tilde{W}_{n}} 
  + \text{(regular)} .
  \label{eq:Ttilde_poles}
\end{equation} 
In contrast, at $W = W_{M B} ( p_{\alpha} )$, the scattering amplitude
becomes zero: $\tilde{T} ( W_{M B} ( p_{\alpha} ) , p_{1} , p_{\alpha}
) = 0$.  Therefore, although $\tilde{\Phi} ( W , \bm{r} )$ contains
the factor $1 / [ W - W_{M B} ( p_{\alpha} ) ]$, the points $W = W_{M
  B} ( p_{\alpha} )$ of the function $\tilde{T} ( W , p_{1} ,
p_{\alpha} ) / [ W - W_{M B} ( p_{\alpha} ) ]$ in $\tilde{\Phi} ( W ,
\bm{r} )$ are removable singularities\footnote{Again, the energy of
free particles $W = W_{M B} ( p_{\alpha} )$ is not the eigenenergy for
physical states.}.  After all, the R-correlator in lattice
spacetime can be explicitly written as
\begin{align}
  \tilde{R} ( t , \bm{r} )
  = & 1 + \tilde{\Phi} ( M_{\rm th} , \bm{r} )
  \notag \\
  & + \sum _{n} \frac{e^{- ( \tilde{W}_{n} - M_{\rm th} ) t}
    \tilde{\gamma}_{n} ( p_{1} )}
  {( \tilde{W}_{n} - M_{\rm th} ) L^{3}} \sum _{\alpha = 1}^{A}
  \frac{v_{a} m_{M} m_{B}}
       {\mathcal{E}_{M} ( p_{\alpha} ) \mathcal{E}_{B} ( p_{\alpha} )}
  \notag \\
  & \phantom{+} \times
  \frac{\tilde{j} ( p_{\alpha} , \bm{r} ) \tilde{\gamma}_{n}
    ( p_{\alpha} ) }{\tilde{W}_{n} - W_{MB} ( p_{\alpha} )} .
  \label{eq:Rtilde_fin}
\end{align}

In a similar manner, we can evaluate $\tilde{\Phi}_{0} ( W , \bm{r} )$
and the R-correlator $\tilde{R}_{0} ( t , \bm{r} )$ in lattice
spacetime from the transition amplitude $\tilde{\Gamma} ( W ,
p^{\prime} )$.  Here, we show only their explicit forms:
\begin{equation}
  \tilde{\Phi}_{0} ( W , \bm{r} )
  = \frac{1}{L^{3}} \sum _{\alpha = 1}^{A}
  \frac{v_{\alpha} m_{M} m_{B}}{\mathcal{E}_{M} ( p_{\alpha} )
    \mathcal{E}_{B} ( p_{\alpha} )}
  \frac{\tilde{j}_{0} ( p_{\alpha} , \bm{r} )
    \tilde{\Gamma} ( W , p_{\alpha} )}{W - W_{M B} ( p_{\alpha} )} ,
\end{equation}
\begin{align}
  \tilde{R}_{0} ( t , \bm{r} )
  = & \sum _{n} \frac{e^{- ( \tilde{W}_{n} - M_{\rm th} ) t} \tilde{\gamma}_{0 n}}
  {( \tilde{W}_{n} - M_{0} ) L^{3}} \sum _{\alpha = 1}^{A}
  \frac{v_{a} m_{M} m_{B}}
       {\mathcal{E}_{M} ( p_{\alpha} ) \mathcal{E}_{B} ( p_{\alpha} )}
  \notag \\
  & \phantom{+} \times
  \frac{\tilde{j} ( p_{\alpha} , \bm{r} ) \tilde{\gamma}_{n}
    ( p_{\alpha} ) }{\tilde{W}_{n} - W_{MB} ( p_{\alpha} )} ,
\end{align}
where $\tilde{\gamma}_{n} ( p_{\alpha} ) \tilde{\gamma}_{0 n}$ is the
residue of the transition amplitude $\tilde{\Gamma} ( W , p_{\alpha}
)$ at the pole $W = \tilde{W}_{n}$.

Once the R-correlator is calculated in lattice spacetime, we can
calculate the HAL QCD local potential according to the following
formula:
\begin{align}
  & \tilde{V}_{\rm HAL} ( \bm{r} )
  \notag \\
  & \equiv
  \frac{1}{\tilde{R} ( t , \bm{r} )} \left ( \frac{\bm{\nabla}^{2}_{\rm lat}}{2 \mu}
  - \frac{\partial}{\partial t}_{\rm lat}
  + \frac{1 + 3 \delta ^{2}}{8 \mu}
  \frac{\partial ^{2}}{\partial t^{2}}_{\rm lat} \right ) \tilde{R} ( t , \bm{r} ) ,
  \label{eq:V_HAL_latt}
\end{align}
where 
\begin{align}
  \bm{\nabla}^{2}_{\rm lat} \tilde{R}
  \equiv &
  \frac{\tilde{R} ( t , \bm{r} + a \bm{e}_{x} ) - 2 \tilde{R} ( t , \bm{r} )
    + \tilde{R} ( t , \bm{r} - a \bm{e}_{x} )}{a^{2}}
  \notag \\
  & + \frac{\tilde{R} ( t , \bm{r} + a \bm{e}_{y} ) - 2 \tilde{R} ( t , \bm{r} )
    + \tilde{R} ( t , \bm{r} - a \bm{e}_{y} )}{a^{2}}
  \notag \\
  & + \frac{\tilde{R} ( t , \bm{r} + a \bm{e}_{z} ) - 2 \tilde{R} ( t , \bm{r} )
    + \tilde{R} ( t , \bm{r} - a \bm{e}_{z} )}{a^{2}} ,
\end{align}
\begin{equation}
  \frac{\partial \tilde{R}}{\partial t}_{\rm lat}
  \equiv \frac{\tilde{R} ( t + a , \bm{r} ) - \tilde{R} ( t - a , \bm{r} )}{2 a} ,
\end{equation}
\begin{equation}
  \frac{\partial ^{2} \tilde{R}}{\partial t^{2}}_{\rm lat}
  \equiv \frac{\tilde{R} ( t + a , \bm{r} ) - 2 \tilde{R} ( t , \bm{r} )
    + \tilde{R} ( t - a , \bm{r} )}{a^{2}} .
\end{equation}
Here, $\bm{e}_{x, y, z} = ( 1 , 0 , 0 )^{\rm T}$, $( 0 , 1 , 0 )^{\rm
  T}$, $( 0 , 0 , 1 )^{\rm T}$ are unit vectors.  As mentioned above,
the HAL QCD potential $\tilde{V}_{\rm HAL} ( \bm{r} )$ in lattice
spacetime is, strictly speaking, not a function of $r \equiv | \bm{r}
|$ but depends only slightly on the angle of $\bm{r}$, as we will see
later.

\section{Model of the scattering amplitude}
\label{sec:amp}

As we have seen in the previous section, we can calculate
R-correlators and HAL QCD potentials once we obtain the meson-baryon
scattering amplitude and the transition amplitude from the bare state to
the meson-baryon state.  Therefore, in this section we construct our
model of the meson-baryon amplitudes to analyze the effectiveness of
the HAL QCD method.  We focus on meson ($M$) -baryon ($B$)
scattering in the $s$ wave, and allow the $s$-wave meson-baryon system
to couple to one bare state $\Lambda _{0}$.

Below, after presenting our interaction terms, we formulate the
scattering amplitudes in continuum spacetime.  We then introduce the
so-called compositeness to quantify the fraction of the hadronic
molecular component in a bound state.  Finally, we formulate the
scattering amplitudes with the lattice discretization of spacetime.

\subsection{Interaction}

\begin{figure}[t]
  \centering
  \includegraphics[scale=0.95]{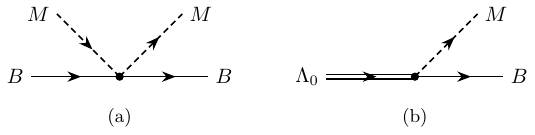} 
  \caption{Diagrams for the meson-baryon ($M B$) interaction: (a)
    elastic interaction and (b) transition of a bare $\Lambda _{0}$ to
    $M B$.}
  \label{fig:interaction}
\end{figure}

\begin{figure*}[t]
  \centering
  \includegraphics[scale=0.95]{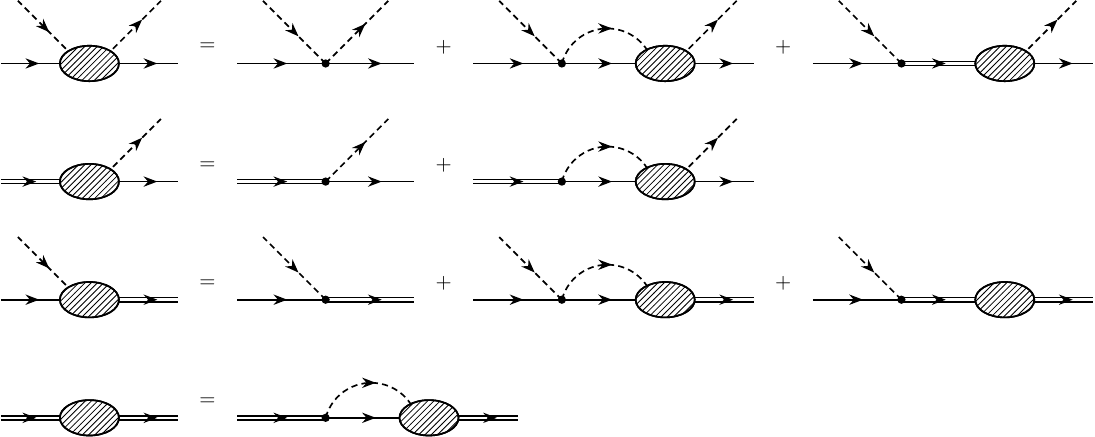}
  \caption{Diagrammatic expression of the Lippmann--Schwinger
    equation.  The dashed, solid, and double lines correspond to the
    meson, baryon, and bare $\Lambda _{0}$, respectively.  The
    vertices represent interactions~\eqref{eq:V2_I},
    \eqref{eq:V2_II}, and \eqref{eq:V1}, while the shaded ellipses
    represent the scattering
    amplitude~\eqref{eq:Tamp}--\eqref{eq:Damp}.}
  \label{fig:LSeq}
\end{figure*}

In the present study, the meson-baryon interaction is governed by the
diagrams in Fig.~\ref{fig:interaction}: the elastic interaction $M B
\to M B$ and the transition $\Lambda _{0} \to M B$.

For the elastic interaction $M B \to M B$
[Fig.~\ref{fig:interaction}(a)], we employ an $s$-wave interaction in
momentum space governed by the interaction strength $V_{\rm L}$ and
range $b_{\rm L}$:
\begin{align}
  V_{\rm int} ( V_{\rm L} , b_{\rm L} ; p , p^{\prime} )
  = & \frac{2 ( \pi b_{\rm L}^{2} )^{3/2} V_{\rm L}}{p p^{\prime} b_{\rm L}^{2}}
  \exp \left [ - \frac{( p^{2} + p^{\prime 2} ) b_{\rm L}^{2}}{4} \right ]
  \notag \\
  & \times
  \sinh \left ( \frac{p p^{\prime} b_{\rm L}^{2}}{2} \right ) ,
  \label{eq:Vint}
\end{align}
where $p^{( \prime )}$ is the relative momentum of the meson-baryon
system in the initial (final) state.  This form is motivated by the
Gaussian local potential
\begin{equation}
  V ( r ) = V_{\rm L} \exp \left ( - \frac{r^{2}}{b_{\rm L}^{2}} \right ) .
  \label{eq:Gaussian}
\end{equation}
Indeed, the Fourier transform of this Gaussian local potential
is given by
\begin{equation}
  \int d^{3} r \, e^{i \bm{r} \cdot \bm{q}} V ( r )
  = ( \pi b_{\rm L}^{2} )^{3/2} V_{\rm L}
  \exp \left ( - \frac{q^{2} b_{\rm L}^{2}}{4} \right ) ,
\end{equation}
and its $s$-wave projection with $\bm{q} = \bm{p} - \bm{p}^{\prime}$
yields exactly the interaction $V_{\rm int}$ in Eq.~\eqref{eq:Vint}.
Then, we describe the meson-baryon interaction in a single term:
\begin{equation}
  V_{( 2 )} ( p , p^{\prime} )
  = V_{\rm int} ( V_{\rm L} , b_{\rm L} ; p , p^{\prime} ) ,
  \label{eq:V2_I}
\end{equation}
where the subscript (2) indicates that this is a two-body
interaction.  This interaction is attractive (repulsive) when $V_{\rm
  L} < 0$ ($V_{\rm L} > 0$).  In addition, we may consider a two-term
interaction
\begin{equation}
  V_{( 2 )} ( p , p^{\prime} )
  = V_{\rm int} ( V_{\rm L} , b_{\rm L} ; p , p^{\prime} )
  - V_{\rm int} ( 2 V_{\rm L} , b_{\rm L} / 2 ; p , p^{\prime} ) ,
  \label{eq:V2_II}
\end{equation}
with $V_{\rm L} < 0$.  This means that the interaction has a repulsive
core in the inner region and an attractive pocket in the outer region.

When we allow the transitions $\Lambda _{0} \to M B$ and $M B \to
\Lambda _{0}$ in the $s$ wave [see Fig.~\ref{fig:interaction}(b)], we
introduce a Gaussian form with range $b_{0}$ for the $\Lambda _{0}$-$M
B$ coupling:
\begin{equation}
  V_{(1)} ( p ) = f_{0} \exp \left ( - \frac{p^{2} b_{0}^{2}}{4} \right ) ,
  \label{eq:V1}
\end{equation}
where $f_{0}$ is a coupling constant and $p$ is the meson-baryon
relative momentum.  The subscript (1) indicates that this is a one-
to two-body transition.

In these interaction terms, we treat $V_{\rm L}$, $b_{\rm L}$, $f_{0}$,
and $b_{0}$ as model parameters.  Therefore, together with the bare
state mass $M_{0}$, the complete set of model parameters is
$(V_{\rm L} , b_{\rm L} , f_{0} , b_{0} , M_{0})$.

\subsection{Scattering amplitude in continuum spacetime}

Based on the $M B$ interaction, we calculate the scattering
amplitude in continuum and lattice spacetimes.  In the present
formulation, we have the $s$-wave $M B ( p ) \to M B ( p^{\prime} )$
amplitude $T ( W , p , p^{\prime} )$, $\Lambda _{0} \to M B ( p )$
amplitude $\Gamma ( W , p )$, and $\Lambda _{0} \to \Lambda _{0}$
amplitude $\Delta ( W )$, where $W$ is the two-body center-of-mass
energy.  Calculations are performed in the center-of-mass frame.  We can
diagrammatically express the Lippmann--Schwinger equation as in
Fig.~\ref{fig:LSeq}.  In the continuum case, we have
\begin{widetext}
\begin{equation}
  T ( W , p , p^{\prime} )
  = V_{(2)} ( p , p^{\prime} )
  + \int _{0}^{\infty} d k \frac{k^{2}}{2 \pi ^{2}}
  \frac{m_{M} m_{B}}{\mathcal{E}_{M} ( k ) \mathcal{E}_{B} ( k )}
  \frac{V_{(2)} ( p , k ) T ( W , k , p^{\prime} )}{W
    - W_{MB} ( k ) + i 0}
  + \frac{V_{(1)} ( p ) \Gamma ( W , p^{\prime} )}{W - M_{0}}
  \label{eq:Tamp}
\end{equation}
\begin{align}
  \Gamma ( W , p )
  = & V_{(1)} ( p )
  + \int _{0}^{\infty} d k \frac{k^{2}}{2 \pi ^{2}}
  \frac{m_{M} m_{B}}{\mathcal{E}_{M} ( k ) \mathcal{E}_{B} ( k )}
  \frac{V_{(1)} ( k ) T ( W , k , p )}{W
    - W_{MB} ( k ) + i 0}
  \notag \\
  = & V_{(1)} ( p )
  + \int _{0}^{\infty} d k \frac{k^{2}}{2 \pi ^{2}}
  \frac{m_{M} m_{B}}{\mathcal{E}_{M} ( k ) \mathcal{E}_{B} ( k )}
  \frac{V_{(2)} ( p , k ) \Gamma ( W , k )}{W
    - W_{MB} ( k ) + i 0}
  + \frac{V_{(1)} ( p ) \Delta ( W )}{W - M_{0}}
  \label{eq:Gamp}
\end{align}
\begin{equation}
  \Delta ( W )
  = \int _{0}^{\infty} d k \frac{k^{2}}{2 \pi ^{2}}
  \frac{m_{M} m_{B}}{\mathcal{E}_{M} ( k ) \mathcal{E}_{B} ( k )}
  \frac{V_{(1)} ( k ) \Gamma ( W , k )}{W
    - W_{MB} ( k ) + i 0} .
  \label{eq:Damp}
\end{equation}
\end{widetext}

The Lippmann--Schwinger equations~\eqref{eq:Tamp}--\eqref{eq:Damp} can
be solved in a standard way.  In particular, when the energy $W$ is
real and below the meson-baryon threshold, we may rewrite the momentum
integrals as
\begin{equation}
  \int _{0}^{\infty} d p f ( p )
  = \sum _{n = 1}^{N} w_{n} f ( p_{n} )
\end{equation}
with discretized momenta $p_{n}$ ($n = 1$, $2$, $\ldots$, $N$)
and weights $w_{n}$, as done in, e.g., Gauss--Legendre quadrature.
Using this discretization, we can calculate the scattering amplitude via
a matrix equation:
\begin{align}
  \mathcal{T} ( W )
  = & \mathcal{V} ( W )
  + \mathcal{V} ( W ) \mathcal{G} ( W ) \mathcal{T} ( W )
  \notag \\
  = & \left [ 1 - \mathcal{V} ( W ) \mathcal{G} ( W ) \right ] ^{-1}
  \mathcal{V} ( W ) ,
  \label{eq:scatt_amp}
\end{align}
where
\begin{equation}
  \mathcal{T} ( W )
  \equiv
  \begin{pmatrix}
    T ( W , p_{1} , p_{1} ) & \cdots & T ( W , p_{1} , p_{N} ) &
    \Gamma ( W , p_{1} )
    \\
    \vdots & \ddots & \vdots & \vdots
    \\
    T ( W , p_{N} , p_{1} ) & \cdots & T ( W , p_{N} , p_{N} ) &
    \Gamma ( W , p_{N} ) \\
    \Gamma ( W , p_{1} ) & \cdots & \Gamma ( W , p_{N} ) & \Delta ( W )
  \end{pmatrix} ,
\end{equation}
\begin{equation}
  \mathcal{V} ( W )
  \equiv
  \begin{pmatrix}
    V_{(2)} ( p_{1} , p_{1} ) & \cdots & V_{(2)} ( p_{1} , p_{N} ) &
    V_{(1)} ( p_{1} )
    \\
    \vdots & \ddots & \vdots & \vdots
    \\
    V_{(2)} ( p_{N} , p_{1} ) & \cdots & V_{(2)} ( p_{N} , p_{N} ) &
    V_{(1)} ( p_{N} ) \\
    V_{(1)} ( p_{1} ) & \cdots & V_{(1)} ( p_{N} ) & 0
  \end{pmatrix} ,
\end{equation}
and
\begin{equation}
  \mathcal{G} ( W )
  \equiv
  \text{diag} ( G_{1} ( W ) , G_{2} ( W ) , \ldots , G_{N+1} ( W ) ) 
\end{equation}
with 
\begin{equation}
  G_{n} ( W )
  = \frac{w_{n} p_{n}^{2}}{2 \pi ^{2}}
  \frac{m_{M} m_{B}}{\mathcal{E}_{M} ( p_{n} ) \mathcal{E}_{B} ( p_{n} )}
  \frac{1}{W - W_{MB} ( p_{n} )}
  \label{eq:Gn_mom}
\end{equation}
for $n = 1$, $2$, $\ldots$, $N$, and 
\begin{equation}
  G_{N + 1} ( W )
  = \frac{1}{W - M_{0}} .
  \label{eq:Gn_M0}
\end{equation}
On the other hand, when the energy $W$ is real and above the
meson-baryon threshold, we must append the on-shell momentum
\begin{equation}
  p_{\rm on} ( W )
  \equiv \frac{\sqrt{\lambda ( W^{2} , m_{M}^{2} , m_{B}^{2} )}}{2 W} ,
  \label{eq:pon}
\end{equation}
where $\lambda ( x , y , z ) \equiv x^{2} + y^{2} + z^{2} - 2 x y - 2
y z - 2 z x$ is the K\"{a}llen function, to the array $p_{n}$, and
append
\begin{equation}
  G_{\rm on} ( W )
  =
  - i \frac{p_{\rm on} ( W ) m_{M} m_{B}}{2 \pi W} ,
\end{equation}
which originates from the $+ i 0$ prescription in the propagator, to the
array $G_{n} ( W )$~\eqref{eq:Gn_mom} as well.

\subsection{Compositeness}
\label{sec:composite}

From the scattering amplitude, we can extract the so-called
compositeness~\cite{Hyodo:2013nka, Sekihara:2014kya, Sekihara:2016xnq}
(see also a recent review~\cite{Kinugawa:2024crb}), which quantifies
the fraction of the hadronic molecular component inside a bound state,
as was done in the 1960s to investigate the internal structure of the
deuteron~\cite{Weinberg:1965zz}.

The key point is the fact that the $T$-matrix $\hat{T}$ is formally
expressed in terms of the free Hamiltonian $\hat{H}_{0}$, the interaction
$\hat{V}$, and the full Hamiltonian $\hat{H} \equiv \hat{H}_{0} +
\hat{V}$:
\begin{equation}
  \hat{T}
  = \hat{V} + \hat{V} \frac{1}{W - \hat{H}_{0}} \hat{T}
  = \hat{V} + \hat{V} \frac{1}{W - \hat{H}} \hat{V} ,
\end{equation}
where $W$ is the two-body center-of-mass energy.  Therefore, near the
bound-state mass $W \approx M_{\rm B}$, the $T$-matrix is dominated by
the bound-state pole contribution as
\begin{equation}
  \hat{T}
  = \hat{V} | \Psi _{\rm B} \rangle \frac{1}{W - M_{\rm B}}
  \langle \Psi _{\rm B} | \hat{V}
  + \text{(regular)} ,
  \label{eq:Tmat}
\end{equation}
where $| \Psi _{\rm B} \rangle$ is the physical bound-state vector
satisfying $\hat{H} | \Psi _{\rm B} \rangle = M_{\rm B} | \Psi _{\rm
  B} \rangle$ and $\langle \Psi _{\rm B} | \Psi _{\rm B} \rangle = 1$.

Equation~\eqref{eq:Tmat} indicates that the residue of the scattering
amplitude at the bound-state pole position contains information on the
two-body component of the bound-state wave function.  Namely, when we
denote the free $s$-wave two-body state of relative momentum $p$ by $|
p \rangle$, the two-body component of the bound-state wave function
can be expressed as
\begin{equation}
  \psi _{\rm B} ( p )
  \equiv \langle p | \Psi _{\rm B} \rangle
  = \langle \Psi _{\rm B} | p \rangle .
\end{equation}
This quantity appears in the scattering amplitude $T ( W , p , p^{\prime}
)$:
\begin{equation}
  T ( W , p , p^{\prime} )
  = \langle p^{\prime} | \hat{T} | p \rangle
  = \frac{\gamma ( p ) \gamma ( p^{\prime} )}{W - M_{\rm B}}
  + \text{(regular)} ,
\end{equation}
with
\begin{align}
  \gamma ( p )
  = & \langle p | \hat{V} | \Psi _{\rm B} \rangle
  = \langle \Psi _{\rm B} | \hat{V} | p \rangle
  = \langle p | ( \hat{H} - \hat{H}_{0} ) | \Psi _{\rm B} \rangle
  \notag \\
  = & [ M_{\rm B} - W_{MB} ( p )]
  \psi _{\rm B} ( p ) .
  \label{eq:gamma_p}
\end{align}
Then, the compositeness $X$ is defined as the norm of the two-body
component of the bound-state wave function $\psi _{\rm B} ( p )
= \gamma ( p ) / [ M_{\rm B} - W_{MB} ( p )]$~\cite{Sekihara:2014kya,
  Sekihara:2016xnq}:
\begin{equation}
  X \equiv \int _{0}^{\infty} d p \frac{p^{2}}{2 \pi ^{2}}
  \frac{m_{M} m_{B}}{\mathcal{E}_{M} ( p ) \mathcal{E}_{B} ( p )}
  \left [ \frac{\gamma ( p )}{M_{\rm B} - W_{MB} ( p )} \right ] ^{2} ,
\end{equation}
where the factor $m_{M} m_{B} / [ \mathcal{E}_{M} ( p )
  \mathcal{E}_{B} ( p ) ]$ originates from the propagators of the
meson and baryon, as in Eq.~\eqref{eq:prop}.  We note that we do not
need to take the complex conjugate for $\psi _{\rm B} ( p )$ or
$\gamma ( p )$ in the residue due to the time-reversal invariance of
the scattering.  Hence, we have the usual square of the wave function
instead of the absolute value squared in the present
formulation~\cite{Sekihara:2014kya}.  Nevertheless, a stable bound
state yields a real-valued compositeness.

We can transform the two-body component of the bound-state wave
function $\psi _{\rm B} ( p )$, which is in momentum space, into that
in coordinate space via the Fourier transform:
\begin{align}
  \langle r | \Psi _{\rm B} \rangle
  = & \int \frac{d^{3} p}{(2 \pi )^{3}}
  \frac{m_{M} m_{B}}{\mathcal{E}_{M} ( p ) \mathcal{E}_{B} ( p )}
  \langle r | p \rangle \langle p | \Psi _{\rm B} \rangle
  \notag \\
  = & \int _{0}^{\infty} d p \frac{p^{2}}{2 \pi ^{2}}
  \frac{m_{M} m_{B}}{\mathcal{E}_{M} ( p ) \mathcal{E}_{B} ( p )}
  j_{0} ( p r ) \psi _{\rm B} ( p ) ,
\end{align}
where $| r \rangle$ is the noninteracting meson-baryon state with
spatial distance $r$ and $\langle r | p \rangle = e^{i \bm{r} \cdot
  \bm{p}}$.  In fact, this form has already appeared in
Eq.~\eqref{eq:Psi_B}.  Together with relation~\eqref{eq:gamma_p},
we can express $\Psi _{\rm B} ( r )$ in Eq.~\eqref{eq:Psi_B} in terms
of the two-body component of the bound-state wave function as
\begin{equation}
  \Psi _{\rm B} ( r )
  = \psi _{\rm B} ( 0 ) \langle r | \Psi _{\rm B} \rangle .
  \label{eq:PsiB_equiv}
\end{equation}
Because $| p = 0 \rangle$ is precisely $| M B ( p = 0 ) \rangle$
introduced in Eq.~\eqref{eq:Rcorr_MB_phys}, we also have $\psi _{\rm
  B} ( 0 ) = \langle \Psi _{\rm B} | M B ( p = 0 ) \rangle$, which
leads to the agreement between the first term in
Eq.~\eqref{eq:Rcorr_MB} and the first term in
Eq.~\eqref{eq:Rcorr_MB_phys}.

In a similar manner, we have
\begin{equation}
  \Gamma ( W , p )
  = \langle p | \hat{T} | \Lambda _{0} \rangle
  = \langle \Lambda _{0} | \hat{T} | p \rangle
  = \frac{\gamma ( p ) \gamma _{0}}{W - M_{\rm B}}
  + \text{(regular)} ,
\end{equation}
\begin{equation}
  \Delta ( W )
  = \langle \Lambda _{0} | \hat{T} | \Lambda _{0} \rangle
  = \frac{\gamma _{0} \gamma _{0}}{W - M_{\rm B}}
  + \text{(regular)} ,
\end{equation}
where the residue contains 
\begin{align}
  \gamma _{0}
  = & \langle \Psi _{\rm B} | \hat{V} | \Lambda _{0} \rangle
  = \langle \Psi _{\rm B} | ( \hat{H} - \hat{H}_{0} ) | \Lambda _{0} \rangle
  \notag \\
  = & ( M_{\rm B} - M_{0} ) 
  \langle \Psi _{\rm B} | \Lambda _{0} \rangle .
  \label{eq:gamma0}
\end{align}
Therefore, we define the bare-state contribution $Z$ as
\begin{equation}
  Z \equiv
  \langle \Psi _{\rm B} | \Lambda _{0} \rangle
  \langle \Lambda _{0} | \Psi _{\rm B} \rangle
  = \left ( \frac{\gamma _{0}}{M_{\rm B} - M_{0}} \right ) ^{2} .
\end{equation}
In terms of the bare state, we can rewrite $\Psi _{\rm B0} ( r )$ in
Eq.~\eqref{eq:Psi_Bzero} as
\begin{equation}
  \Psi _{\rm B0} ( r )
  = \langle r | \Psi _{\rm B} \rangle
  \langle \Psi _{\rm B} | \Lambda _{0} \rangle .
  \label{eq:PsiBzero_equiv}
\end{equation}
This proves the agreement between the first term in
Eq.~\eqref{eq:Rcorr_bare} and the first term in
Eq.~\eqref{eq:Rcorr_bare_phys}.

When the interaction has no intrinsic energy dependence, we
have the normalization condition for the compositeness and the bare-state
contribution for the bound state~\cite{Sekihara:2014kya}:
\begin{equation}
  X + Z = 1 .
  \label{eq:XplusZ}
\end{equation}
On the other hand, introducing an intrinsic energy dependence
into the interaction, as discussed at the end of Sec.~\ref{sec:singular},
would violate the sum rule~\eqref{eq:XplusZ}.

\subsection{Scattering amplitude in lattice spacetime}
\label{eq:amp_lat}

When considering lattice spacetime, both the relative distance
$\bm{r}$ and relative momentum $\bm{p}$ are discretized as in
Eqs.~\eqref{eq:r_an} and \eqref{eq:p_2Ln}.  We can then translate the
scattering amplitude into that in lattice spacetime using the
formulae in Sec.~\ref{sec:lattice} and Appendix~\ref{app:projection}
(see also Ref.~\cite{Doring:2011vk} for scattering amplitudes in
lattice spacetime).

We can use the same $V_{(1)}$ as in the continuum
spacetime case, but we must replace the $M B \to M B$ interaction term
$V_{\rm int}$~\eqref{eq:Vint} with $\tilde{V}_{\rm int}$:
\begin{align}
  & \tilde{V}_{\rm int} ( V_{\rm L} , b_{\rm L} ; p_{\alpha} , p_{\alpha ^{\prime}} )
  \notag \\
  & = 
  \frac{( \pi b_{\rm L}^{2} )^{3/2} V_{\rm L}}{24}
  \sum _{g \in G}  
  \exp \left ( - \frac{| g \bm{p}_{\alpha}
    - \bm{p}_{\alpha ^{\prime}} |^{2} b_{\rm L}^{2}}{4} \right ) .
\end{align}
We note that the summation over $g \in G$ yields a function
that depends only on the labels $\alpha$ and $\alpha ^{\prime}$,
where $\alpha$, $\alpha ^{\prime} = 1$, $2$, $\ldots$, $A$ are the
labels in Table~\ref{tab:A1plus}, as shown in
Appendix~\ref{app:projection}.  With this interaction term, we
calculate the scattering amplitude $\tilde{\mathcal{T}}$ in
lattice spacetime as a solution of the Lippmann--Schwinger equation:
\begin{equation}
  \tilde{\mathcal{T}} ( W )
  = \left [ 1 - \tilde{\mathcal{V}} ( W )
    \tilde{\mathcal{G}} ( W ) \right ] ^{-1}
  \tilde{\mathcal{V}} ( W ) ,
  \label{eq:Ttilde}
\end{equation}
where
\begin{equation}
  \tilde{\mathcal{T}} ( W )
  \equiv
  \begin{pmatrix}
    \tilde{T} ( W , p_{1} , p_{1} ) & \cdots & \tilde{T} ( W , p_{1} , p_{A} ) &
    \tilde{\Gamma} ( W , p_{1} )
    \\
    \vdots & \ddots & \vdots & \vdots
    \\
    \tilde{T} ( W , p_{A} , p_{1} ) & \cdots & \tilde{T} ( W , p_{A} , p_{A} ) &
    \tilde{\Gamma} ( W , p_{A} ) \\
    \tilde{\Gamma} ( W , p_{1} ) & \cdots & \tilde{\Gamma} ( W , p_{A} ) &
    \tilde{\Delta} ( W )
  \end{pmatrix} ,
  \label{eq:barT}
\end{equation}
\begin{equation}
  \tilde{\mathcal{V}} ( W )
  \equiv
  \begin{pmatrix}
    \tilde{V}_{(2)} ( p_{1} , p_{1} ) & \cdots &
    \tilde{V}_{(2)} ( p_{1} , p_{A} ) &
    V_{(1)} ( p_{1} )
    \\
    \vdots & \ddots & \vdots & \vdots
    \\
    \tilde{V}_{(2)} ( p_{A} , p_{1} ) & \cdots &
    \tilde{V}_{(2)} ( p_{A} , p_{A} ) &
    V_{(1)} ( p_{A} ) \\
    V_{(1)} ( p_{1} ) & \cdots & V_{(1)} ( p_{A} ) & 0
  \end{pmatrix} ,
\end{equation}
\begin{equation}
  \tilde{\mathcal{G}} ( W )
  \equiv
  \text{diag} ( \tilde{G}_{1} ( W ) , \tilde{G}_{2} ( W ) , \ldots ,
  \tilde{G}_{A + 1} ( W ) ) ,
\end{equation}
\begin{equation}
  \tilde{G}_{\alpha} ( W )
  \equiv \frac{v_{\alpha}}{L^{3}}
  \frac{m_{M} m_{B}}{\mathcal{E}_{M} ( p_{\alpha} ) \mathcal{E}_{B} ( p_{\alpha} )}
  \frac{1}{W - W_{MB} ( p_{\alpha} )} 
  \label{eq:Gtilde}
\end{equation}
for $\alpha = 1$, $2$, $\ldots$, $A$, and
\begin{equation}
  \tilde{G}_{A + 1} ( W )
  = \frac{1}{W - M_{0}} .
\end{equation}
The $+ i 0$ prescription in the propagator is unnecessary in
lattice spacetime.

\section{Model analysis: meson-baryon elastic scattering}
\label{sec:modelA}

Now we start the model analysis using the formulations provided in the
previous sections.  Namely, with a given meson-baryon interaction, we
calculate the R-correlator~\eqref{eq:Rcorr_MB} and the HAL QCD
potential~\eqref{eq:V_HAL_cont} or their lattice
counterparts~\eqref{eq:Rtilde_fin}, \eqref{eq:V_HAL_latt}.  Then, we
examine whether the obtained HAL QCD potential can reproduce the
physical quantities evaluated directly from the original meson-baryon
interaction and scattering amplitude.  In this section, we focus on
meson-baryon elastic scattering, so we fix $f_{0} = 0$.  The meson and
baryon masses are fixed according to Ref.~\cite{Murakami:2023phq}:
$m_{M} = \SI{671.2}{MeV}$ and $m_{B} = \SI{1488.8}{MeV}$.  The other
model parameters are set so as to cover several physical conditions
(see Secs.~\ref{sec:modelA_NR} and \ref{sec:modelA_lat}).  Throughout
this study, the time $t$ for evaluating the R-correlators and HAL QCD
potentials is fixed at $t = 7 a$ with $a = \SI{0.121}{fm}$, as done in
Ref.~\cite{Murakami:2023phq}.  We note that the evaluation time $t =
7a$ only controls the strength of the elastic meson-baryon scattering
states in our model, whereas it is tuned to suppress inelastic
contamination in actual lattice QCD simulations.

\subsection{Nonrelativistic formulation}
\label{sec:modelA_NR}

\begin{table}[!t]
  \caption{Parameters $( V_{\rm L} , b_{\rm L} )$, number of ({\#})
    interaction terms, and resulting binding energies $B_{\rm E}$ in
    the meson-baryon elastic scattering case.  In the binding energy
    rows, ``NR'', ``SR'', ``HAL'', ``lattice'', and ``N${}^{2}$LO''
    indicate the binding energies in the nonrelativistic formulation,
    in the semirelativistic formulation, from the HAL QCD potentials,
    in lattice spacetime with periodic boundary conditions, and from
    the HAL QCD potentials at the next-to-next-to-leading order
    (N${}^{2}$LO) analysis, respectively.}
  \label{tab:elastic}
  \centering
  \begin{ruledtabular}
    \begin{tabular}{llcccc}
      & & Set I & Set II & Set III & Set IV
      \\
      \hline
      $V_{\rm L}$ [MeV] & &
      $- 645.9$ & $- 300.0$ & $300.0$ & $- 709.1$
      \\
      $b_{\rm L}$ [fm] & &
      $0.50$ & $0.50$ & $0.50$ & $0.60$
      \\
      \multicolumn{2}{l}{\# interaction term}
      & 
      1 & 1 & 1 & 2
      \\
      $B_{\rm E}$ [MeV] & NR &
      $20.0$ & --- & --- & $10.0$
      \\
      & SR &
      $\phantom{0}6.4$ & --- & --- & $\phantom{0}5.4$
      \\
      & HAL &
      $\phantom{0}3.6$ & --- & --- & $\phantom{0}3.8$
      \\
      & lattice &
      $20.0$ & ($5.1$) & --- & 20.6
      \\
      & N${}^{2}$LO &
      $\phantom{0}5.5$ & --- & --- & $\phantom{0}4.6$
    \end{tabular}
  \end{ruledtabular}
\end{table}

\subsubsection{Setup}

First of all, to grasp the features of our framework, we perform the
nonrelativistic reduction in this subsection, where the factor in the
momentum integrals becomes
\begin{equation}
  \frac{m_{M} m_{B}}{\mathcal{E}_{M} ( k ) \mathcal{E}_{B} ( k )}
  \to 1 ,
\end{equation}
and hence the Lippmann--Schwinger equation simplifies to
\begin{align}
  & T ( E , p , p^{\prime} )
  \notag \\ &
  = V_{(2)} ( p , p^{\prime} )
  + \int _{0}^{\infty} d k \frac{k^{2}}{2 \pi ^{2}}
  \frac{V_{(2)} ( p , k ) T ( E , k , p^{\prime} )}{E
    - E_{\rm NR} ( k ) + i 0} ,
  \label{eq:LS_NR}
\end{align}
with $E_{\rm NR} ( k ) \equiv k^{2} / ( 2 \mu )$.  The evaluation of
the R-correlator~\eqref{eq:Rcorr_MB} is performed with the replacement
$W - M_{\rm th} \to E$.  Finally, the evaluation of the HAL QCD
potentials~\eqref{eq:V_HAL_cont} is performed without the second
derivative $\partial ^{2} / \partial t^{2}$ term.

In this study, we employ four sets of model parameters as follows.
The first set (I) provides a bound state with binding energy $B_{\rm
  E} = \SI{20.0}{MeV}$ using a single interaction term in the
nonrelativistic formulation.  The second set (II) is attractive but
not strong enough to generate a bound state with a single interaction
term, while the third set (III) is repulsive with a single interaction
term.  The fourth set (IV) provides a bound state of $B_{\rm E} =
\SI{10.0}{MeV}$ with two interaction terms: a repulsive core plus an
attractive pocket potential~\eqref{eq:V2_II}.  The interaction range
is fixed at $b = \SI{0.50}{fm}$ ($b = \SI{0.60}{fm}$) for sets I--III
(set IV).  The parameters are summarized in Table~\ref{tab:elastic}.

\subsubsection{Coincidence of HAL QCD potentials with input potentials}

\begin{figure}[t]
  \centering
  \includegraphics[width=8.6cm]{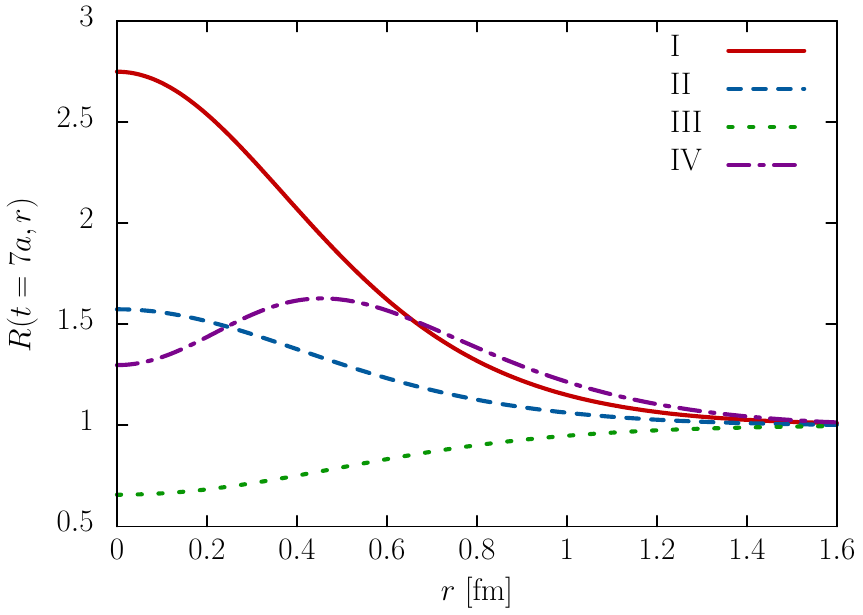}
  \caption{R-correlators in the nonrelativistic formulation of
    meson-baryon elastic scattering.  The time $t$ for the evaluation
    is fixed at $t = 7 a$.  We employ the model parameters of sets I--IV
    in Table~\ref{tab:elastic}.}
  \label{fig:Rcorr_NR}
\end{figure}

With the formulation and setup above, we can calculate the
R-correlators~\eqref{eq:Rcorr_MB}.  The results are shown in
Fig.~\ref{fig:Rcorr_NR}.  As can be seen, an attractive (repulsive)
interaction leads to an increase (decrease) of the R-correlator near
the origin.  This behavior is reasonable, as the wave function prefers
to reside in the attractive region, affecting the R-correlator via the
factor $\langle r | \Psi _{\rm B} \rangle$ or $\langle r | p_{\rm
  phys} \rangle$ in Eq.~\eqref{eq:Rcorr_MB_phys}.  In this sense, it
is worth noting the slight decrease at the origin for parameter set
IV, where a repulsive core is present.

\begin{figure}[t]
  \centering
  \includegraphics[width=8.6cm]{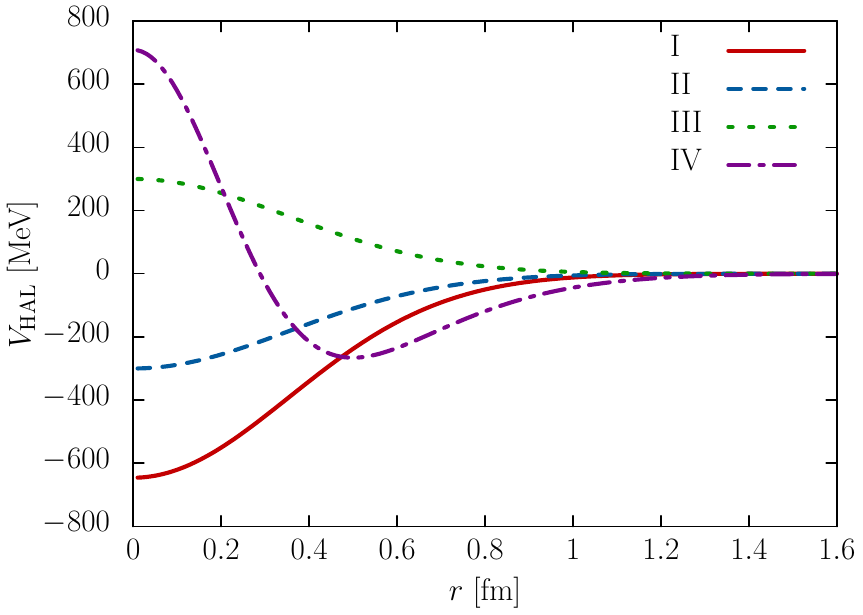}
  \caption{HAL QCD potentials in the nonrelativistic formulation of
    meson-baryon elastic scattering.  The time $t$ for the evaluation
    is fixed at $t = 7 a$.  We employ the model parameters of sets I--IV
    in Table~\ref{tab:elastic}.}
  \label{fig:VHAL_NR}
\end{figure}

From the R-correlators, using formula~\eqref{eq:V_HAL_cont} while
neglecting the second derivative $\partial ^{2} / \partial t^{2}$
term, we can evaluate the HAL QCD potentials.  The results are shown
in Fig.~\ref{fig:VHAL_NR}.  Most remarkably, the HAL QCD potentials
exactly coincide with the original input potentials, which are local
and described by single (I, II, and III) and double (IV) Gaussians.
This is almost trivial because the R-correlators contain the factors
$\langle r | \Psi _{\rm B} \rangle$ and $\langle r | p_{\rm phys}
\rangle$, both of which automatically satisfy the Schr\"{o}dinger
equation in the nonrelativistic formulation:
\begin{equation}
  \left [ \frac{\partial}{\partial t}
    - \frac{1}{2 \mu r} \left ( \frac{d}{d r} \right ) ^{2} r
    + V ( r ) \right ] \langle r | \Psi _{E} \rangle e^{- E t} 
  = 0 ,
  \label{eq:r_PsiE}
\end{equation}
where $| \Psi _{E} \rangle$ denotes $| \Psi _{\rm B} \rangle$ or $| p_{\rm
  phys} \rangle$.

\begin{figure}[t]
  \centering
  \includegraphics[width=8.6cm]{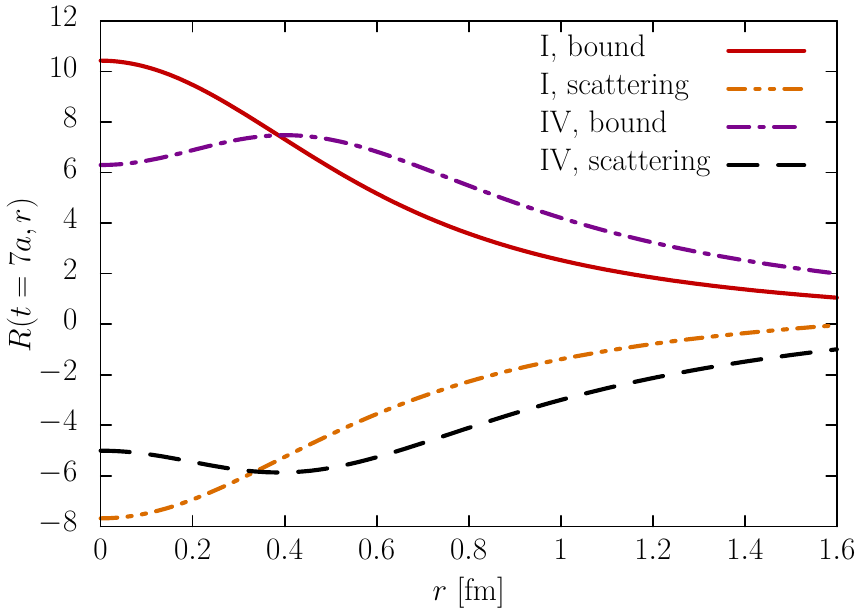}
  \caption{Decomposition of the R-correlators in the nonrelativistic
    formulation of meson-baryon elastic scattering.  The time $t$
    for the evaluation is fixed at $t = 7 a$.  We employ the model
    parameters of sets I and IV in Table~\ref{tab:elastic}.}
  \label{fig:Rcorr_NR_each}
\end{figure}

To inspect this point more closely and to understand the HAL QCD
potentials more clearly, we decompose the R-correlators for parameter
sets I and IV into contributions from the bound state ($E = - B_{\rm
  E}$) and those from scattering states ($E \ge 0$).  The decomposed
R-correlators are plotted in Fig.~\ref{fig:Rcorr_NR_each}.  Comparing
them with the full R-correlators in Fig.~\ref{fig:Rcorr_NR}, we see
that while the bound-state contributions are indeed large for both
sets I and IV, cancellations with the scattering states are
nonnegligible.  Nevertheless, owing to the dominant bound-state
contributions, especially originating from the eigenenergy factor
$e^{- E t}$, there are no nodes in the full R-correlators.  We note
here that even if we restrict the energy range used to evaluate the
R-correlators (e.g., considering only the bound-state or
scattering-state contributions as in Fig.~\ref{fig:Rcorr_NR_each}), we
obtain the same HAL QCD potentials~\eqref{eq:V_HAL_cont} in the
nonrelativistic formulation.  This can be understood from the
Schr\"{o}dinger equation~\eqref{eq:r_PsiE}.  By virtue of this
equation, any superposition of $\langle r | \Psi _{E} \rangle e^{- E
  t}$ satisfies the same Schr\"{o}dinger equation~\eqref{eq:r_PsiE},
leading to identical HAL QCD potentials via
formula~\eqref{eq:V_HAL_cont}.

\subsection{Semirelativistic formulation}
\label{sec:modelA_SR}

\subsubsection{R-correlators and HAL QCD potentials}

We extend our discussion to the semirelativistic formulation, which
corresponds to the one developed in the previous sections.  We use the same
parameters as in Table~\ref{tab:elastic}.  The binding energies of the
bound state in parameter sets I and IV, which correspond to the pole
positions of the scattering amplitudes $T ( W , p , p^{\prime} )$
measured from the threshold, are $\SI{6.4}{MeV}$ and $\SI{5.4}{MeV}$,
respectively; these are listed in Table~\ref{tab:elastic} as ``$B_{\rm
  E}$ SR''.  They are smaller than those in the nonrelativistic
formulation.  This reduction is mainly due to the factor $m_{M} m_{B} / [
  \mathcal{E}_{M} ( k ) \mathcal{E}_{B} ( k ) ]$ in the momentum
integrals, which suppresses the attractive strength of the interaction.

\begin{figure}[t]
  \centering
  \includegraphics[width=8.6cm]{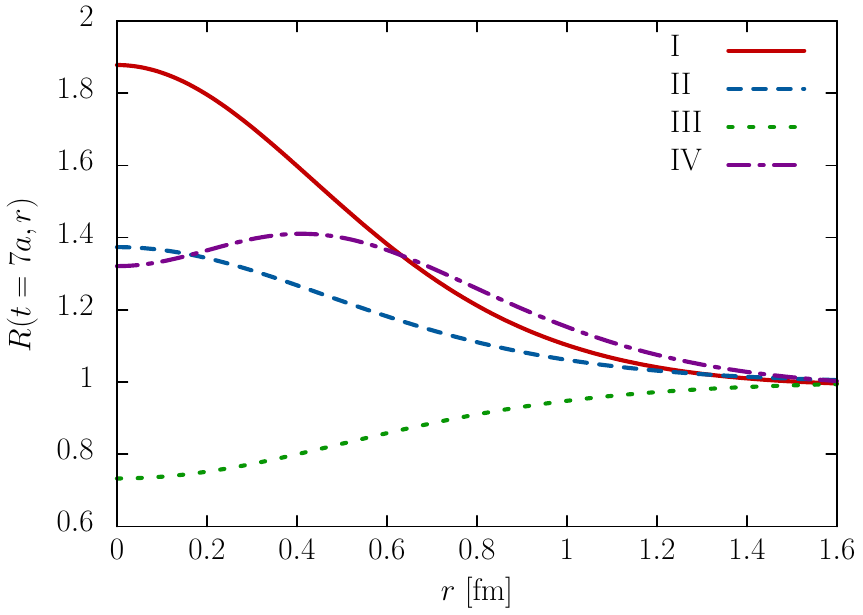}
  \caption{R-correlators in the semirelativistic formulation of
    meson-baryon elastic scattering.  The time $t$ for the evaluation
    is fixed at $t = 7 a$.  We employ the model parameters of sets I--IV
    in Table~\ref{tab:elastic}.}
  \label{fig:Rcorr_SR}
\end{figure}

The resulting R-correlators are shown in Fig.~\ref{fig:Rcorr_SR}.
Compared with the nonrelativistic case, the R-correlators exhibit
quite similar behavior, although they are slightly suppressed.  In
fact, the scattering wave functions, bound-state wave functions, and
binding energies are all modified in the semirelativistic formulation
compared to the nonrelativistic case.  Nevertheless, the mechanism of
cancellation between the ground-state and higher-state contributions
remains essentially the same.

\begin{figure}[t]
  \centering
  \includegraphics[width=8.6cm]{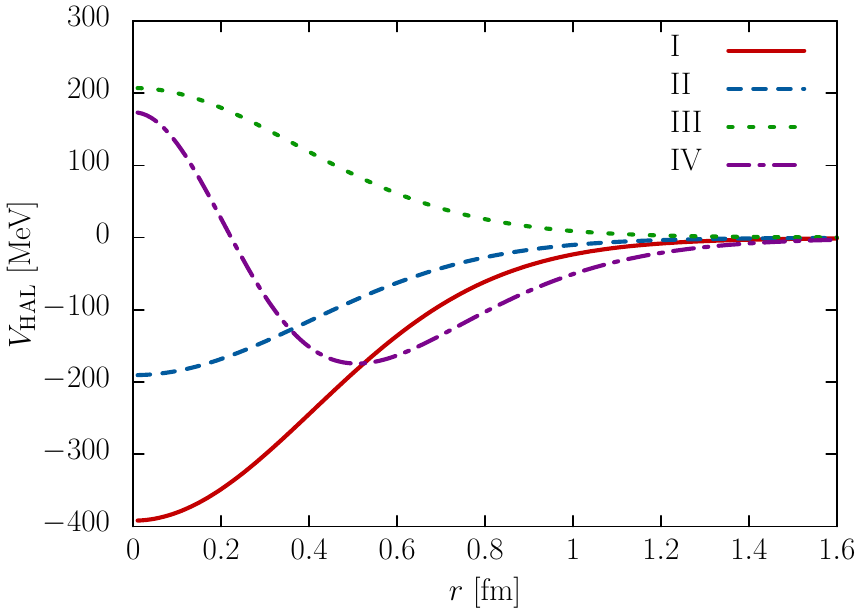}
  \caption{HAL QCD potentials in the semirelativistic formulation of
    meson-baryon elastic scattering.  The time $t$ for the evaluation
    is fixed at $t = 7 a$.  We employ the model parameters of sets I--IV
    in Table~\ref{tab:elastic}.}
  \label{fig:VHAL_SR}
\end{figure}

From the R-correlators, we calculate the HAL QCD potentials as shown in
Fig.~\ref{fig:VHAL_SR}.  Compared with the nonrelativistic case, while
the HAL QCD potentials are quantitatively very similar, their
strength near the origin becomes more moderate owing to the factor $m_{M}
m_{B} / [ \mathcal{E}_{M} ( k ) \mathcal{E}_{B} ( k ) ]$, as
mentioned above.

\subsubsection{Reproduction of physical quantities}

Next, let us examine the effectiveness of the HAL QCD method by
considering binding energies and phase shifts.  These are evaluated in
two ways: directly from the scattering amplitude $T ( W , p ,
p^{\prime} )$~\eqref{eq:Tamp}, and by solving the Schr\"{o}dinger
equation with the HAL QCD potentials~\eqref{eq:V_HAL_cont}.  We note
that we treat the former as the primary benchmark because the
scattering amplitude $T ( W , p , p^{\prime} )$ contains comprehensive
information on the two-body system.  Furthermore, while the former
approach uses the semirelativistic formulation, the latter relies on
the nonrelativistic formulation once the HAL QCD potentials are
evaluated.

The binding energies evaluated via the former method (i.e., from the
pole positions of the scattering amplitudes relative to the threshold)
are $\SI{6.4}{MeV}$ (I) and $\SI{5.4}{MeV}$ (IV).  On the other hand,
those obtained via the latter method are $3.6$ (I) and $\SI{3.8}{MeV}$
(IV), as shown in the row ``$B_{\rm E}$ HAL'' of
Table~\ref{tab:elastic}, which are obtained by numerically solving the
Schr\"{o}dinger equation with the HAL QCD potentials.  Furthermore,
parameter sets II and III do not generate any bound states in either
approach.  These results indicate that the HAL QCD potentials
qualitatively capture the existence or absence of bound states
correctly.  However, the binding energies differ by a few MeV.  This
is reasonable, as we are measuring a binding energy of a few MeV while
probing a HAL QCD potential on the order of several hundred MeV near
the origin.  Consequently, a small percentage error in the potential,
arising, e.g., from the derivative expansion or nonrelativistic
reduction in the HAL QCD method, can shift the binding energy
significantly.

\begin{figure}[t]
  \centering
  \includegraphics[width=8.6cm]{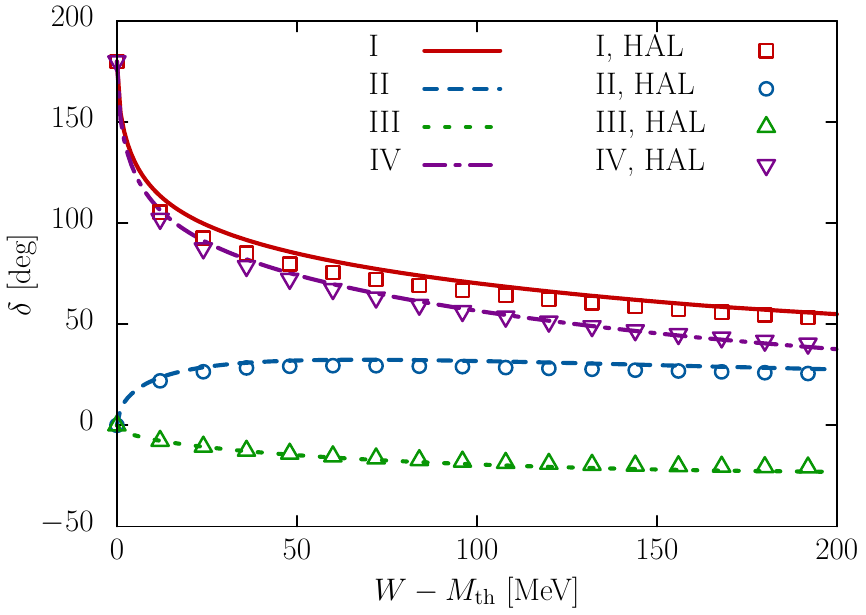}
  \caption{Phase shifts of meson-baryon elastic scattering.  The
    lines indicate the phase shifts evaluated directly from the
    scattering amplitude~\eqref{eq:phase_shift}, while the points indicate
    those obtained by solving the Schr\"{o}dinger equation with the HAL QCD
    potentials.  We employ the model parameters of sets I--IV in
    Table~\ref{tab:elastic}.}
  \label{fig:delta_SR}
\end{figure}

Next, we evaluate the phase shifts using both approaches.  The results
are shown in Fig.~\ref{fig:delta_SR} up to an energy where
nonrelativistic treatment remains valid: $W - M_{\rm th} = \sqrt{p^{2}
  + m_{M}^{2}} + \sqrt{p^{2} + m_{B}^{2}} - M_{\rm th} \simeq p^{2} /
( 2 \mu ) \approx \SI{200}{MeV}$, corresponding to a center-of-mass
momentum $p \approx \SI{430}{MeV}$.  As can be seen, the phase shifts
obtained from the HAL QCD potentials reproduce those calculated
directly from the original scattering amplitude,
\begin{equation}
  \delta ( W )
  = \arg [ - T ( W , p_{\rm on} ( W ) , p_{\rm on} ( W ) ) ]
  ,
  \label{eq:phase_shift}
\end{equation}
fairly well.

These findings demonstrate that the HAL QCD method is quantitatively
effective when meson-baryon scattering is elastic and the interaction
is predominantly local, as represented by the interaction
term~\eqref{eq:Gaussian}.  Of course, one should clarify how
nonlocality affects the HAL QCD potentials.  This point will be
examined in Sec.~\ref{sec:N2LO} in terms of the N${}^{2}$LO
potentials.

\subsection{Lattice discretization and finite-volume effects of spacetime}
\label{sec:modelA_lat}

Now, let us discretize spacetime and examine how our results above are
modified in a lattice spacetime of finite volume.  The formulations
for the HAL QCD method in lattice spacetime were developed in
Sec.~\ref{sec:lattice}, and those for the scattering amplitude in
Sec.~\ref{eq:amp_lat}.  In the present study, we employ $a =
\SI{0.121}{fm}$ and $N = 32$ following Ref.~\cite{Murakami:2023phq},
and adopt periodic boundary conditions.  We first use the same
parameter sets I--IV listed in Table~\ref{tab:elastic}.

An interesting feature is that, owing to periodic boundary conditions
and interactions with particles in neighboring boxes, a ``bound
state'' appears for parameter set II at $\SI{5.1}{MeV}$ below
threshold as a pole of the scattering amplitude.  However, this state
shifts back into the scattering state at the threshold in the
infinite-volume limit, since the attraction in set II is insufficient
to form a bound state in infinite volume.  Furthermore, periodic
boundary conditions make the bound states in parameter sets I and IV
deeper~\cite{Sekihara:2013wlq}.  The binding energy values evaluated
from the pole positions of the scattering amplitudes are summarized in
Table~\ref{tab:elastic} under ``$B_{\rm E}$ lattice''.

\begin{figure}[t]
  \centering
  \includegraphics[width=8.6cm]{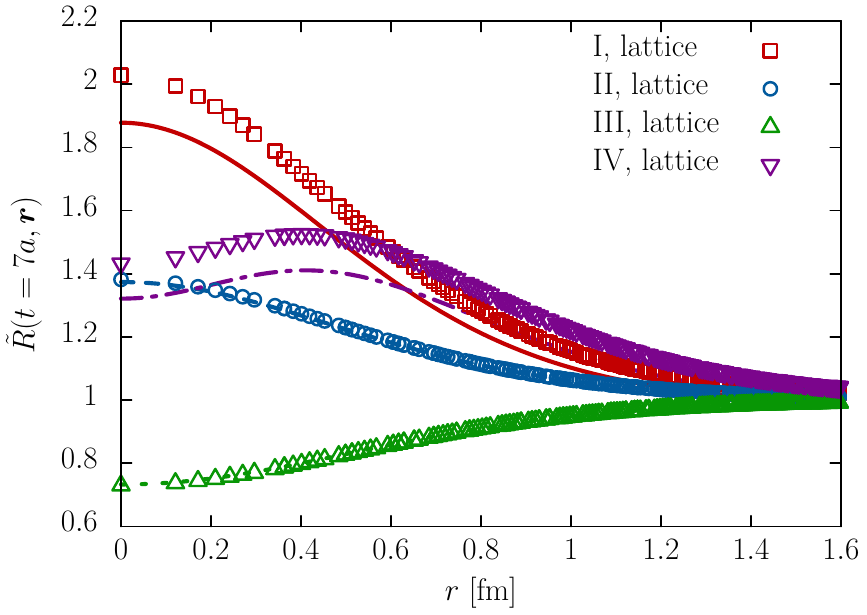}
  \caption{R-correlators in the semirelativistic formulation of
    meson-baryon elastic scattering in lattice spacetime with periodic
    boundary conditions.  The time $t$ for evaluation is fixed at $t =
    7 a$.  We employ the model parameters of sets I--IV in
    Table~\ref{tab:elastic}.  The lines represent the results in
    infinite-volume continuum spacetime shown in
    Fig.~\ref{fig:Rcorr_SR}.}
  \label{fig:Rcorr_SR_lat}
\end{figure}

Using formula~\eqref{eq:Rtilde_fin}, we obtain the R-correlators in
lattice spacetime $\tilde{R} ( t , \bm{r} )$, shown as points in
Fig.~\ref{fig:Rcorr_SR_lat}.  We note here that, strictly speaking,
the R-correlators (and HAL QCD potentials described later as well) are
multi-valued functions of $r \equiv | \bm{r} |$ rather than
single-valued functions.  Indeed, at certain values of $r$ (e.g., $r =
3 a = \sqrt{3^{2} + 0^{2} + 0^{2}} \cdot a = \sqrt{2^{2} + 2^{2} +
  1^{2}} \cdot a$), the R-correlators can take multiple values for
different grid vectors $\bm{r}$.  However, such differences at
identical $r$ are extremely small and invisible in
Fig.~\ref{fig:Rcorr_SR_lat}, because only higher partial waves ($l \ge
4$) contribute to anisotropy under $A_{1}^{+}$ projection.

\begin{figure}[t]
  \centering
  \includegraphics[width=8.6cm]{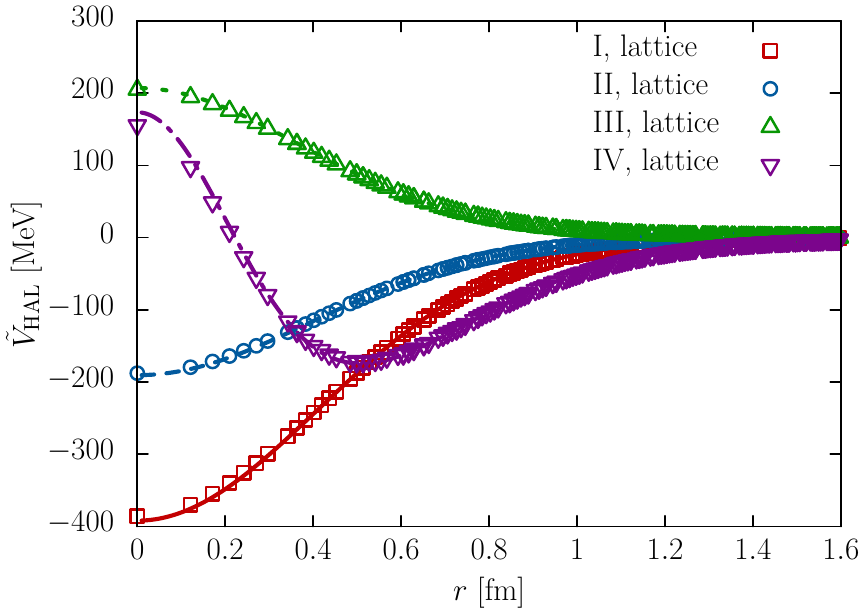}
  \caption{HAL QCD potentials in the semirelativistic formulation of
    meson-baryon elastic scattering in lattice spacetime with periodic
    boundary conditions.  The time $t$ for evaluation is fixed at $t =
    7 a$.  We employ the model parameters of sets I--IV in
    Table~\ref{tab:elastic}.  The lines represent the results in
    infinite-volume continuum spacetime shown in
    Fig.~\ref{fig:VHAL_SR}.}
  \label{fig:VHAL_SR_lat}
\end{figure}

For parameter sets II and III, which do not generate a bound state in
infinite-volume continuum spacetime, the R-correlators show no visible
discrepancy compared to the continuum results shown as lines in
Fig.~\ref{fig:Rcorr_SR_lat}.  On the other hand, for sets I and IV,
slight discrepancies appear relative to continuum spacetime, stemming
from delicate cancellations between bound-state and scattering-state
contributions.  Nevertheless, even though spatial sampling points for
$r$ are discretized due to the lattice grid, the points match the
continuum lines almost quantitatively across all parameter sets.  This
indicates that discretization effects on the R-correlators remain
below roughly 10 percent.

\begin{table}[!t]
  \caption{Parameters $( V_{\rm L} , b_{\rm L} )$ and resulting
    binding energies $B_{\rm E}$ in the meson-baryon elastic
    scattering case.  In all parameter sets, the number of the
    interaction term is $1$.  In the binding energy rows, ``NR'',
    ``SR'', ``HAL'', and ``lattice'' indicate the binding energies in
    the nonrelativistic formulation, in the semirelativistic
    formulation, from the HAL QCD potentials, and in lattice spacetime
    with periodic boundary conditions, respectively.}
  \label{tab:para_ranges}
  \centering
  \begin{ruledtabular}
    \begin{tabular}{llcccc}
      & & Set V & Set VI & Set VII & Set VIII
      \\
      \hline
      $V_{\rm L}$ [MeV] & &
      $- 3279.5$ & $- 943.6$ & $- 305.2$ & $- 118.4$
      \\
      $b_{\rm L}$ [fm] & &
      $0.20$ & $0.40$ & $0.80$ & $1.60$
      \\
      $B_{\rm E}$ [MeV] & NR &
      $20.0$ & $20.0$ & $20.0$ & $20.0$
      \\
      & SR &
      --- & $\phantom{0}2.6$ & $13.2$ & $17.9$
      \\
      & HAL &
      --- & $\phantom{0}0.6$ & $10.8$ & $17.0$
      \\
      & lattice &
      ($6.2$) & $16.0$ & $28.2$ & $50.5$
      \\
    \end{tabular}
  \end{ruledtabular}
\end{table}

By computing finite differences in lattice
spacetime~\eqref{eq:V_HAL_latt}, we obtain the HAL QCD potentials
$\tilde{V}_{\rm HAL} ( \bm{r} )$.  The results are displayed in
Fig.~\ref{fig:VHAL_SR_lat} as points, together with the continuum
results shown as lines.  Comparing the points and lines, we observe
that the HAL QCD potentials on the lattice deviate only slightly from
their continuum counterparts, except near the origin, where
discretization effects become significant.  Specifically, the points
adjacent to the origin $( 0 , 0 , 0 )$ along the $x$-axis are $( \pm 1
, 0 , 0 )$, whereas those near $( 3 , 2 , 1 )$ include $( 4 , 2 , 1 )$
and $( 2 , 2 , 1 )$, which probe differences over a smaller radial
interval: the former interval corresponds to $\pm 1$, whereas the
latter corresponds to $\sqrt{14} - \sqrt{21} \approx - 0.84$ or
$\sqrt{14} - 3 \approx 0.74$.

\begin{figure}[t]
  \centering
  \includegraphics[width=8.6cm]{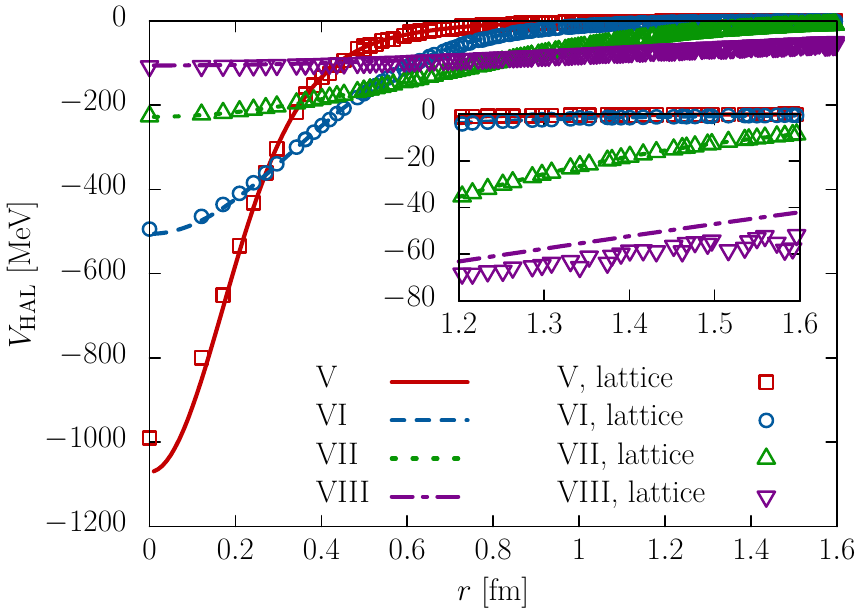}
  \caption{HAL QCD potentials with various interaction ranges in
    Table~\ref{tab:para_ranges}.  The inset represents an enlarged
    figure.}
  \label{fig:VHAL_ranges}
\end{figure}

\begin{figure*}[t]
  \centering
  \includegraphics[width=8.6cm]{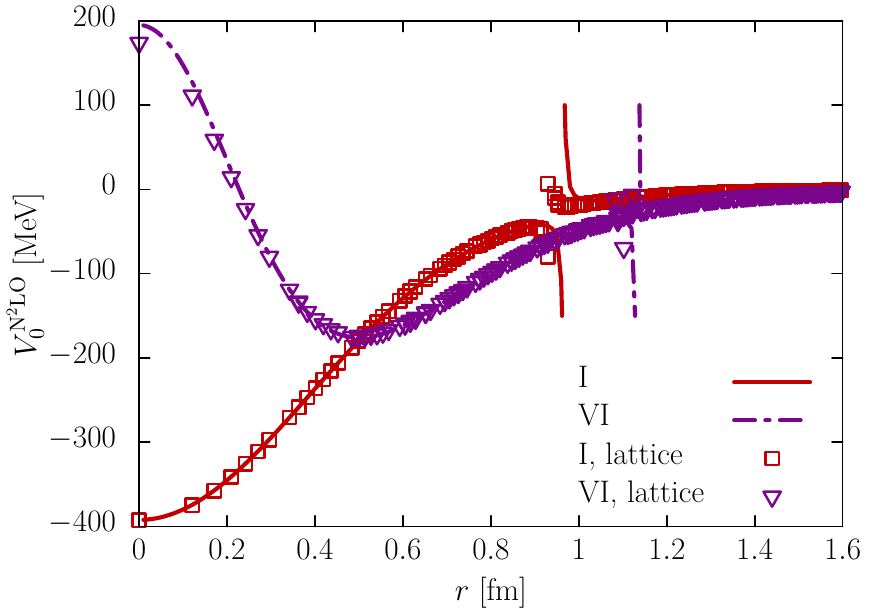}
  ~
  \includegraphics[width=8.6cm]{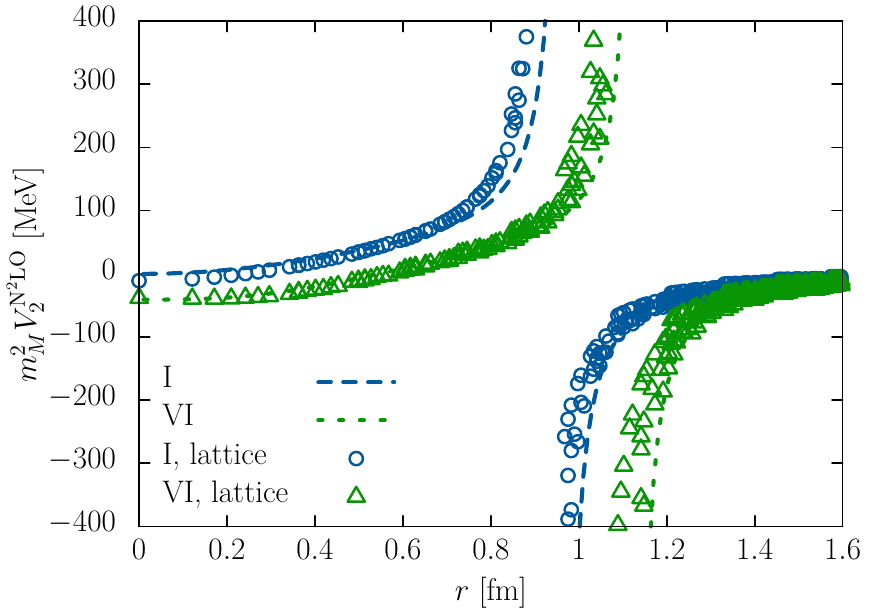}  
  \caption{HAL QCD next-to-next-to-leading order (N${}^{2}$LO)
    potentials in the derivative expansion.  We take the parameter
    sets I and IV in Table~\ref{tab:elastic}.}
  \label{fig:VHAL_N2LO}
\end{figure*}

To analyze finite-volume and lattice discretization effects on the HAL
QCD potentials more deeply, we employ further model parameter sets
with various interaction ranges $b_{\rm L} = \SI{0.20}{fm}$,
$\SI{0.40}{fm}$, $\SI{0.80}{fm}$, and $\SI{1.60}{fm}$ of the single
Gaussian interaction term, as in Table~\ref{tab:para_ranges}.  In each
parameter set, the interaction strength $V_{\rm L}$ is fixed so as to
generate a bound state at the binding energy $B_{\rm E} =
\SI{20.0}{MeV}$ in the nonrelativistic formulation.  We refer to the
parameter sets as V, VI, VII, and VIII, respectively.  Although the
binding energy is $\SI{20.0}{MeV}$ in the nonrelativistic formulation,
it becomes less bound in the semirelativistic one mainly due to the
factor $m_{M} m_{B} / [ \mathcal{E}_{M} ( k ) \mathcal{E}_{B} ( k ) ]$
in the momentum integrals, as summarized in
Table~\ref{tab:para_ranges}.

The HAL QCD potentials $V_{\rm HAL} ( r )$ obtained in the present
formula are shown in Fig.~\ref{fig:VHAL_ranges} both in continuum and
lattice spacetimes.  As can be seen, as the interaction range
decreases, the discrepancy between the potentials in continuum and
lattice spacetimes at the origin becomes larger.  In particular, for
set V, in which the interaction range $b_{\rm L} = \SI{0.20}{fm}$ is
less than twice the lattice spacing ($2 a = \SI{0.242}{fm}$), the
discrepancy amounts to about 8 percent at the origin.  This behavior
is already seen in Fig.~\ref{fig:VHAL_SR_lat}, in particular for the
case of set IV, where the shorter interaction range is $b_{\rm L} / 2
= \SI{0.30}{fm}$.  On the other hand, the interaction range $b_{\rm L}
= \SI{1.60}{fm}$ in parameter set VIII, which is more than 2/5 of the
length of the box $L = N a = \SI{3.872}{fm}$, provides a clear
discrepancy between potentials in continuum and lattice spacetimes at
$r > \SI{1}{fm}$.  This is caused by periodic boundary conditions of
the finite box, as the attraction is largely enhanced by the
interactions with particles in neighboring boxes.  However, such a
discrepancy at $r > \SI{1}{fm}$ cannot be seen in parameter set VII
with $b_{\rm L} = \SI{0.80}{fm}$, which is about 1/5 of the length of
the box.  Therefore, we expect that the maximum range of the
interaction which the HAL QCD method can probe is around 1/5 of the
length of the box.

From these observations, we conclude that the lattice discretization
and finite-volume effects of spacetime modify the results of the HAL
QCD potential only slightly, as long as their scale is not comparable
to the interaction range.

Finally, combined with the conclusions of the previous subsection ---
that HAL QCD potentials in continuum spacetime correctly reproduce the
original phase shifts and accurately indicate the existence or absence
of bound states with binding energies on the order of $\si{MeV}$ ---
we conclude that HAL QCD potentials extracted from lattice QCD
simulations yield physical predictions quantitatively consistent with
actual scattering observables when the interaction is predominantly
local.

\subsection{Next-to-next-to-leading order potentials}
\label{sec:N2LO}

Up to this point, we have considered only the leading order of the
derivative expansion, in which the HAL QCD local potential $V_{\rm
  HAL} ( r )$ arises from the nonlocal potential $U ( \bm{r} ,
\bm{r}^{\prime} )$ as in Eq.~\eqref{eq:U_LO}. The HAL QCD potentials
$V_{\rm HAL} ( r )$ in various cases can quantitatively reproduce
phase shifts of hadron-hadron scattering and correctly indicate the
existence/absence of the bound states with its binding energies of
$\sim \SI{}{\MeV}$.  However, the HAL QCD potentials $V_{\rm HAL} ( r
)$ underestimate the binding energies of the bound states in sets I
and IV, which would be critical when extracting, e.g., the nuclear
force in the HAL QCD method and searching for the deuteron as the
proton-neutron bound state located $\SI{2.2}{MeV}$ below the
threshold.

To investigate this point, we examine the HAL QCD N${}^{2}$LO
potentials via Eq.~\eqref{eq:U_N2LO} in the present framework.
Because we have two potential terms $V_{0, 2}^{\mathrm{N}^{2}
  \mathrm{LO}} ( r )$ at N${}^{2}$LO, a single R-correlator evaluated
in Sec.~\ref{sec:modelA_SR} or \ref{sec:modelA_lat} is insufficient to
determine both terms $V_{0, 2}^{\mathrm{N}^{2} \mathrm{LO}} ( r )$ via
Eq.~\eqref{eq:LO_N2LO}.  In Ref.~\cite{HALQCD:2018gyl}, the authors
used two R-correlators with different quark sources (wall and smear)
in lattice QCD simulations.  In the present study of model analysis,
on the other hand, we evaluate another R-correlator by truncating the
integral of the energy $W$ in Eq.~\eqref{eq:Rcorr_MB} to be $( M_{\rm
  th} , W_{\rm cut} )$ instead of $( M_{\rm th} , \infty )$.  With
this R-correlator $R_{\rm cut} ( t , r )$, we can calculate $V_{\rm
  HAL}^{\rm cut} ( r )$ by using the formula~\eqref{eq:V_HAL_cont} and
then formulate the equation
\begin{equation}
  V_{0}^{\mathrm{N}{}^{2}\mathrm{LO}} ( r )
  + V_{2}^{\mathrm{N}{}^{2}\mathrm{LO}} ( r )
  \frac{\bm{\nabla}^{2} R_{\rm cut} ( t , r )}{R_{\rm cut} ( t , r )}
  = V_{\rm HAL}^{\rm cut} ( r ) .
  \label{eq:LO_N2LO_II}
\end{equation}
Finally, by combining Eqs.~\eqref{eq:LO_N2LO} and
\eqref{eq:LO_N2LO_II}, we can determine the HAL QCD N${}^{2}$LO
potentials $V_{0, 2}^{\mathrm{N}^{2} \mathrm{LO}} ( r )$.  The
counterpart of $V_{0, 2}^{\mathrm{N}^{2} \mathrm{LO}} ( r )$ in
lattice spacetime can be evaluated in a straightforward manner.  In
the present study, we fix $W_{\rm cut} = M_{\rm th} + \SI{100}{MeV}$
so as to weight lower-energy contributions.

We choose parameter sets I and IV, which generate bound states in
continuum spacetime, and calculate the potentials $V_{0,
  2}^{\mathrm{N}^{2} \mathrm{LO}} ( r )$ both in continuum and lattice
spacetimes.  The results are shown in Fig.~\ref{fig:VHAL_N2LO}, where
lines and points denote results in continuum and lattice spacetimes,
respectively.  Here, $V_{2}^{\mathrm{N}^{2} \mathrm{LO}} ( r )$ is
multiplied by $m_{M}^{2}$ so as to show it in units of MeV.  As can be
seen, owing to the nonlocal nature of the present interaction, we
obtain nonzero values of $V_{2}^{\mathrm{N}^{2} \mathrm{LO}} ( r )$ in
parameter sets I and IV, although the term $V_{0}^{\mathrm{N}^{2}
  \mathrm{LO}} ( r )$ is almost the same as the leading-order term
$V_{\rm HAL} ( r )$ shown in Figs.~\ref{fig:VHAL_SR} and
\ref{fig:VHAL_SR_lat}.  Besides, $V_{0, 2}^{\mathrm{N}^{2}
  \mathrm{LO}} ( r )$ potentials diverge at $r \approx \SI{1}{fm}$
both in continuum and lattice spacetimes.  This behavior is caused by
the division by $\bm{\nabla}^{2} R / R - \bm{\nabla}^{2} R_{\rm cut} /
R_{\rm cut}$, which becomes nearly zero at a certain value of $r$, and
was indeed observed in the HAL QCD analysis of lattice QCD simulations
as well~\cite{HALQCD:2018gyl, Lyu:2025ncq}.

\begin{figure}[t]
  \centering
  \includegraphics[width=8.6cm]{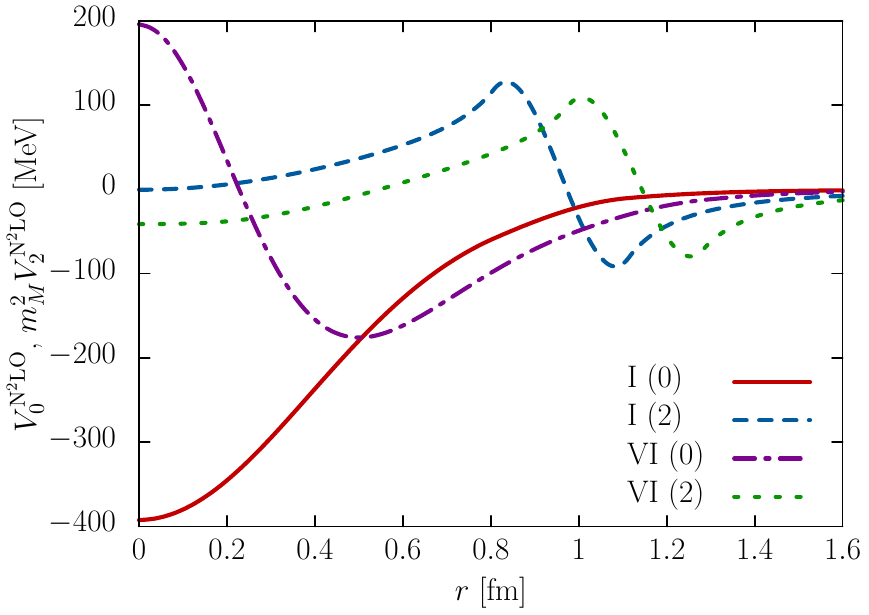}
  \caption{HAL QCD next-to-next-to-leading order (N${}^{2}$LO)
    potentials in the derivative expansion with the divergence
    removal.  We take the parameter sets I and IV in
    Table~\ref{tab:elastic}.  In the legend, ``(0)'' and ``(2)''
    indicate $V_{0}^{\mathrm{N}^{2}\mathrm{LO}}$ and $m_{M}^{2}
    V_{2}^{\mathrm{N}^{2}\mathrm{LO}}$, respectively.}
  \label{fig:VHAL_N2LO_subt}
\end{figure}

Now we want to evaluate the binding energies of the bound states at
N${}^{2}$LO.  To avoid the divergence of the $V_{0, 2}^{\mathrm{N}^{2}
  \mathrm{LO}} ( r )$ potentials, fits using Gaussian-type functions
were performed in Refs.~\cite{HALQCD:2018gyl, Lyu:2025ncq}.  Instead,
in the present study, we simply discard the potential values around
the diverging point in the range $\pm \SI{0.15}{fm}$ and smoothly
connect them by spline interpolation.  This is justified by the fact
that the energy $E ( = W - M_{\rm th} )$-dependent local potential
from the $V_{0, 2}^{\mathrm{N}^{2} \mathrm{LO}} ( r )$
terms~\cite{HALQCD:2018gyl}
\begin{equation}
  V_{\rm eff} ( r ; E )
  = \frac{V_{0}^{\mathrm{N}^{2} \mathrm{LO}} ( r ) - 2 \mu E
    V_{2}^{\mathrm{N}^{2} \mathrm{LO}} ( r )}{1
    - 2 \mu V_{2}^{\mathrm{N}^{2} \mathrm{LO}} ( r )}
\end{equation}
is, except for the diverging point, smooth enough as a function of $r$
at energies around the threshold $E \simeq 0$.  The potentials
obtained via this prescription are shown in
Fig.~\ref{fig:VHAL_N2LO_subt}.  Then, we solve the Schr\"{o}dinger
equation with the $V_{0, 2}^{\mathrm{N}^{2} \mathrm{LO}} ( r )$
potentials:
\begin{equation}
  \left [ - \frac{1}{2 \mu} \bm{\nabla}^{2}
    + V_{0}^{\mathrm{N}^{2} \mathrm{LO}} ( r )
    + V_{2}^{\mathrm{N}^{2} \mathrm{LO}} ( r ) \bm{\nabla}^{2} \right ] \psi ( r )
  = E \psi ( r )
\end{equation}
As a result, we obtain bound states with binding energies of
$\SI{5.5}{MeV}$ for set I and $\SI{4.6}{MeV}$ for set IV (see the last
line of Table~\ref{tab:elastic}).  These values are closer to the
original values of the binding energies ($\SI{6.4}{MeV}$ and
$\SI{5.4}{MeV}$, respectively) compared to those from the HAL QCD
potential $V_{\rm HAL} ( r )$ at the leading-order analysis
($\SI{3.6}{MeV}$ and $\SI{3.8}{MeV}$, respectively).  Therefore, the
HAL QCD N${}^{2}$LO potentials in the derivative expansion can provide
a better description of the binding energies of the bound states.

We expect some hadron-hadron interactions to be nonlocal.  For
instance, the $u$-channel crossed Born term contributes to the $K N$
interaction, although its contribution is expected to be
small~\cite{Aoki:2017hel, Aoki:2018wug}.  Understanding such
nonlocality may offer clues to interpreting the $K N$ potential in HAL
QCD analyses, particularly regarding the underestimation of phase
shifts at higher momentum seen in recent
results~\cite{Murakami:2025owk}.  In such cases, one can utilize the
present approach to evaluate the HAL QCD potential starting from the
scattering amplitude $T ( W , p , p^{\prime} )$.  Specifically, given
a model, one calculates the scattering amplitude, R-correlator, and
HAL QCD potential, and then checks whether the HAL QCD potential
faithfully reproduces the properties of the original scattering
amplitude.

\section{Model analysis: bare to meson-baryon transition}
\label{sec:modelB}

Next, we turn to the application of the HAL QCD method to the bare
($\Lambda _{0}$) to meson-baryon transition.  We calculate the
$\Lambda _{0}$-$M B$ transition amplitude using a set of model
parameters, evaluate the R-correlator and HAL QCD potential for the
meson-baryon system from it, and then examine whether the HAL QCD
potential can reproduce the original quantities.

In this section, we concentrate on the case where the $s$-wave
meson-baryon system couples to the bare $\Lambda _{0}$ and forms a bound
state of mass $M_{\rm B} = \SI{1923.4}{MeV}$ in continuum
spacetime.  This bound-state mass is the same as that in
Ref.~\cite{Murakami:2023phq}.

\subsection{Setup}

\begin{table}[!t]
  \caption{Parameters $( V_{\rm L} , b_{\rm L} , f_{0} , b_{0} , M_{0}
    )$ and the number of ({\#}) meson-baryon interaction terms in the bare
    to meson-baryon transition case.  The meaning of the additional
    parameter $V_{\rm L}^{\prime}$ is explained in
    Sec.~\ref{sec:singular}.}
  \label{tab:para_Lam0}
  \centering
  \begin{ruledtabular}
    \begin{tabular}{lcccc}
      & Set IX & Set X & Set XI & Set XII
      \\
      \hline
      $V_{\rm L}$ [MeV] & $-1717.5$ & $0.0$ & $-2563.1$ & $-3894.1$
      \\
      $b_{\rm L}$ [fm] & $0.50$ & $0.50$ & $0.60$ & $0.30$
      \\
      $f_{0}$ [GeV${}^{-1/2}$] & $1.361$ & $1.361$ & $1.361$ & $0.100$
      \\
      $b_{0}$ [fm] & $0.50$ & $0.50$ & $0.50$ & $0.10$
      \\
      $M_{0}$ [MeV] & $3000.0$ & $1933.4$ & $3000.0$ & $2500.0$
      \\
      \# interaction term 
      & 1 & 1 & 2 & 1
      \\
      $V_{\rm L}^{\prime}$ & --- & --- & --- & $-50.0$
      \\
    \end{tabular}
  \end{ruledtabular}
\end{table}

To this end, we set the model parameters such that the bound state is
(i) almost composite (i.e., composed of a meson-baryon molecule), (ii)
almost a bare state (i.e., predominantly $\Lambda _{0}$), or (iii) almost
composite but generated by a repulsive core plus an attractive pocket in the
meson-baryon interaction.

The parameter setting is performed in continuum spacetime as follows.
First, we fix the typical interaction range $b_{\rm L} = b_{0} =
\SI{0.50}{fm}$ and generate the bound state with $M_{\rm B} =
\SI{1923.4}{MeV}$ solely from the $\Lambda _{0}$-$M B$ coupling
$V_{(1)}$~\eqref{eq:V1}, setting $V_{\rm L} = 0$.  The bare $\Lambda
_{0}$ mass is chosen to be $M_{0} = M_{\rm B} + \SI{10.0}{MeV} =
\SI{1933.4}{MeV}$, which yields $f_{0} = \SI{1.361}{GeV^{-1/2}}$.
This corresponds to case (ii), the almost bare state.  Second, we
generate the bound state with a larger bare mass $M_{0} = \SI{3}{GeV}$
while keeping the same $\Lambda _{0}$-$MB$ coupling constant and
interaction range, yielding $V_{\rm L} = \SI{-1717.5}{MeV}$.  This
corresponds to case (i), the almost composite state.  Third, by using
the same parameters $f_{0} = \SI{1.361}{GeV^{-1/2}}$, $b_{0} =
\SI{0.50}{fm}$, and $M_{0} = \SI{3}{GeV}$ for the bare $\Lambda _{0}$
as in case (i), but introducing double Gaussians to represent a
repulsive core plus an attractive pocket in the meson-baryon
interaction, we generate a bound state.  With the same interaction
range for the double Gaussians as in parameter set IV (see
Table~\ref{tab:elastic}), $b_{\rm L} = \SI{0.60}{fm}$, we obtain
$V_{\rm L} = \SI{-2563.1}{MeV}$.  This corresponds to case (iii),
which is almost composite but generated by a repulsive core plus an
attractive pocket in the meson-baryon interaction.  We denote the
parameter sets for cases (i), (ii), and (iii) as sets IX, X, and XI,
respectively.  The parameters are summarized in
Table~\ref{tab:para_Lam0}.  We note that no other bound states are
found below the threshold.

The composite or bare-state nature of the bound state can be
quantified in terms of the compositeness $X$.  By applying the formula
developed in Sec.~\ref{sec:composite}, we obtain the compositeness of
the bound state as $X = 0.996$, $0.023$, and $0.997$ for parameter
sets IX, X, and XI, respectively.  These results explicitly confirm
the composite (bare-state) nature of sets IX and XI (set
X)\footnote{In our model, we neglect the internal structure of each
hadron.  Therefore, even in such a deeply bound case, the two
constituents retain their individualities.  However, this aspect
should be refined in future studies, as hadrons are actually composed
of several quarks and are expected to deform in deeply bound states.}.
We also note that the binding energy of the bound state, which is
fixed to $\SI{1923.4}{MeV}$ in continuum spacetime, is modified in a
finite volume depending on the physical nature of the bound
state~\cite{Sekihara:2013wlq}.  Specifically, in lattice spacetime
with periodic boundary conditions, the binding energies become
$\SI{1923.0}{MeV}$ and $\SI{1922.4}{MeV}$ for parameter sets IX and
XI, respectively, whereas the value remains unchanged for parameter
set X.  This is because only the meson-baryon molecular component
experiences additional binding under periodic boundary conditions due
to interactions with particles in neighboring boxes.

\subsection{Numerical results}

\begin{figure}[t]
  \centering
  \includegraphics[width=8.6cm]{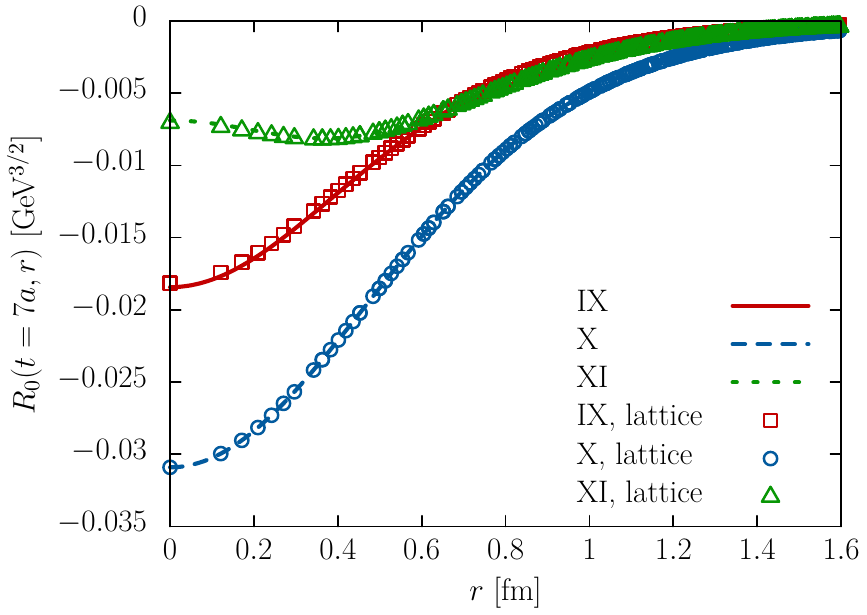}
  \caption{R-correlators in continuum and lattice spacetimes.
    The evaluation time is fixed at $t = 7 a$.
    We employ the model parameters of sets IX, X, and XI in
    Table~\ref{tab:para_Lam0}.}
  \label{fig:Rcorr_Lam0}
\end{figure}

Using parameter sets IX, X, and XI, we calculate the
R-correlators~\eqref{eq:Rcorr_bare}, as shown in
Fig.~\ref{fig:Rcorr_Lam0}.  The time $t$ for evaluation is fixed at $t
= 7 a$ with $a = \SI{0.121}{fm}$.  The solid, dashed, and dotted lines
in Fig.~\ref{fig:Rcorr_Lam0} represent the results for sets IX, X, and
XI, respectively, in continuum spacetime.  As can be seen, although
the physical nature of the bound state differs between sets IX and X,
the overall shapes of the R-correlators in the two cases are nearly
identical.  This occurs because, among the terms in
Eq.~\eqref{eq:Rcorr_bare_phys}, the leading term
\begin{equation}
  e^{- (M_{\rm B} - M_{\rm th}) t } \langle r | \Psi _{\rm B} \rangle
  \langle \Psi _{\rm B} | \Lambda _{0} \rangle
\end{equation}
dominates the R-correlators.  This term is the product of the two-body
component of the bound-state wave function $\langle r | \Psi _{\rm B}
\rangle$ and the overlap between the physical bound state and the bare
state $\langle \Psi _{\rm B} | \Lambda _{0} \rangle$.  In set IX,
while the two-body wave function $\langle r | \Psi _{\rm B} \rangle$
is large owing to its composite nature, the overlap $\langle \Psi
_{\rm B} | \Lambda _{0} \rangle$ is small.  In set X, on the other
hand, this relation is reversed due to its predominant bare-state
nature.  Therefore, when taking the product $\langle r | \Psi _{\rm B}
\rangle \langle \Psi _{\rm B} | \Lambda _{0} \rangle$, the resulting
R-correlators in sets IX and X turn out to be similar.

In parameter set XI, by contrast, the absolute value of the
R-correlator decreases near the origin.  A similar behavior was
observed in the R-correlator of set IV (see Figs.~\ref{fig:Rcorr_NR},
\ref{fig:Rcorr_SR}, and \ref{fig:Rcorr_SR_lat}).  This stems from the
repulsive core of the meson-baryon interaction at short distances.

Next, we examine the R-correlators in lattice
spacetime~\eqref{eq:Rtilde_fin}.  The results are shown in
Fig.~\ref{fig:Rcorr_Lam0} as discrete points.  Interestingly, although
the sampling points of $r$ are discretized due to the lattice
structure, the points quantitatively match the continuum lines across
all parameter sets.  This indicates that the R-correlators for the
bare to meson-baryon transition do not suffer significantly from
lattice discretization under periodic boundary conditions in the
present formulation.

\begin{figure}[t]
  \centering
  \includegraphics[width=8.6cm]{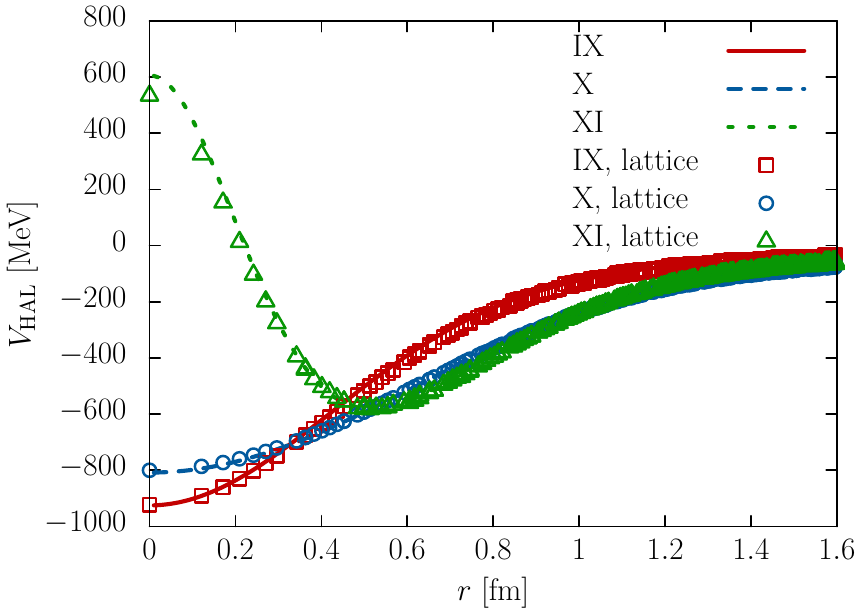}
  \caption{HAL QCD potentials in continuum and lattice spacetimes.
    The evaluation time is fixed at $t = 7 a$.  We employ the model
    parameters of sets IX, X, and XI in Table~\ref{tab:para_Lam0}.}
  \label{fig:VHAL_Lam0}
\end{figure}

We then calculate the HAL QCD potentials $V_{\rm HAL} ( r
)$~\eqref{eq:V_HAL_cont} from the R-correlators.  The continuum
spacetime results are displayed in Fig.~\ref{fig:VHAL_Lam0} as lines.
The HAL QCD potentials across all parameter sets exhibit attraction.
Indeed, we have verified that these HAL QCD potentials yield bound
states with binding energies of $\sim \SI{220}{MeV}$, $\SI{300}{MeV}$,
and $\SI{230}{MeV}$ for sets IX, X, and XI, respectively, as
eigenenergies of the Schr\"{o}dinger equation.  These values are
reasonably close to the original binding energy of $\SI{236.6}{MeV}$
from the bound-state mass $M_{\rm B} = \SI{1923.4}{MeV}$.

The potential curves for both sets IX and X display a similar
attractive region extending to a range of $\sim \SI{1}{fm}$,
reflecting the similar shapes of the two-body bound-state wave
functions $\langle r | \Psi _{\rm B} \rangle$ in sets IX and X.
Therefore, in the absence of a repulsive core at the origin in the
meson-baryon interaction, one obtains similar HAL QCD potentials
regardless of the physical nature of the bound state.  This implies
that the HAL QCD method alone cannot uniquely determine the internal
structure of a bound state.  This point was previously raised in
Ref.~\cite{Murakami:2022cez}, where the HAL QCD method was extended to
$p$-wave $\pi N$ and $\bar{K} \Xi$ interactions for the $\Delta$ and
$\Omega$ states, respectively.  Indeed, in
Ref.~\cite{Murakami:2022cez}, the authors argued that, in view of
their spatial extent, the calculated $\Delta$ and $\Omega$ states
correspond to compact three-quark states rather than meson-baryon
molecular states.

The HAL QCD potential for parameter set XI, on the other hand,
clearly exhibits a repulsive core plus attractive pocket structure,
consistent with the original meson-baryon interaction modeled by
double Gaussians.  Thus, even in bare to meson-baryon transition
cases, the HAL QCD method can accurately reconstruct the meson-baryon
potential when the bound state is dominated by a molecular character.
However, it should be noted that one might fail to extract such a
potential featuring a repulsive core and an attractive pocket if the
bound state is a significant mixture of molecular and bare-state
components.  Because the HAL QCD potential reduces to a simple
attractive form in bare-state-dominated cases (as in set X), a
mixture of molecular and bare-state components would yield a HAL QCD
potential corresponding to an average of the dashed and dotted lines
in Fig.~\ref{fig:VHAL_Lam0}.

We also calculate the HAL QCD potentials in lattice
spacetime~\eqref{eq:V_HAL_latt} using a similar procedure.  The
results are shown in Fig.~\ref{fig:VHAL_Lam0} as points.  For all
parameter sets, the lattice HAL QCD potentials deviate only slightly
from their continuum counterparts near the origin, where
discretization effects become relatively noticeable.

\subsection{\boldmath Can we explain the singular HAL QCD results of $\bar{K} N$?}
\label{sec:singular}

The $\bar{K} N$ interaction remains a subject of fundamental interest
in hadronic physics, as its strongly attractive nature is expected to
generate the $\Lambda (1405)$ resonance.  Consequently, alongside
effective model approaches (e.g., Refs.~\cite{Kaiser:1995eg,
  Oset:1997it, Sekihara:2010uz, Hyodo:2011ur, Mai:2020ltx,
  Conde-Correa:2024qzh, Mutuk:2025tex, Pittler:2025upn}), lattice QCD
simulations have been conducted to unravel the dynamics of the
$\bar{K} N$ system and the structure of the $\Lambda
(1405)$~\cite{Hall:2014uca, Liu:2016wxq,
  BaryonScatteringBaSc:2023zvt}.

In a recent HAL QCD analysis~\cite{Murakami:2023phq} performed in the
flavor SU(3) limit with meson and baryon masses set to $m_{M} =
\SI{671.2}{MeV}$ and $m_{B} = \SI{1488.8}{MeV}$, respectively, the
$\bar{K} N$ potential was extracted from the correlation function
describing the transition from a negative-parity $\Lambda$ baryon to a
$\bar{K} N$ state.  The resulting potential exhibited a singular
behavior at $r \sim 0.2$--$\SI{0.4}{fm}$.  The underlying cause of
this singular potential is a node in the R-correlator, which leads to
division by zero in Eq.~\eqref{eq:V_HAL_latt}.

It is nontrivial that a node in the R-correlator induces a singular
HAL QCD potential.  In nonrelativistic quantum mechanics with a local
potential, a radial component of a bound-state wave function $R ( r )$
possessing a node (such as the $2s$ state of the hydrogen atom)
reproduces the exact underlying local potential in the HAL QCD
formulation~\eqref{eq:V_HAL_cont}.  This occurs because the second
derivative of $r R ( r )$ vanishes at the node $r = r_{0}$, so the
derivative term
\begin{equation}
  \frac{1}{2 \mu r R ( r )} \frac{d^{2}}{d r^{2}} [ r R ( r ) ]
  \label{eq:R_node}
\end{equation}
remains finite and does not induce a singularity even at $r = r_{0}$
where $R ( r_{0} ) = 0$.  In the present framework, however, there are
at least two reasons why a node can generate a singular HAL QCD
potential.  First, introducing a bare state $\Lambda _{0}$ inherently
introduces nonlocality into the $M B \to M B$ interaction.  Second,
lattice discretization replaces the continuous second derivative with
finite differences $\bm{\nabla}^{2}_{\rm lat}$, preventing exact
cancellation between the denominator $R$ and the numerator
$\bm{\nabla}^{2}_{\rm lat} R$.  Consequently, if the R-correlator
evaluated from a $\Lambda _{0} \to M B$ correlation function possesses
a node, a singular local potential becomes unavoidable in the lattice
HAL QCD analysis.

In this subsection, we investigate whether such singular behavior in
the HAL QCD potential can be reproduced within our model.  Employing
the model with one bare state coupled to a single- or double-Gaussian
meson-baryon interaction developed in Sec.~\ref{sec:amp}, we attempt
to fit model parameters to the $\bar{K} N$ potential obtained in the
HAL QCD analysis~\cite{Murakami:2023phq}.  We constrain the parameters
to reproduce a bound state of mass $M_{\rm B} = \SI{1923.4}{MeV}$ as
the ground state of the system in continuum spacetime.  However, this
fitting procedure yields no parameter set capable of quantitatively
reproducing the singular $\bar{K} N$ potential of the HAL QCD
analysis.  The physical reason is the same as that observed in the
decomposition shown in Fig.~\ref{fig:Rcorr_NR_each} for parameter sets
I and IV: owing to dominant bound-state contributions (located more
than $\SI{200}{MeV}$ below the meson-baryon threshold in this setup),
the full R-correlator develops no node.  Therefore, as long as the
ground state, which is node-free, dominates the R-correlator, one
inevitably obtains a non-singular local potential.

To accommodate more complicated dynamics, we can extend our model by
incorporating contributions omitted in the model space, specifically
by adding an energy $W$-dependent term to the single-Gaussian
meson-baryon interaction:
\begin{align}
  V_{(2)} ( p , p^{\prime} )
  = & \frac{2 ( \pi b_{\rm L}^{2} )^{3/2}
    [ V_{\rm L} + V_{\rm L}^{\prime} ( W - M_{\rm B} ) ]}{p p^{\prime} b_{\rm L}^{2}}
  \notag \\
  & \times
  \exp \left [ - \frac{( p^{2} + p^{\prime 2} ) b_{\rm L}^{2}}{4} \right ]
  \sinh \left ( \frac{p p^{\prime} b_{\rm L}^{2}}{2} \right ) .
  \label{eq:Vprime}
\end{align}
Here, $V_{\rm L}^{\prime}$ governs the energy dependence of the
interaction.  We again constrain the parameters to yield a bound state
of mass $M_{\rm B} = \SI{1923.4}{MeV}$ as the ground state in
continuum spacetime.  However, even with this extension, we find no
parameter set that quantitatively reproduces the singular $\bar{K} N$
potential reported in the HAL QCD analysis.

\begin{figure}[t]
  \centering
  \includegraphics[width=8.6cm]{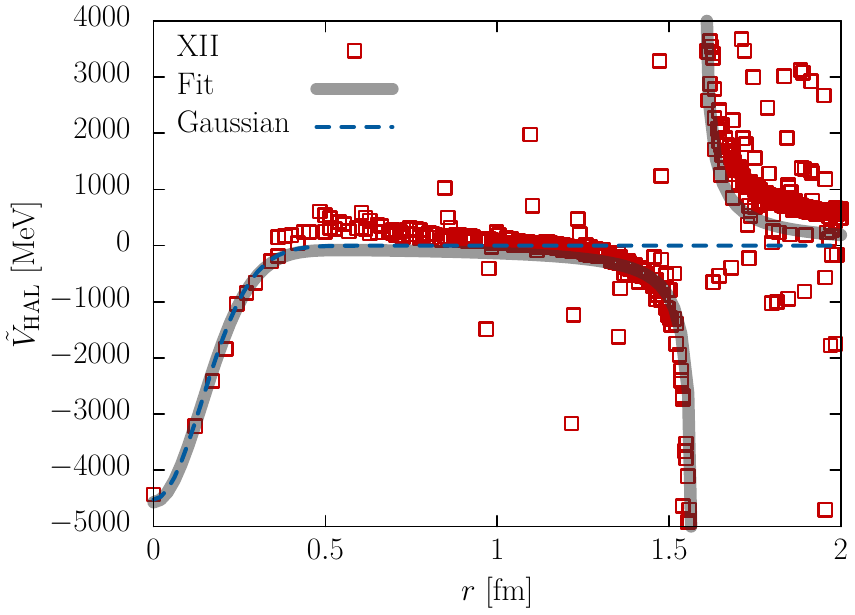}
  \caption{HAL QCD potential for parameter set XII evaluated in
    lattice spacetime with periodic boundary conditions.  The
    evaluation time is fixed at $t = 7 a$.  The thick line labeled
    ``Fit'' shows the fitted potential~\eqref{eq:VHAL_fit}, whereas
    the dashed line labeled ``Gaussian'' represents the
    single-Gaussian component of the fit~\eqref{eq:VHAL_fit_prime}.}
  \label{fig:VHAL_singular}
\end{figure}

Nevertheless, using the energy-dependent
interaction~\eqref{eq:Vprime}, we identify a parameter set that
induces a node in the R-correlator at $r \approx \SI{1.6}{fm}$.  We
designate this as parameter set XII, with values listed in the fifth
column of Table~\ref{tab:para_Lam0}.  We compute the HAL QCD potential
in lattice spacetime for parameter set XII and plot the results in
Fig.~\ref{fig:VHAL_singular} as points.  The singular structure near
$r \approx \SI{1.6}{fm}$ indeed originates from the node in the
R-correlator.  In this parameter set, scattering-state contributions
are enhanced by the energy dependence, becoming comparable in
magnitude to the bound-state contribution to the R-correlator.  As a
consequence, the total R-correlator acquires a node.  Along these
lines, we expect that coupled channels and excited hyperon states
omitted in this study would play a role in full HAL QCD
simulations~\cite{Murakami:2023phq}.  Furthermore, explicitly
incorporating the internal quark structure of hadrons may be essential
for enhancing scattering-state contributions.  Because our current
approach treats hadrons as point-like particles, interaction forms are
restricted to those in Fig.~\ref{fig:interaction}.  Conversely, a
quark-level treatment would allow for more variety of interaction
forms, enabling fine control over couplings such as $\Lambda _{0}$-$M
B$ scattering state.

While the HAL QCD potential for parameter set XII exhibits a
divergence, we observe that the potential outside the singular region,
particularly near the origin, can be effectively described by a single
Gaussian.  To verify this, we fit the HAL QCD potential in
Fig.~\ref{fig:VHAL_singular} using the functional form
\begin{equation}
  \tilde{V}_{\rm HAL} ( r )
  = V_{\rm F} \exp \left ( - \frac{r^{2}}{b_{\rm F}^{2}} \right )
  + \frac{C}{r - r_{0}} .
  \label{eq:VHAL_fit}
\end{equation}
The optimal fit yields $V_{\rm F} = \SI{-4520.0}{MeV}$, $b_{\rm F} =
\SI{0.201}{fm}$, $C = \SI{78.63}{MeV.fm}$, and $r_{0} =
\SI{1.585}{fm}$.  The fitted curve is displayed as a thick line in
Fig.~\ref{fig:VHAL_singular}.  As shown, the fit quantitatively
reproduces the HAL QCD potential for parameter set XII.

Furthermore, the smooth Gaussian component
\begin{equation}
  \tilde{V}_{\rm HAL} ( r ) |_{\rm Gaussian}
  = V_{\rm F} \exp \left ( - \frac{r^{2}}{b_{\rm F}^{2}} \right ) ,
  \label{eq:VHAL_fit_prime}
\end{equation}
plotted as a dashed line in Fig.~\ref{fig:VHAL_singular}, matches the
calculated HAL QCD potential remarkably well away from the
singularity.  Notably, using only the single Gaussian
potential~\eqref{eq:VHAL_fit_prime} with the fitted parameters $V_{\rm
  F} = \SI{-4520.0}{MeV}$ and $b_{\rm F} = \SI{0.201}{fm}$, the
Schr\"{o}dinger equation yields a bound state with a binding energy of
$\SI{234.5}{MeV}$, which is in good agreement with the original value
of $\SI{236.6}{MeV}$ obtained from the pole position of the transition
amplitude.

In this fitting procedure, the singular portion of the HAL QCD
potential was effectively discarded.  Physically, this corresponds to
regularizing the wave function $R_{0} ( r )$ near its node to avoid
non-zero second-derivative term in Eq.~\eqref{eq:R_node}.  Such a
regularization approach appears physically justified even in full HAL
QCD analyses~\cite{Murakami:2023phq}, provided the underlying
R-correlator faithfully captures the physical $\bar{K} N$ wave
function (comprising both bound and scattering states) generated by
the system dynamics.

\section{Summary and outlook}
\label{sec:sum}

The HAL QCD method is a powerful tool to evaluate hadron-hadron
interactions directly from lattice QCD simulations.  To clarify the
effectiveness of the HAL QCD method, we derived a formula to calculate
R-correlators, which are essential to evaluate HAL QCD potentials, in
terms of the meson-baryon scattering amplitude.  This formula enables
us to calculate the R-correlator and HAL QCD potential within a given
model of meson-baryon scattering.

By using this formula, we examined whether the original meson-baryon
interaction is correctly encoded into the HAL QCD potential within an
effective model.  In the elastic scattering case, the scattering phase
shifts were quantitatively reproduced by the HAL QCD potential, and
the existence/absence of a bound state with its binding energy $(\sim
\si{MeV})$ were correctly captured when the meson-baryon interaction
was predominantly local.  The HAL QCD N${}^{2}$LO potentials in the
derivative expansion can provide a better description of the binding
energies of the bound states.  Lattice discretization and
finite-volume effects of spacetime modified the results only slightly,
as long as their scale is not comparable to the interaction range.
The HAL QCD method was applied to the bare to meson-baryon transition
case as well.  As a result, the HAL QCD potential could produce a
bound state equivalent to that generated by the original interaction;
however, we confirmed that the HAL QCD method cannot uniquely
determine the internal structure of the bound state.

The formula was also applied to the $\bar{K} N$ interaction in the
bare $\Lambda _{0}$ to meson-baryon transition amplitude, for which a
recent HAL QCD analysis~\cite{Murakami:2023phq} showed a singular
behavior.  In our model analysis, we could not identify the precise
origin of the singular potential.  However, we confirmed that a node
in the R-correlator indeed induces singular behavior in the HAL QCD
potential.  We expect that, to enhance the scattering-state
contributions to the R-correlator, which are the origin of the node,
the coupled channels and hyperon resonances, as well as the internal
structure of each hadron, would be essential.  Furthermore, our
results implied that, as long as the R-correlator contains the
bound-state wave function as well as scattering wave functions
generated by the $\bar{K} N$ dynamics, one might obtain a nonsingular
local potential if one could properly subtract the singular behavior
from the HAL QCD potential.

Because we derived the formula to calculate the R-correlator from the
scattering amplitude in a general form, one can independently apply
this formula to examine how the R-correlator and HAL QCD potential
reflect the properties of the scattering amplitude in various models.
For example, one can test the effectiveness of the derivative
expansion for nonlocal potentials.  In addition, one can check the
consistency of the HAL QCD potential extracted from lattice QCD
simulations with experimental observables via scattering amplitudes
that reproduce those observables.  In this regard, the $K N$
interaction would be a suitable target for further investigation.

\begin{acknowledgments}
 
  The authors acknowledge Kotaro~Murakami for fruitful discussions
  regarding the HAL QCD $\bar{K} N$ potential derived from lattice QCD
  simulations.  The authors also thank Sinya~Aoki, Takumi~Doi, and
  Noriyoshi~Ishii for helpful discussions regarding the general
  properties of the HAL QCD method.  Sinya~Aoki, Takumi~Doi, and
  Kotaro~Murakami are acknowledged for their careful reading of the
  present manuscript as well.
  This work was supported in part by a Grant-in-Aid for Scientific
  Research from JSPS (25H00655).

\end{acknowledgments}

\appendix

\section{Emergence of the physical states}
\label{app:phys}

In this Appendix, we prove the equivalence between
Eqs.~\eqref{eq:Rcorr_MB} and \eqref{eq:Rcorr_MB_phys}, and between
Eqs.~\eqref{eq:Rcorr_bare} and \eqref{eq:Rcorr_bare_phys}.  Because
the bound-state contribution is the same as explained in the main
text, specifically in Eqs.~\eqref{eq:PsiB_equiv} and
\eqref{eq:PsiBzero_equiv}, we focus here on the scattering-state
contribution.

To prove the equivalence, we define several quantities and establish key
equations.  We first introduce the $s$-wave free meson-baryon state
with relative momentum $p$ as $| p \rangle$.  This is related to the
bra vector $| p_{\rm phys} \rangle$ for the physical meson-baryon
state in the following manner (see, e.g., Ref.~\cite{Taylor:1972pty}):
\begin{equation}
  | p_{\text{phys}} \rangle
  = | p \rangle + \frac{1}{W_{M B} ( p ) - \hat{H}_{0} + i 0}
  \hat{T} ( W_{M B} ( p ) ) | p \rangle ,
  \label{eqA:pphys}
\end{equation}
where $W_{M B} ( p ) \equiv \mathcal{E}_{M} ( p ) + \mathcal{E}_{B} (
p )$ is the on-shell center-of-mass energy of the meson-baryon system
and $\mathcal{E}_{M, B} ( p ) \equiv \sqrt{m_{M, B}^{2} +
  \bm{p}^{2}}$.  The Hamiltonian of free particles $\hat{H}_{0}$ has the
eigenstate $| p \rangle$:
\begin{equation}
  \hat{H}_{0} | p \rangle = W_{M B} ( p ) | p \rangle .
\end{equation}
The $T$-matrix operator $\hat{T} ( W )$ for meson-baryon scattering
depends on the center-of-mass energy $W$ and generates the scattering
amplitude as
\begin{equation}
  \langle p^{\prime} | \hat{T} ( W ) | p \rangle
  = T ( W , p , p^{\prime} )
  = T ( W , p^{\prime} , p ) ,
\end{equation}
where time-reversal invariance guarantees the second equality.
The scattering amplitude satisfies the optical theorem, which in our
notation gives
\begin{align}
  & T ( W_{M B} ( p ) , 0 , p^{\prime} )
  - T ( W_{M B} ( p ) , p^{\prime} , 0 )^{\ast}
  \notag \\
  & = - \frac{i p m_{M} m_{B}}{\pi W_{M B} ( p )}
  T ( W_{M B} ( p ) , p , p^{\prime} )
  T ( W_{M B} ( p ) , p , 0 )^{\ast} 
  \label{eqA:opt}
\end{align}
for meson-baryon elastic scattering, and 
\begin{align}
  & \Gamma ( W_{MB} ( p ) , p^{\prime} )
  - \Gamma ( W_{MB} ( p ) , p^{\prime} )^{\ast}
  \notag \\
  & = - \frac{i p m_{M} m_{B}}{\pi W_{MB} ( p )}
  T ( W_{MB} ( p ) , p , p^{\prime} ) \Gamma ( W_{MB} ( p ) , p )^{\ast} 
  \label{eqA:opt_Gamma}
\end{align}
for the bare to meson-baryon transition.

To simplify the integrals, we introduce a new measure
\begin{equation}
  D p
  \equiv d p \frac{p^{2}}{2 \pi ^{2}}
  \frac{m_{M} m_{B}}{\mathcal{E}_{M} ( p ) \mathcal{E}_{B} ( p )} 
\end{equation}
and a new delta function 
\begin{equation}
  \Delta ( p^{\prime} - p )
  \equiv \frac{2 \pi ^{2}}{p^{2}}
  \frac{\mathcal{E}_{M} ( p ) \mathcal{E}_{B} ( p )}{m_{M} m_{B}}
  \delta ( p^{\prime} - p ) ,
\end{equation}
so that for an arbitrary function $f ( p )$ we have
\begin{equation}
  \int _{0}^{\infty} D p f ( p ) \Delta ( p^{\prime} - p )
  = f ( p^{\prime} ) .
  \label{eqA:Delta}
\end{equation}
The delta function appears as the normalization of the $| p \rangle$
states:
\begin{equation}
  \langle p^{\prime} | p \rangle
  = \Delta ( p^{\prime} - p ) .
\end{equation}
By using the measure $D p^{\prime}$, we can decompose identity as
\begin{equation}
  1
  = \int _{0}^{\infty} D p^{\prime} | p^{\prime} \rangle \langle p^{\prime} |
  + | \Lambda _{0} \rangle \langle \Lambda _{0} | ,
  \label{eqA:unity}
\end{equation}
where the bare state $| \Lambda _{0} \rangle$ plays a role in only a
limited portion of the following discussion.  Finally, the $s$-wave
free meson-baryon state with relative distance $r$ is defined as $| r
\rangle$, which yields
\begin{equation}
  \langle r | p \rangle = j_{0} ( p r ) .
\end{equation}

\subsection{Meson-baryon elastic scattering}

In the meson-baryon elastic scattering case, we start with
Eq.~\eqref{eq:Rcorr_MB_phys}, specifically the term containing
$\langle r | p_{\rm phys} \rangle \times \langle p_{\rm phys} | M B
\rangle$, and arrive at Eq.~\eqref{eq:Rcorr_MB} by rewriting
Eq.~\eqref{eq:Rcorr_MB_phys}.

By using the formula~\eqref{eqA:pphys} and inserting
identity~\eqref{eqA:unity}, we evaluate $\langle r | p_{\rm phys}
\rangle$ as
\begin{equation}
  \langle r | p_{\rm phys} \rangle
  = j_{0} ( p r ) + \int _{0}^{\infty} D p^{\prime} 
  \frac{j_{0} ( p^{\prime} r ) T ( W_{M B} ( p ) , p , p^{\prime} )}{W_{M B} ( p )
    - W_{M B} ( p^{\prime} ) + i 0} ,
  \label{eqA:r_pphys}
\end{equation}
where the bare state $| \Lambda _{0} \rangle$ in Eq.~\eqref{eqA:unity}
gives no contribution.  Furthermore, in Eq.~\eqref{eq:Rcorr_MB_phys} we
introduced the initial meson-baryon state $| M B \rangle$, which
corresponds to zero-momentum meson and baryon in the present
study.  Thus, $| M B \rangle = | p = 0 \rangle$ and $\hat{H}_{0} | M B
\rangle = M_{\rm th} | M B \rangle$.  Then, using
formula~\eqref{eqA:pphys}, we obtain
\begin{equation}
  \langle p_{\rm phys} | M B \rangle
  = \Delta ( p )
  + \frac{T ( W_{M B} ( p ) , p , 0 )^{\ast}}{W_{M B} ( p ) - M_{\rm th}} .
  \label{eqA:pphys_MB}
\end{equation}
With this equation, we can rewrite Eq.~\eqref{eq:Rcorr_MB_phys} as
\begin{align}
  & \int _{0}^{\infty} D p \langle r | p_{\rm phys} \rangle
  \langle p_{\rm phys} | M B \rangle e^{- [ W_{M B} ( p ) - M_{\rm th} ] t}
  \notag \\
  & = \int _{0}^{\infty} D p \langle r | p_{\rm phys} \rangle
  \Delta ( p ) e^{- [ W_{M B} ( p ) - M_{\rm th} ] t}
  \notag \\
  & \phantom{=}
  + \int _{0}^{\infty} D p \langle r | p_{\rm phys} \rangle
  \frac{T ( W_{M B} ( p ) , p , 0 )^{\ast}}{W_{M B} ( p ) - M_{\rm th}}
  e^{- [ W_{M B} ( p ) - M_{\rm th} ] t} .
  \label{eqA:rppMB}
\end{align}
The first term on the right-hand side becomes, owing to the property
of the delta function~\eqref{eqA:Delta} and the expression in
Eq.~\eqref{eqA:r_pphys},
\begin{align}
  & \int _{0}^{\infty} D p \langle r | p_{\rm phys} \rangle
  \Delta ( p ) e^{- [ W_{M B} ( p ) - M_{\rm th} ] t}
  \notag \\
  & = j_{0} ( 0 ) + \int _{0}^{\infty} D p^{\prime} 
  \frac{j_{0} ( p^{\prime} r ) T ( M_{\rm th} , 0 , p^{\prime} )}{M_{\rm th}
    - W_{M B} ( p^{\prime} ) + i 0}
  \notag \\
  & = 1 + \Phi ( M_{\rm th} , r ) ,
\end{align}
which coincides with the second and third terms of
Eq.~\eqref{eq:Rcorr_MB}.  Furthermore, the second term in
Eq.~\eqref{eqA:rppMB} becomes
\begin{align}
  & \int _{0}^{\infty} D p \langle r | p_{\rm phys} \rangle
  \frac{T ( W_{M B} ( p ) , p , 0 )^{\ast}}{W_{M B} ( p ) - M_{\rm th}}
  e^{- [ W_{M B} ( p ) - M_{\rm th} ] t}
  \notag \\
  & = \int _{0}^{\infty} D p \, j_{0} ( p r )
  \frac{T ( W_{M B} ( p ) , p , 0 )^{\ast}}{W_{M B} ( p ) - M_{\rm th}}
  e^{- [ W_{M B} ( p ) - M_{\rm th} ] t}
  \notag \\
  & \phantom{=} + \int _{0}^{\infty} D p \int _{0}^{\infty} D p^{\prime} 
  \frac{j_{0} ( p^{\prime} r ) T ( W_{M B} ( p ) , p , p^{\prime} )}{W_{M B} ( p )
    - W_{M B} ( p^{\prime} ) + i 0}
  \notag \\
  & \phantom{= +} \times
  \frac{T ( W_{M B} ( p ) , p , 0 )^{\ast}}{W_{M B} ( p ) - M_{\rm th}}
  e^{- [ W_{M B} ( p ) - M_{\rm th} ] t} .
  \label{eqA:2nd_term}
\end{align}
To evaluate the right-hand side of this equation further, we change
the integral variable from $p$ to $W$:
\begin{equation}
  D p = \frac{d p}{d W} \frac{p^{2}}{2 \pi ^{2}}
  \frac{m_{M} m_{B}}{\mathcal{E}_{M} ( p ) \mathcal{E}_{B} ( p )} d W
  = \frac{p_{\rm on} ( W ) m_{M} m_{B}}{2 \pi ^{2} W} d W ,
  \label{eqA:p_to_W}
\end{equation}
where $p_{\rm on} ( W )$ is the on-shell momentum defined in
Eq.~\eqref{eq:pon}.  Together with the relation for the delta function
\begin{equation}
  \Delta ( p^{\prime} - p )
  = \frac{2 \pi ^{2} W}{p m_{M} m_{B}}
  \delta ( W_{M B} ( p^{\prime} ) - W_{M B} ( p ) ) ,
\end{equation}
we transform the first term on the right-hand side of
Eq.~\eqref{eqA:2nd_term} into
\begin{align}
  & \int _{0}^{\infty} D p \, j_{0} ( p r )
  \frac{T ( W_{M B} ( p ) , p , 0 )^{\ast}}{W_{M B} ( p ) - M_{\rm th}}
  e^{- [ W_{M B} ( p ) - M_{\rm th} ] t}
  \notag \\
  & = \int _{M_{\rm th}}^{\infty} d W
  \frac{p_{\rm on} ( W ) m_{M} m_{B}}{2 \pi ^{2} W}
  j_{0} ( p_{\rm on} ( W ) r ) 
  \notag \\
  & \phantom{=} \times
  \frac{T ( W , p_{\rm on} ( W ) , 0 )^{\ast}}{W - M_{\rm th}}
  e^{- ( W - M_{\rm th} ) t}
  \notag \\
  & = \int _{M_{\rm th}}^{\infty} d W
  \frac{p_{\rm on} ( W ) m_{M} m_{B}}{2 \pi ^{2} W}
  \frac{e^{- ( W - M_{\rm th} ) t}}{W - M_{\rm th}}
  \notag \\
  & \phantom{=} \times
  \int _{0}^{\infty} D p^{\prime}
  \Delta ( p_{\rm on} ( W ) - p^{\prime} )
  j_{0} ( p^{\prime} r ) 
  T ( W , p^{\prime} , 0 )^{\ast}
  \notag \\
  & = \frac{1}{\pi} \int _{M_{\rm th}}^{\infty} d W
  \frac{e^{- ( W - M_{\rm th} ) t}}{W - M_{\rm th}}
  \int _{0}^{\infty} D p^{\prime} j_{0} ( p^{\prime} r )
  \notag \\
  & \phantom{=} \times  
  \pi \delta ( W - W_{M B} ( p^{\prime} ) )
  T ( W , p^{\prime} , 0 )^{\ast} .
  \label{eqA:2nd1st}
\end{align}
Similarly, by using the optical theorem~\eqref{eqA:opt}, we can
transform the second term on the right-hand side of
Eq.~\eqref{eqA:2nd_term} into
\begin{align}
  & \int _{0}^{\infty} D p \int _{0}^{\infty} D p^{\prime} 
  \frac{j_{0} ( p^{\prime} r ) T ( W_{M B} ( p ) , p , p^{\prime} )}{W_{M B} ( p )
    - W_{M B} ( p^{\prime} ) + i 0}
  \notag \\
  & \phantom{=} \times
  \frac{T ( W_{M B} ( p ) , p , 0 )^{\ast}}{W_{M B} ( p ) - M_{\rm th}}
  e^{- [ W_{M B} ( p ) - M_{\rm th} ] t}
  \notag \\
  & = \int _{M_{\rm th}}^{\infty} d W
  \frac{e^{- ( W - M_{\rm th} ) t}}{W - M_{\rm th}}
  \int _{0}^{\infty} D p^{\prime} j_{0} ( p^{\prime} r )
  \notag \\
  & \phantom{=} \times
  \frac{i}{2 \pi}
  \frac{T ( W , 0 , p^{\prime} ) - T ( W , p^{\prime} , 0 )^{\ast}}{W
    - W_{MB} ( p^{\prime} ) + i 0} .
  \label{eqA:2nd2nd}
\end{align}
In Eqs.~\eqref{eqA:2nd1st} and \eqref{eqA:2nd2nd}, combining the common factor yields
\begin{align}
  & \pi \delta ( W - W_{M B} ( p^{\prime} ) ) T^{\ast}
  + \frac{i}{2} \frac{T - T^{\ast}}{W - W_{M B} ( p^{\prime} ) + i 0}
  \notag \\
  & = \frac{i}{2} \mathcal{P} \frac{1}{W - W_{M B} ( p^{\prime} )}
  ( T - T^{\ast} )
  \notag \\
  & \phantom{=}
  + \frac{\pi}{2} \delta ( W - W_{M B} ( p^{\prime} ) ) ( T + T^{\ast} )
  \notag \\
  & = - \mathcal{P} \frac{1}{W - W_{M B} ( p^{\prime} )} \Im T
    + \pi \delta ( W - W_{M B} ( p^{\prime} ) ) \Re T
  \notag \\
  & = - \Im \left [ \frac{T}{W - W_{M B} ( p^{\prime} ) + i 0} \right ] ,
  \label{eqA:ImT}
\end{align}
where we abbreviated $T ( W , 0 , p^{\prime} ) = T ( W , p^{\prime} ,
0 )$ as $T$.  Therefore, we obtain
\begin{align}
  & \int _{0}^{\infty} D p^{\prime} \, j_{0} ( p^{\prime} r )
  \left [ \pi \delta ( W - W_{M B} ( p^{\prime} ) )
    T ( W , p^{\prime} , 0 )^{\ast} 
    \phantom{\frac{T}{W}} \right .
  \notag \\
  & \phantom{=}
  \left . + \frac{i}{2} \frac{T ( W , 0 , p^{\prime} )
    - T ( W , p^{\prime} , 0 )^{\ast}}{W - W_{M B} ( p^{\prime} ) + i 0}
  \right ]
  \notag \\
  & = - \int _{0}^{\infty} D p^{\prime} \, j_{0} ( p^{\prime} r )
  \Im \left [ \frac{T ( W , 0 , p^{\prime} )}{W - W_{M B} ( p^{\prime} )
      + i 0} \right ]
  \notag \\
  & = - \Im \Phi ( W , r ) .
\end{align}
Finally, the right-hand side of Eq.~\eqref{eqA:2nd_term} simplifies to
\begin{align}
  & \int _{0}^{\infty} D p \langle r | p_{\rm phys} \rangle
  \frac{T ( W_{M B} ( p ) , p , 0 )^{\ast}}{W_{M B} ( p ) - M_{\rm th}}
  e^{- [ W_{M B} ( p ) - M_{\rm th} ] t}
  \notag \\
  & = - \frac{1}{\pi} \int _{M_{\rm th}}^{\infty} d W
  \frac{e^{- ( W - M_{\rm th} ) t}}{W - M_{\rm th}}
  \Im \Phi ( W , r ) ,
\end{align}
which coincides with the fourth term of Eq.~\eqref{eq:Rcorr_MB}.  This proves the equivalence between Eqs.~\eqref{eq:Rcorr_MB} and
\eqref{eq:Rcorr_MB_phys}, apart from the bound-state contribution.

\subsection{Bare to meson-baryon transition}

Next, we consider the transition from the bare state to the
meson-baryon state and prove the equivalence between
Eqs.~\eqref{eq:Rcorr_bare} and \eqref{eq:Rcorr_bare_phys}.  Again, we
start with Eq.~\eqref{eq:Rcorr_bare_phys}, specifically the term
containing $\langle r | p_{\rm phys} \rangle \langle p_{\rm phys} |
\Lambda _{0} \rangle$, and arrive at Eq.~\eqref{eq:Rcorr_bare} by
rewriting Eq.~\eqref{eq:Rcorr_bare_phys}.

For $\langle r | p_{\rm phys} \rangle$, we use
formula~\eqref{eqA:r_pphys}.  On the other hand, we calculate $\langle
p_{\text{phys}} | \Lambda _{0} \rangle$ using
formula~\eqref{eqA:pphys}, which results in
\begin{equation}
  \langle p_{\text{phys}} | \Lambda _{0} \rangle
  = \frac{\Gamma ( W_{MB} ( p ) , p )^{\ast}}{W_{MB} ( p ) - M_{0}} .
\end{equation}
Therefore, we can transform Eq.~\eqref{eq:Rcorr_bare_phys} into
\begin{align}
  & \int _{0}^{\infty} D p \langle r | p_{\rm phys} \rangle
  \langle p_{\rm phys} | \Lambda _{0} \rangle e^{- [ W_{M B} ( p ) - M_{\rm th} ] t}
  \notag \\
  & = \int _{0}^{\infty} D p \, j_{0} ( p r )
  \frac{\Gamma ( W_{M B} ( p ) , p )^{\ast}}{W_{M B} ( p ) - M_{0}}
  e^{- [ W_{M B} ( p ) - M_{\rm th} ] t}
  \notag \\
  & \phantom{=} + \int _{0}^{\infty} D p \int _{0}^{\infty} D p^{\prime} 
  \frac{j_{0} ( p^{\prime} r ) T ( W_{M B} ( p ) , p , p^{\prime} )}{W_{M B} ( p )
    - W_{M B} ( p^{\prime} ) + i 0}
  \notag \\
  & \phantom{= +} \times
  \frac{\Gamma ( W_{M B} ( p ) , p )^{\ast}}{W_{M B} ( p ) - M_{0}}
  e^{- [ W_{M B} ( p ) - M_{\rm th} ] t} .
  \label{eqA:pphys_Lam}
\end{align}
The first term can be rewritten in the same manner as in the
meson-baryon elastic scattering case [see Eq.~\eqref{eqA:2nd1st}]:
\begin{align}
  & \int _{0}^{\infty} D p \, j_{0} ( p r )
  \frac{\Gamma ( W_{M B} ( p ) , p )^{\ast}}{W_{M B} ( p ) - M_{0}}
  e^{- [ W_{M B} ( p ) - M_{\rm th} ] t}
  \notag \\
  & = \frac{1}{\pi} \int _{M_{\rm th}}^{\infty} d W
  \frac{e^{- ( W - M_{\rm th} ) t}}{W - M_{0}}
  \int _{0}^{\infty} D p^{\prime} j_{0} ( p^{\prime} r )
  \notag \\
  & \phantom{=} \times  
  \pi \delta ( W - W_{M B} ( p^{\prime} ) )
  \Gamma ( W , p^{\prime} )^{\ast} .
\end{align}
Furthermore, for the second term of Eq.~\eqref{eqA:pphys_Lam}, we can
utilize the optical theorem in Eq.~\eqref{eqA:opt_Gamma}.  With this
optical theorem and the change of variable from $p$
to $W$~\eqref{eqA:p_to_W}, we rewrite the second term of
Eq.~\eqref{eqA:pphys_Lam} as [see Eq.~\eqref{eqA:2nd2nd}]
\begin{align}
  & \int _{0}^{\infty} D p \int _{0}^{\infty} D p^{\prime} 
  \frac{j_{0} ( p^{\prime} r ) T ( W_{M B} ( p ) , p , p^{\prime} )}{W_{M B} ( p )
    - W_{M B} ( p^{\prime} ) + i 0}
  \notag \\
  & \phantom{=} \times
  \frac{\Gamma ( W_{M B} ( p ) , p )^{\ast}}{W_{M B} ( p ) - M_{0}}
  e^{- [ W_{M B} ( p ) - M_{\rm th} ] t}
  \notag \\
  & = 
  \int _{M_{\rm th}}^{\infty} d W
  \frac{e^{- ( W - M_{\rm th} ) t}}{W - M_{0}}
  \int _{0}^{\infty} D p^{\prime} j_{0} ( p^{\prime} r )
  \notag \\
  & \phantom{=} \times
  \frac{i}{2 \pi}
  \frac{\Gamma ( W , p^{\prime} ) - \Gamma ( W , p^{\prime} )^{\ast}}{W
    - W_{MB} ( p^{\prime} ) + i 0} .
\end{align}
Now, following the same steps as in Eq.~\eqref{eqA:ImT} but with
$T$ replaced by $\Gamma$, we obtain
\begin{align}
  & \int _{0}^{\infty} D p^{\prime} \, j_{0} ( p^{\prime} r )
  \left [ \pi \delta ( W - W_{M B} ( p^{\prime} ) )
    \Gamma ( W , p^{\prime} )^{\ast} 
    \phantom{\frac{T}{W}} \right .
  \notag \\
  & \phantom{=}
  \left . + \frac{i}{2} \frac{\Gamma ( W , p^{\prime} )
    - \Gamma ( W , p^{\prime} )^{\ast}}{W - W_{M B} ( p^{\prime} ) + i 0}
  \right ]
  \notag \\
  & = - \int _{0}^{\infty} D p^{\prime} \, j_{0} ( p^{\prime} r )
  \Im \left [ \frac{\Gamma ( W , p^{\prime} )}{W - W_{M B} ( p^{\prime} )
      + i 0} \right ]
  \notag \\
  & = - \Im \Phi _{0} ( W , r ) .
\end{align}
Consequently, we finally obtain
\begin{align}
  & \int _{0}^{\infty} D p
  \langle r | p_{\text{phys}} \rangle
  \langle p_{\text{phys}} | \Lambda _{0} \rangle
  e^{- [ W_{MB} ( p ) - M_{\rm th} ] t}
  \notag \\
  & = - \frac{1}{\pi} \int _{M_{\rm th}}^{\infty} d W 
  e^{- (W - M_{\rm th} ) t} \Im \Phi _{0} ( W , r ) .
\end{align}
This completes the proof of equivalence between Eqs.~\eqref{eq:Rcorr_bare} and
\eqref{eq:Rcorr_bare_phys}, except for the bound-state contribution.

\section{\boldmath $A_{1}^{+}$ projection}
\label{app:projection}

On the lattice, we perform the $A_{1}^{+}$ projection.  In this
projection, we employ the 24 elements of the cubic symmetry group $G = ( g_{1} , g_{2} , \ldots , g_{24} )$:
\begin{widetext}
  \begin{equation*}
    \begin{split}
  & g_{1} =
  \begin{pmatrix}
    1 & 0 & 0
    \\
    0 & 1 & 0
    \\
    0 & 0 & 1
  \end{pmatrix} ,
  \quad 
  g_{2} =
  \begin{pmatrix}
    1 & 0 & 0
    \\
    0 & 0 & -1
    \\
    0 & 1 & 0 
  \end{pmatrix} ,
  \quad 
  g_{3} =
  \begin{pmatrix}
    1 & 0 & 0
    \\
    0 & -1 & 0 
    \\
    0 & 0 & -1 
  \end{pmatrix} ,
  \quad 
  g_{4} =
  \begin{pmatrix}
    1 & 0 & 0
    \\
    0 & 0 & 1
    \\
    0 & -1 & 0 
  \end{pmatrix} ,
  \quad 
  g_{5} =
  \begin{pmatrix}
    0 & 0 & 1
    \\
    0 & 1 & 0 
    \\
    -1 & 0 & 0
  \end{pmatrix} ,
  \\ & 
  g_{6} =
  \begin{pmatrix}
    -1 & 0 & 0
    \\
    0 & 1 & 0 
    \\
    0 & 0 & -1 
  \end{pmatrix} ,
  \quad 
  g_{7} =
  \begin{pmatrix}
    0 & 0 & -1
    \\
    0 & 1 & 0 
    \\
    1 & 0 & 0
  \end{pmatrix} ,
  \quad 
  g_{8} =
  \begin{pmatrix}
    0 & -1 & 0 
    \\
    1 & 0 & 0
    \\
    0 & 0 & 1
  \end{pmatrix} ,
  \quad 
  g_{9} =
  \begin{pmatrix}
    -1 & 0 & 0
    \\
    0 & -1 & 0 
    \\
    0 & 0 & 1 
  \end{pmatrix} ,
  \quad 
  g_{10} =
  \begin{pmatrix}
    0 & 1 & 0 
    \\
    -1 & 0 & 0
    \\
    0 & 0 & 1
  \end{pmatrix} ,
  \\ & 
  g_{11} =
  \begin{pmatrix}
    -1 & 0 & 0
    \\
    0 & 0 & -1
    \\
    0 & -1 & 0 
  \end{pmatrix} ,
  \quad 
  g_{12} =
  \begin{pmatrix}
    -1 & 0 & 0
    \\
    0 & 0 & 1 
    \\
    0 & 1 & 0 
  \end{pmatrix} ,
  \quad 
  g_{13} =
  \begin{pmatrix}
    0 & 0 & -1
    \\
    0 & -1 & 0 
    \\
    -1 & 0 & 0
  \end{pmatrix} ,
  \quad 
  g_{14} =
  \begin{pmatrix}
    0 & 0 & 1
    \\
    0 & -1 & 0 
    \\
    1 & 0 & 0
  \end{pmatrix} ,
  \quad 
  g_{15} =
  \begin{pmatrix}
    0 & -1 & 0 
    \\
    -1 & 0 & 0
    \\
    0 & 0 & -1 
  \end{pmatrix} ,
  \\ &
  g_{16} =
  \begin{pmatrix}
    0 & 1 & 0 
    \\
    1 & 0 & 0
    \\
    0 & 0 & -1
  \end{pmatrix} ,
  \quad 
  g_{17} =
  \begin{pmatrix}
    0 & -1 & 0 
    \\
    0 & 0 & -1
    \\
    1 & 0 & 0
  \end{pmatrix} ,
  \quad 
  g_{18} =
  \begin{pmatrix}
    0 & 0 & 1 
    \\
    -1 & 0 & 0
    \\
    0 & -1 & 0 
  \end{pmatrix} ,
  \quad 
  g_{19} =
  \begin{pmatrix}
    0 & 0 & -1
    \\
    -1 & 0 & 0
    \\
    0 & 1 & 0 
  \end{pmatrix} ,
  \quad 
  g_{20} =
  \begin{pmatrix}
    0 & -1 & 0 
    \\
    0 & 0 & 1
    \\
    -1 & 0 & 0
  \end{pmatrix} ,
 \\     & g_{21} =
  \begin{pmatrix}
    0 & 0 & 1 
    \\
    1 & 0 & 0
    \\
    0 & 1 & 0 
  \end{pmatrix} ,
  \quad 
  g_{22} =
  \begin{pmatrix}
    0 & 1 & 0 
    \\
    0 & 0 & 1
    \\
    1 & 0 & 0
  \end{pmatrix} ,
  \quad 
  g_{23} =
  \begin{pmatrix}
    0 & 1 & 0 
    \\
    0 & 0 & -1
    \\
    -1 & 0 & 0
  \end{pmatrix} ,
  \quad 
  g_{24} =
  \begin{pmatrix}
    0 & 0 & -1 
    \\
    1 & 0 & 0
    \\
    0 & -1 & 0 
  \end{pmatrix} .
  \end{split}
\end{equation*}
\end{widetext}
These elements comprise the identity ($g_{1}$), rotations by
$\ang{90}$, $\ang{180}$, and $\ang{270}$ about axes passing through
the centers of face planes ($g_{2}$--$g_{10}$), rotations by
$\ang{180}$ around axes passing through the midpoints of edges
($g_{11}$--$g_{16}$), and rotations by $\ang{120}$ and $\ang{240}$
around axes passing through diagonal vertices ($g_{17}$--$g_{24}$).

Because $g_{i}$ forms a group $G$, for any arbitrary $g_{j}$ with fixed $j$,
both $g_{i} g_{j}$ and $g_{j} g_{i}$ span all 24 elements as
$i$ varies.  Therefore, for a function $F_{\bm{n}}$ depending on
an integer vector $\bm{n}$, we have
\begin{equation}
  \sum _{g \in G} F_{g \bm{n}}
  = \sum _{g \in G} F_{g_{j} g \bm{n}}
  = \sum _{g \in G} F_{g g_{j} \bm{n}} ,
  \label{eqA:fg}
\end{equation}
where $g_{j}$ is an element of $G$.  We define the $A_{1}^{+}$
projection $\tilde{\mathcal{H}}$ for a scalar function $H_{\bm{n} ,
  \bm{n}^{\prime}}$ containing the inner product $\bm{n} \cdot
\bm{n}^{\prime}$ as
\begin{equation}
  \tilde{\mathcal{H}}_{\alpha , \alpha ^{\prime}}
  \equiv
  \frac{1}{24} \sum _{g \in G} H_{g \bm{n} , \bm{n}^{\prime}} .
\end{equation}
This $\tilde{\mathcal{H}}$ can be labeled by $\alpha$ and
$\alpha ^{\prime}$, to which the vectors $\bm{n}$ and $\bm{n}^{\prime}$
belong, respectively, because Eq.~\eqref{eqA:fg} implies the
relation
\begin{align}
  \sum _{g \in G} H_{g \bm{n} , \bm{n}^{\prime}}
  & = \sum _{g \in G} H_{g g_{i} \bm{n} , \bm{n}^{\prime}}
  = \sum _{g \in G} H_{g_{j}^{-1} g g_{i} \bm{n} , \bm{n}^{\prime}}
  \notag \\
  & = \sum _{g \in G} H_{g g_{i} \bm{n} , g_{j} \bm{n}^{\prime}} ,
  \label{eqA:formula_2}
\end{align}
for fixed $g_{i}$ and $g_{j}$ in $G$.

By using these formulae for the $A_{1}^{+}$ projection, we can
simplify the Lippmann--Schwinger equation in lattice space, which is
expressed in a general form as
\begin{equation}
  \tilde{T}_{\bm{n} , \bm{n}^{\prime}}
  = \tilde{V}_{\bm{n} , \bm{n}^{\prime}}
  + \frac{1}{L^{3}} \sum _{\bm{n}^{\prime \prime}}
  \frac{\tilde{V}_{\bm{n} , \bm{n}^{\prime \prime}}
    \tilde{T}_{\bm{n}^{\prime \prime} , \bm{n}^{\prime}}}
       {W - W_{MB} ( 2 \pi \bm{n}^{\prime \prime} / L )} .
  \label{eqA:LSeq_lat}
\end{equation}
Here, we have neglected the bare state $\Lambda _{0}$ and omitted an
irrelevant factor $m_{M} m_{B} / ( \mathcal{E}_{M} \mathcal{E}_{B} )$.
To restore this factor in the equations below, one multiplies $m_{M}
m_{B} / ( \mathcal{E}_{M} \mathcal{E}_{B} )$ immediately after $1 /
L^{3}$ (or the summation symbol following $1 / L^{3}$).  The
$A_{1}^{+}$-projected scattering amplitude and interaction are given
by
\begin{equation}
  \tilde{\mathcal{T}}_{\alpha , \alpha ^{\prime}}
  \equiv \frac{1}{24} \sum _{g \in G} \tilde{T}_{g \bm{n} , \bm{n}^{\prime}} ,
  \quad
  \tilde{\mathcal{V}}_{\alpha , \alpha ^{\prime}}
  \equiv \frac{1}{24} \sum _{g \in G} \tilde{V}_{g \bm{n} , \bm{n}^{\prime}} .
  \label{eqA:Taaprime}
\end{equation}
Owing to formula~\eqref{eq:formula_1}, the second term on the
right-hand side of Eq.~\eqref{eqA:LSeq_lat} becomes
\begin{align}
  & \frac{1}{L^{3}} \sum _{\bm{n}^{\prime \prime}}
  \frac{\tilde{V}_{\bm{n} , \bm{n}^{\prime \prime}}
    \tilde{T}_{\bm{n}^{\prime \prime} , \bm{n}^{\prime}}}{W - W_{MB} ( 2 \pi \bm{n}^{\prime \prime} / L )}
  \notag \\
  & = 
  \frac{1}{L^{3}} \sum _{\alpha ^{\prime \prime}} \frac{v_{\alpha ^{\prime \prime}}}{24}
  \sum _{g^{\prime \prime} \in G}
  \frac{\tilde{V}_{\bm{n} , g^{\prime \prime} \bm{n}^{\prime \prime}}
    \tilde{T}_{g^{\prime \prime} \bm{n}^{\prime \prime} , \bm{n}^{\prime}}}{W
    - W_{MB} ( p_{\alpha ^{\prime \prime}} )}
\end{align}
where $W_{MB} ( 2 \pi \bm{n}^{\prime \prime} / L )$ is replaced by
$W_{MB} ( p_{\alpha ^{\prime \prime}} )$.  Then, we take the average with
respect to the angle of $\bm{n}$ for the $A_{1}^{+}$-projected
interaction~\eqref{eqA:Taaprime}.  By using
formula~\eqref{eqA:formula_2}, we obtain
\begin{align}
  & \frac{1}{24} \sum _{g \in G}
  \frac{1}{L^{3}} \sum _{\alpha ^{\prime \prime}} \frac{v_{\alpha ^{\prime \prime}}}{24}
  \sum _{g^{\prime \prime} \in G}
  \frac{\tilde{V}_{g \bm{n} , g^{\prime \prime} \bm{n}^{\prime \prime}}
    \tilde{T}_{g^{\prime \prime} \bm{n}^{\prime \prime} , \bm{n}^{\prime}}}{W
    - W_{MB} ( p_{\alpha ^{\prime \prime}} )}
  \notag \\
  & = 
  \frac{1}{L^{3}} \sum _{\alpha ^{\prime \prime}} \frac{v_{\alpha ^{\prime \prime}}}{24}
  \sum _{g^{\prime \prime} \in G}
  \frac{\tilde{\mathcal{V}}_{\alpha , \alpha ^{\prime \prime}}
    \tilde{T}_{g^{\prime \prime} \bm{n}^{\prime \prime} , \bm{n}^{\prime}}}{W
    - W_{MB} ( p_{\alpha ^{\prime \prime}} )}
  \notag \\
  & = 
  \frac{1}{L^{3}} \sum _{\alpha ^{\prime \prime}}
  \frac{v_{\alpha ^{\prime \prime}} \tilde{\mathcal{V}}_{\alpha , \alpha ^{\prime \prime}}}{W
    - W_{MB} ( p_{\alpha ^{\prime \prime}} )}
  \cdot \frac{1}{24}
  \sum _{g^{\prime \prime} \in G}
  \tilde{T}_{g^{\prime \prime} \bm{n}^{\prime \prime} , \bm{n}^{\prime}}
  \notag \\
  & = 
  \frac{1}{L^{3}} \sum _{a^{\prime \prime}}
  \frac{v_{\alpha ^{\prime \prime}} \tilde{\mathcal{V}}_{\alpha , \alpha ^{\prime \prime}}
    \tilde{\mathcal{T}}_{\alpha ^{\prime \prime}, \alpha ^{\prime}}}{W
    - W_{MB} ( p_{\alpha ^{\prime \prime}} )} .
\end{align}
Finally, we obtain
\begin{equation}
  \tilde{\mathcal{T}}_{\alpha , \alpha ^{\prime}}
  = \tilde{\mathcal{V}}_{\alpha, \alpha ^{\prime}} + \sum _{\alpha ^{\prime \prime}} 
  \tilde{\mathcal{V}}_{\alpha , \alpha ^{\prime \prime}} \tilde{G}_{\alpha ^{\prime \prime}}
  \tilde{\mathcal{T}}_{\alpha ^{\prime \prime}, \alpha ^{\prime}}
\end{equation}
\begin{equation}
  \tilde{\mathcal{G}}_{\alpha}
  \equiv \frac{v_{\alpha}}{L^{3}} \frac{1}{W - W_{MB} ( p_{\alpha} )}
\end{equation}
which corresponds to Eqs.~\eqref{eq:Ttilde} and \eqref{eq:Gtilde}.

\end{document}